\documentstyle[12pt,epsf,epsfig]{article}
\def\be{\begin{equation}}
\def\ee{\end{equation}}
\def\ba{\begin{eqnarray}}
\def\ea{\end{eqnarray}}

\def\bdm{\begin{displaymath}}
\def\edm{\end{displaymath}}
\def\la{~\mbox{\raisebox{-.6ex}{$\stackrel{<}{\sim}$}}~}
\def\ga{~\mbox{\raisebox{-.6ex}{$\stackrel{>}{\sim}$}}~}
\def\bq{\begin{quote}}
\def\eq{\end{quote}}

 at 10truept

\newcommand{\beq}{\begin{equation}}
\newcommand{\eeq}{\end{equation}}
\newcommand{\beqa}{\begin{eqnarray}}
\newcommand{\eeqa}{\end{eqnarray}}

\def\la{~\mbox{\raisebox{-.6ex}{$\stackrel{<}{\sim}$}}~}
\def\ga{~\mbox{\raisebox{-.6ex}{$\stackrel{>}{\sim}$}}~}

\def\ltap{\ \raise.3ex\hbox{$<$\kern-.75em\lower1ex\hbox{$\sim$}}\ }
\def\gtap{\ \raise.3ex\hbox{$>$\kern-.75em\lower1ex\hbox{$\sim$}}\ }
\def\gl{\ \raise.5ex\hbox{$>$}\kern-.8em\lower.5ex\hbox{$<$}\ }
\def\roughly#1{\raise.3ex\hbox{$#1$\kern-.75em\lower1ex\hbox{$\sim$}}}

\begin{document}

\thispagestyle{empty}
\begin{flushright}
hep-th/0703190\\ March 2007
\end{flushright}
\vspace*{.8cm}
\begin{center}
{\Large \bf Charting the Landscape of Modified Gravity}\\

\vspace*{1.5cm} {\large Nemanja Kaloper\footnote{\tt
kaloper@physics.ucdavis.edu} and
Derrick Kiley\footnote{\tt dtkiley@physics.ucdavis.edu}}\\
\vspace{.5cm} {\em Department of Physics, University of
California, Davis,
CA 95616}\\
\vspace{.15cm} \vspace{1.2cm} ABSTRACT
\end{center}

We explore brane induced gravity on a 3-brane in six locally flat
dimensions. To regulate the short distance singularities in the
brane core, we resolve the thin brane by a cylindrical 4-brane,
with the geometry of $4D$ Minkowski $\times$ a circle, which has
an axion flux to cancel the vacuum pressure in the compact
direction. We discover a large diversity of possible solutions
controlled by the axion flux, as governed by its boundary
conditions. Hence brane induced gravity models really give rise to
a {\it landscape} of vacua, at least semiclassically. For sub-critical tensions, the
crossover scale, below which gravity may look $4D$, and the
effective $4D$ gravitational coupling are sensitive to vacuum
energy. This shows how the vacuum energy problem manifests in
brane induced gravity: instead of tuning the $4D$ curvature,
generically one must tune the crossover scale. On the other hand,
in the near-critical limit, branes live inside very deep throats
which efficiently compactify the angular dimension. In there, $4D$
gravity first changes to $5D$, and only later to $6D$. The
crossover scale saturates at the gravitational see-saw scale,
independent of the tension. Using the fields of static loops on a
wrapped brane, we check the perturbative description of long range
gravity below the crossover scale. In sub-critical cases the
scalars are strongly coupled already at the crossover scale even
in the vacuum, because the brane bending is turned on by the axion
flux. Near the critical limit, linearized perturbation theory
remains under control below the crossover scale, and we find that
linearized gravity around the vacuum looks like a scalar-tensor
theory.

\vfill \setcounter{page}{0} \setcounter{footnote}{0}
\newpage

\section{Introduction}

\subsection{Prologue}

The mystery of the cosmological constant is arguably the most
pressing and puzzling problem in contemporary fundamental physics.
To date, the attempts to explain a small cosmological constant in
the framework of effective field theory formulation of matter
coupled to gravity, based on conventional lore of naturalness
alone, have not yielded an answer (see the classic work
\cite{Wein} which still provides a state-of-the-art review). On
the other hand, cosmological observations \cite{sne} strongly
suggesting a small cosmological constant, and the discovery of the
landscape of string vacua \cite{landscape}, have lent support to
the idea of statistical selection of vacua, and the value of the
cosmological constant in them \cite{andrei}-\cite{tom}, prompting
a new debate about anthropic reasoning in physics.

One may try to alter the problem by embedding our universe in a
fundamentally higher-dimensional space-time in some way
\cite{rusha}-\cite{selftun}. If, for example, our universe is a
brane in extra dimensions, it would also have extrinsic curvature,
and then one may divert the brane vacuum energy into the extrinsic
curvature \cite{selftun}. In this way, the vacuum energy could
remain invisible to long distance $4D$ gravity, yielding a very
weakly curved universe. However, while attempting to recover $4D$
General Relativity at large distances by compactifying the bulk,
with covariantly conserved  sources, one needed additional branes
in the bulk, that restore a fine-tuning similar to the standard
$4D$ one \cite{selftun,nillest}. The resulting picture was again
that of a landscape: a theory admitting multiple $4D$ vacua,
classified by the free integration constants for some of the bulk
fields (see, e.g.. \cite{polstra}). Using these integration
constants, one could hope to derive a framework where the
effective cosmological constant could change in small amounts from
one background to another. But one gains no more than that: the
landscape reemerges, and one must retreat to either statistical or
anthropic arguments to select a phenomenologically viable ground
state.

A different idea for recovering $4D$ General Relativity at large
distances has been pursued in the so-called brane induced gravity
theory \cite{DGP}. In more interesting variants of brane-induced
gravity, the extra-dimensional space has infinite volume. When the
bulk volume is infinite, the $4D$ graviton zero mode is not
normalizable, and hence it completely decouples. Thus $4D$ gravity
ought to emerge from the exchange of the continuum of bulk modes,
which are all massive from the $4D$ viewpoint. To ensure this,
\cite{DGP} introduced induced curvature terms on the brane,
arguing that they will be generated by brane quantum corrections
anyway. If the scale $M_4$ that normalizes them is much greater
than the bulk Planck scale, the brane kinetic terms would pull the
bulk gravitons with wavelengths shorter than a certain crossover
scale $r_c$ very close to the brane, yielding the momentum
transfer due to the scattering of virtual bulk gravitons $\propto
1/p^{2}$ for the momenta $p > r_c{}^{-1}$, and a very small
coupling of the three-point vertex $\sim 1/M_4$. This gives a
force which scales as $1/r^{2}$, simulating $4D$ behavior
\cite{DGP}, and is nicely illustrated in the exact gravitational
shock waves of \cite{shocks}.

This force contains admixtures of longitudinal, helicity-0
gravitons, and sometimes also radions. They couple gravitationally
and their long range contributions to the force never completely
disappears in perturbation theory \cite{vdvz}. However it has been
argued, for massive gravity \cite{veinsh} and similarly for brane
induced gravity \cite{strongcouplings}, that the theory goes
nonlinear at large distances from the source, and that the
nonlinearities might screen away the extra scalar modes. On the
other hand, this also indicates \cite{niges}-\cite{nira} that
brane induced gravity runs into a strong coupling regime at
macroscopic distances $r_{\rm strong} \sim (r_c^2/M_4)^{1/3}$,
requiring very low scale UV completions, and forcing the question
of whether such completions even exist! The extra scalars may also
turn into a perturbative ghost on self-accelerating backgrounds in
$5D$ bulks \cite{Luty,koyama,cgkp,giigl,tanko,minjoon}. This suggests that
the self-accelerating backgrounds are unstable, but the strong
coupling at subhorizon scales obstruct  the exploration of their
fate in perturbation theory. Still, the shock waves
\cite{shocks,cgkp} can identify energy leaks from a
self-accelerating brane into the bulk, that render the dynamics
manifestly different from a $4D$ one.

Similar mechanisms may also operate on higher-codimension defects
\cite{DGP,highercod}. The authors of \cite{highercod} have pursued
ideas how UV regularizations of gravity might affect the crossover
scale and the IR spectrum of the theory, hoping to suppress the
extra graviton helicites. Subsequent discussions involved
arguments as to how some regularizations may suffer from ghosts
\cite{durub}, which precisely cancel the extra graviton
helicities, and how to avoid them \cite{koporo,gashif}. The
explicit constructions are hindered by the short-distance
singularities in the core of the defect, which must be regulated
before one can reliably calculate the low energy behavior of
gravity.

We feel that exploring brane induced gravity on codimension-2
brane models is particularly interesting. In this case we can find
the exact background solutions that play the role of the $4D$
vacua directly, thanks to the magic of gravity in $3D$. It is well
known from early braneworld analysis that $3$-branes in $6D$ need
not locally curve the bulk even if they carry tension
\cite{raman}. Although this leads to flat $4D$ geometries for
nonvanishing tensions, it does not help with the cosmological
constant problem after one compactifies the bulk to get long range
$4D$ gravity. This was immediately realized already in
\cite{raman} to be in agreement with Weinberg's no-go theorem
\cite{Wein}, and was discussed in more detail in \cite{nilles}.
Still, many interesting properties of gravity localized on
codimension-2 defects, flat or not, in a $6D$ bulk, compact and
not, have been since explored in \cite{popo}-\cite{papakoba},
including exact black holes and gravitational shock waves
straddling the brane \cite{kalkil}. The sheer proliferation of
such configurations points to a structure as rich as the landscape
of supergravity solutions, albeit more exotic. Sketching a chart
of this realm is the primary aim of the present paper, as we now
elaborate.

\subsection{The Landscape of Brane Induced Gravity}

We begin our foray into the landscape of brane induced gravity by
exploring thin static codimension-2 branes, which exist for
sub-critical tensions $\lambda < \lambda_c = 2\pi M^4_6$ and have
a flat $4D$ induced metric. Because the brane induced curvature
terms identically vanish, these solutions are identical to the
ones in $6D$ General Relativity with an empty bulk. We then
construct {\it exact} gravitational shock waves of relativistic
particles on the brane, generalizing the solutions of
\cite{kalkil} to the framework of brane induced gravity. The
shocks confirm that brane induced gravity on singular defects
remains $4D$ all the way to infinity \cite{DGP,highercod}.

Yet, these solutions are pathological. Even an infinitesimal
displacement of the probe from the brane into the bulk reveals
that the thin brane theory is singular, since the gravitational
field jumps from a finite value to exactly zero everywhere in the
bulk \cite{DGP,highercod}. Mathematically, the discontinuities on
thin branes can be attributed to an {\it ill-posed exterior
boundary value problem}: modelling brane sources by points on an
infinite cone introduces singularities both near and far. One
turns it into a well-posed problem simply by replacing the point
sources with finite rings, and considering the combined interior
and exterior problems. Physically, this amounts to smearing the
source over a finite space. We do it by replacing the thin 3-brane
by a 4-brane wrapped on a circle, by adding an axionic field on
the brane, whose {\it vev} breaks translations along the circle
\cite{msled,pesota}. This is similar to the Scherk-Schwarz
mechanism \cite{schsch} for supersymmetry breaking and mass
generation by dimensional reduction, an important ingredient of
the string landscape model-building \cite{landscape}. Much like in
the string landscape, the axion `charge' determines the radius of
the compact circle $r_0$ for a given value of the tension, and we
must choose it carefully to get the desired value of $r_0$.
Alternatively, if some other dynamics were to fix $r_0$, we need
to tune the axion charge to precisely cancel the pressure of the
brane tension along the circle, because otherwise the background
could not simulate the $4D$ Minkowski vacuum from the get go.

The thick branes will still have a flat static geometry with an
infinite bulk, but only for tensions smaller than the critical
value. When the tension exceeds the critical value, our solutions
show how the bulk compactifies and develops a naked singularity
that soaks up the brane's gravitational field lines, just like the
supercritical cosmic string in $4D$ \cite{naked}. These solutions
look like singular teardrop compactifications analyzed by
Gell-Mann and Zwiebach in \cite{gell}. As with $4D$ supercritical
defects \cite{naked}, one expects that for supercritical branes,
as well as for the cases with a mismatch between the tension and
the axion charge, there exist nonsingular solutions which describe
some curved, nonstationary backgrounds. Examples which describe
topologically inflating defects have been found recently in
\cite{oriol}.

We next calculate the crossover scale where gravity changes from
$4D$, again using the shock wave solutions, sourced by
relativistic strings moving on the wrapped brane. Below the
crossover scale, the shock wave indeed approximates the $4D$
Aichelburg-Sexl solution \cite{aichsexl} down to $r_0$, whereas at much larger
distances it changes over to the $6D$ shock of
\cite{higherdwaves}, for sub-critical branes. The physics of
crossover is {\it very intricate}: it is controlled by the deficit
angle of the vacuum solution. This is because of the `lightning
rod' amplification of the bulk gravitational coupling, leading to
$M_{6 \, {\rm eff}}^4 = (1-b) M_6^4$ \cite{kalkil}, where $1-b =
1-\frac{2\lambda_5 r_0}{M_6^4}$ measures the deficit angle of the
brane. It is intuitively clear why it affects the gravitational
force: on a cone, gravitational field lines spread more slowly
than on a plane, and hence gravity must look stronger
\cite{kalkil}. For sub-critical strings with deficit angle smaller
than $2\pi$, our exact crossover scale is qualitatively very
different from the see-saw scale of \cite{highercod}, yielding
instead $r^2_c  \sim \frac{M^2_4}{M_{6 \, {\rm eff}}^4}$, up to a log,
where $M_4$ is the effective $4D$ Planck scale found by compactifying
the wrapped 4-brane theory on the circle of radius $r_0$.

On the other hand, in the near-critical limit the bulk compactifies  to a
cylinder, which opens up into a cone only very far from the brane, at
distances $\ga \frac{r_0}{1-b}$.
In the cylindrical throat gravity is in a $5D$ gravity regime,
separating the $4D$ and $6D$ ones. Thus the crossover scale beyond
which gravity is not $4D$ is really the demarcation between the
$4D$ regime and $5D$ gravity which lives inside the throat
\cite{kalwall}. To see where this happens, we can use the naive
crossover formula on a codimension-1 brane, realized as a wrapped
4-brane in a $6D$ flat space with a compact circle: by Gauss
formulas for Planck masses $M_{4 \, eff}^2 = M_5^3 r_0$ and $M_{5
\, eff}^3 = M_6^4 r_0$, we find exactly $r_c \sim \frac{M_{4 \,
eff}^2}{M_{5 \, eff}^3} \sim \frac{M^3_5}{M^4_{6 }}$, which is our
crossover formula from the exact shock waves, and which is the
same as the see-saw scale $r_c \sim \frac{M^2_4}{M^4_{6}r_0}$ of
\cite{highercod}. Thus we see precisely {\it how} the see-saw
mechanism emerges.

Our analysis reveals the presence of a large topographic diversity
of solutions. The axion flux which cancels the pressure in the
compact direction sets the radius of the circle by $q^2 =
2\lambda_5 r_0^2$. Since the $4D$ Planck scale is $M^2_4 \sim
M_5^3 r_0 \sim \frac{M_5^3q}{\sqrt{\lambda_5}}$, and, for
sub-critical branes, the crossover scale $r_c \sim \frac{M_5^3
q}{\sqrt{2\lambda_5} M_6^4 - 2\lambda_5 q}$ we can find the charge
$q$ that will produce any desired $M_4$ for any values of $M_5$
and $\lambda_5$. But then, the crossover scale is uniquely fixed
in terms of these parameters. The vacuum energy problem {\it
reappears} in brane induced gravity: although the brane is flat,
when we fix $M_4$, we must finely tune both the axion charge and
the tension of the brane to get $r_c$ to take some desired value
on a sub-critical brane. On the other hand, once we fix
$\lambda_5$, $M_5$ and $M_6$, and let the charge $q$ vary, we will
find that the thickness, the effective $4D$ Planck scale, and, for
sub-critical branes, the crossover scale of the brane also vary,
spanning a wide range of theories below the crossover scale. The
near-critical branes, where the deficit angle approaches $2\pi$,
are more interesting since their crossover scale saturates at $r_c
\sim \frac{M_5^3}{M_6^4}$, which is completely independent of the
brane tension in the leading order. This is because their conical
throat \cite{kalwall} shields them from the asymptotic infinity.
Only their effective $4D$ Planck mass is sensitive to the brane
tension, and changes when the tension varies and the axion charge
is fixed. The charge and the tension still need to be tuned right
to get a desired value of $M_4$, but this may be substantially
easier.

This is a signature of a landscape: low energy parameters of a
theory depend on boundary conditions, instead of being uniquely
fixed by a symmetry principle. We could also imagine a brane where
$q$ and $\lambda_5$ might slowly vary along the brane, leading to
regions of $4D$ space with a different strength of simulated $4D$
gravity, but always with a flat static $4D$ vacuum for
sub-critical vacuum energies! Thus, we unveil a {\it classical
landscape} of codimension-2 brane induced gravity, where the brane
may remain flat since the curvature is soaked up by the deficit
angle, but $M_4$ and generically $r_c$ depend on the vacuum
energy.

Including quantum mechanics in this picture requires a lot of care.
An immediate, semiclassical issue is that since the axion is a phase of some complex
$5D$ scalar field, its charge is quantized in the units of a new UV mass
scale $\mu$. The quantization law, $q = \mu^{3/2} n$, where $n$
is an integer winding number, yields $M_4^2 \sim \frac{M_5^3
\mu^{3/2}}{\sqrt{\lambda_5}} n$, so that for a given set of
dimensional parameters we must choose the right winding number.
The value of $M_4$ changes discretely with it. Similarly, for
sub-critical branes, the crossover scale can only change in
discrete jumps on static backgrounds. Thus the static
configurations will only exist at discrete locations of the
landscape, most of which is covered by nonstationary solutions,
like in the string landscape with a discretuum of vacua
\cite{landscape}. Finding the nonstationary solutions is outside
of the scope of this work; we hope to return to it elsewhere. Nevertheless
as long as gravity is treated classically and field theory
is consistently cut off at some scale in the UV, this can be done in principle,
showing that the classical landscape which we glimpse at can be extended
semiclassically. However, there remains the looming question about UV completions of 
brane induced gravity  models, and their embeddings into the 
theories of quantum gravity \cite{DGP,kiritsis,lowe,ignatios}. 
We will have nothing new on this to add here, taking a more pedestrian approach of
simply charting out the (semi)classical countenance of what the 
quantum landscape might be.

The shock wave solutions have taught us that at distances between
$r_0$ and $r_c$, the theory contains $4D$ General Relativity. To
see what else is there, we develop, in painstaking detail,
linearized perturbation theory about a wrapped brane. For
sub-critical tension the scalars are strongly coupled essentially
exactly at the crossover scale $r_c$, which is the Vainshtein
scale \cite{veinsh} of the vacuum itself. This happens because the
scalar gravitons probe the asymmetric vacuum energy distribution, set up
by the background axion flux, which triggers brane bending.
However in the near-critical limit linearized perturbation theory
around the vacuum remains under control below the crossover scale.
Solving for the linearized gravitational field of a static uniform
ring of mass on the brane, we find that below the crossover scale
the theory contains helicity-0 and radion-like scalars in addition
to the helicity-2 modes. The radion modes decouple but the
helicity-0 modes remain active at the linearized level. Thus the
linearized theory around near-critical vacua approximates a Brans-Dicke
gravity with $\omega=0$. This would disagree with the classic
tests of General Relativity \cite{vdvz}, but perhaps
non-linearities or dynamics on vacua of more complex structure
could come to the rescue. We however find that rings of brane
matter built of the lightest KK states on the wrapped brane do not
entice any fast instabilities in the leading order of perturbation
theory. While we suspect that fast instabilities might remain absent
in perturbation theory beyond linear order, we have not proven it,
and it remains a question for future studies.

The paper is organized as follows: we begin with a lightning
review of thin codimension-2 flat branes and their short distance
singularities, as revealed by shock waves, in section 2. We then
construct the regulated backgrounds, with a 4-brane wrapped on a
circle, in section 3. We also discuss how critical and
supercritical branes compactify the bulk, the latter inducing a
naked singularity far away. We derive regulated shock wave
solutions in section 4, to explore the long range properties of
gravity and compute the crossover scale. In section 5 we set up
linear perturbation theory, and give the
linearized solution of a static ring of mass on the 4-brane. 
We end with conclusions.

\section{Tensional Thin Branes on Bulk Cones}

\subsection{Vacua}

The field equations describing a 3-brane in an empty $6D$ bulk in
brane induced gravity are \cite{DGP,highercod}, in a brane fixed
Gaussian-normal gauge,
\be M_6^4 G_6{}^A{}_B + M_4^2 G_{4}{}^\mu{}_\nu \, \delta^A{}_\mu
\delta^\nu{}_B \, \delta^{(2)}(\vec{y}) = T^\mu{}_\nu \,
\delta^A{}_\mu \delta^\nu{}_B \, \delta^{(2)}(\vec{y}) \, .
\label{fieldeqs} \ee
Our conventions are that the indices $A, B, \ldots$ count all
spacetime dimensions, while $\mu, \nu, \ldots$ run over the brane
worldvolume. The $\delta$-function is the tensor
$\delta^{(2)}(\vec{y}) = \frac{\sqrt{g_4}}{\sqrt{g_6}} \Pi
\delta(y_i)$, given by the normalization of the usual tensor
density by the metric determinant of the extra space,
coordinatized by $(y_1,y_2)$ (we will use $x^\mu$ for the
coordinates along the brane worldvolume). As in
\cite{DGP,highercod} we take the stress-energy tensor
$T^\mu{}_\nu$ to be localized on the brane, separating the tension
$\lambda$ from the matter sources by $T^\mu{}_\nu = -\lambda
\delta^\mu{}_\nu + \tau^\mu{}_\nu$.

\begin{figure}[thb]
\vskip1cm
\centerline{\includegraphics[width=0.4\hsize,width=0.4\vsize,angle=0]{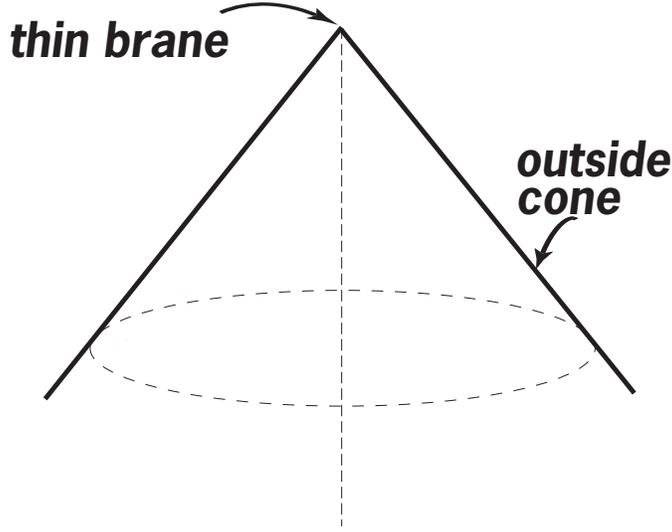}}
\caption{{$2D$ conical bulk geometry of a thin brane vacuum.}
\label{fig:cone}} \vskip.5cm
\end{figure}

The vacua of this theory are maximally symmetric solutions with
$\tau^\mu{}_\nu = 0$. Substituting this in (\ref{fieldeqs}),
imposing a flat brane $g_{\mu\nu} = \eta_{\mu\nu}$ and tracing
over, we find $R_2 = \frac{2\lambda}{M_6^4}
\delta^{(2)}(\vec y)$, where $R_2$ is the curvature in the spatial
dimensions transverse to the brane. The solution is a conical
space \cite{raman}, which can be seen most simply as follow. Since
any metric in two dimensions is conformally flat, we can write
$ds_2{}^2 = e^{- \vartheta} d\vec y^2$. Then by conformal flatness
of the metric, $R_2 = e^{\vartheta} \vec \nabla_y^2 \vartheta$,
and recalling the Euclidean $2D$ Green's function $\ln(|\vec
y|/\ell)$, which obeys $\vec \nabla_y^2 \ln(|\vec y|/\ell)= 2\pi
\Pi \delta(y_i)$, we see that to get the solution we can set
$\vartheta = 2 b \ln(|\vec y|/\ell)$ to find $R_2 = 4\pi b e^{2b
\ln(|\vec y|/\ell)}  \Pi \delta(y_i)$. We next compare this to
$R_2 = \frac{2\lambda}{M_6^4} \delta^{(2)}(\vec y) =
\frac{2\lambda}{M_6^{4}} e^{2b \ln(|\vec y|/\ell)} \Pi
\delta(y_i)$ from above, from which we finally obtain
\be
b = \frac{\lambda}{2\pi M_6^{4}} \, . \label{tendef} \ee
The arbitrary length scale $\ell$, needed for dimensional reasons,
completely drops out of these equations. Then substituting the
solution for $\vartheta$, going to spherical polar coordinates and
changing the radial coordinate to $\rho = \frac{1}{1-b} \ell^{b}
|\vec y|^{1-b}$, we get
\be
ds_2{}^2 = d\rho^2 + (1-b)^2 \rho^2 d\phi^2 \, .
\label{2dmetricdef}
\ee
Since the range of the angular variable $\phi$ is $2\pi$, $1-b$ is
the deficit angle induced by the tension, as per Eq.
(\ref{tendef}). The full $6D$ vacuum solution is
\be ds_6^2 = \eta_{\mu\nu} dx^\mu dx^\nu +  d\rho^2 + (1-b)^2
\rho^2 d\phi^2 \, . \label{6dcone} \ee
This is identical to the thin 3-brane in a flat bulk found in
models with large extra dimensions \cite{raman} because the induced
curvature identically vanishes on flat branes. The bulk geometry
is a cone, given in Fig. (\ref{fig:cone}).
It is well known, however, that for backgrounds with matter
sources on a thin brane in higher-dimensional gravity the short
distance structure of the geometry near the brane is more involved
\cite{clinvin}-\cite{tbha,kalkil,pesota}. In general, the length
scale $\ell$ does not drop out but needs to be promoted into a UV
regulator, smoothing the tip of the cone in Fig. (\ref{fig:cone}).
As we will see below this becomes even more important in brane
induced gravity.

Clearly, we see that the solution (\ref{6dcone}) is not well
defined at $b=1$, corresponding to the critical value of the
tension $\lambda_{cr} = 2\pi M_6^4$. While the metric
(\ref{6dcone}) makes sense, as it stands, for $b>1$, it is not
clear what it represents since we can't smoothly deform it to
relate it to sub-critical cases. Hence in the critical limit, and
beyond it, we cannot tell what happens to the brane until we sort
out how to match the brane's core to the exterior, in precisely
the same manner as for the critical and super-critical strings
\cite{naked}.

\subsection{Shocks on Thin Branes}

A simple way to probe the nature of gravity, as argued in
\cite{shocks,kalkil}, once a vacuum solution is known, is to shock
it. By this we mean, put a relativistic particle on a background,
use its stress energy tensor as a source, and solve the equations
for its gravitational field. Because of Lorentz boosts, the field
becomes completely confined to the space transverse to the
direction of motion, and hence the field equations, however
formidable they may be, will linearize. A straightforward
technique to solve them is to use the cutting and pasting
technique of Dray and 't Hooft \cite{thooft,kostas}, which works
in our case since the sub-critical vacuum metric Eq.
(\ref{6dcone}), with $b<1$, is locally flat. So: the shocked
metric will be
\be ds_6{}^2 = 4dudv - 4 \delta(u) f(\vec x_\bot,\vec y)du^2 +
d\vec{x}^2_\bot + d\rho^2 + (1-b)^2 \rho^2 d\phi^2 \, ,
\label{shockedcone} \ee
where $\vec x_\bot$ coordinatize the two directions along the
brane transverse to the particle that moves along the null
geodesic $u=0$, and $\vec y$ are the coordinates on the cone. The
function $f$ is the shock wave profile that we need to solve for.
Then, using Eq. (\ref{shockedcone}) as an ansatz, we substitute it
into the field equations Eq. (\ref{fieldeqs}), with the
relativistic stress energy tensor $\tau^\mu{}_\nu =
\frac{2p}{\sqrt{g_4}}g_{4uv}
\delta(u)\delta^{(2)}(\vec{x}_\bot)\delta^\mu{}_v \delta^u{}_\nu$
of the source with 4-momentum $p^\mu = (p,0,0,p)$ included, and
determine the equation controlling the shock wave profile $f$.

Since for relativistic particles $\tau^\mu{}_\mu = 0$, we still
have $R_6 = \frac{2\lambda}{M_6^4} \delta^{(2)}(\vec y)$ and $R_4
= 0$ regardless of the source. We substitute this in the 
field equations along the brane worldvolume,
\be M_6^4 R_6{}^\mu{}_\nu + M_4^2 R_4{}^\mu{}_\nu \, \delta^{(2)}
(\vec{y}) = \tau^\mu{}_\nu \, \delta^{(2)}(\vec{y}) \, .
\label{riccishocks} \ee
The field equations in the bulk,  $R_6 = \frac{2\lambda}{M_6^4}
\delta^{(2)}(\vec y)$, are trivially solved by Eq.
(\ref{shockedcone}) and so all the new information is completely
contained in (\ref{riccishocks}). The solution of these equations
will be consistent, since one can easily verify that $\nabla_\mu
\tau^\mu{}_\nu = 0$. Now, the only nontrivial components of the
Ricci tensors for the metric (\ref{shockedcone}) are
\ba && R_6{}^v {}_u = \delta(u) \, \nabla_4{}^2 f \, ,  \nonumber \\
&& R_4{}^v {}_u = \delta(u) \, \nabla_2{}^2 f  \, , \label{riccis}
\ea
where $\nabla_4{}^2$ is the full transverse Laplacian, and
$\nabla_2{}^2$ its restriction on the $2D$ transverse plane on the
brane, coordinatized by $\vec{x}_\bot$. Substituting this and the
formula for $\tau^\mu{}_\nu$ in Eq. (\ref{riccishocks}) we finally
obtain the field equation for $f$:
\be \nabla_4{}^2 f + \frac{M_4^2}{M_6^4} \nabla_2{}^2 f \,
\delta^{(2)} (\vec{y}) =
\frac{2p}{M_6^4}\delta^{(2)}(\vec{x}_\bot) \,
\delta^{(2)}(\vec{y}) \, . \label{profile} \ee
All other equations in (\ref{riccishocks}) are trivially satisfied
on (\ref{shockedcone}). Thus we see that the problem of
determining the configuration which solves (modified) gravity
equations for relativistic sources, as before, maps onto a much
simpler problem of solving a (modified) Poisson equation for a
static charge \cite{thooft}-\cite{kostas},\cite{shocks}. This
illustrates the power of shock therapy as a diagnostic method.
Once we have the solution of Eq. (\ref{profile}), we can compare
it to the Aichelburg-Sexl solution \cite{aichsexl} to see if, and
how, the theory mimics $4D$ General Relativity.

Let us now solve Eq. (\ref{profile}), at least formally.  Fourier
transforming in the brane transverse space
\be \delta^2(\vec{x}_\bot) = \frac{1}{(2\pi)^2}\int d^2\vec{k}
e^{i \vec{k}\cdot \vec{x}_\bot} \, , ~~~~~~~~~ f =
\frac{1}{(2\pi)^2}\int d^2\vec{k} \, \varphi_k(\vec{y}) \, e^{i
\vec{k}\cdot \vec{x}_\bot} \, , \label{fourier} \ee
using $\nabla_4{}^2 = \nabla_{\vec{y}}{}^2 + \nabla_2{}^2$, where
the operator $\nabla_{\vec{y}}{}^2$ is the Laplacian on the $2D$
cone, and replacing $\delta^{(2)}(\vec y) = \frac{1}{2\pi (1-b)
\rho} \delta(\rho)$ by using the axial symmetry of
(\ref{shockedcone}) and the null source $\tau^\mu{}_\nu$, after a
little algebra we obtain
\be \big(\nabla_{\vec{y}}^2 - k^2\big)\varphi_k = \frac{1}{2\pi
(1-b) M_6^4}\big(2p + M_4^2 k^2 \varphi_k
)\frac{\delta(\rho)}{\rho} \, , \label{helmholtz} \ee
which is very similar to the field equation of a bulk scalar with
kinetic terms induced on the brane \cite{DGP}. Note that the bulk
coupling $M_6^{-4}$ has been replaced by the effective coupling
$[(1-b) M_6^4]^{-1}$ which is stronger on the cone because of its
deficit angle, representing the `lightning rod' amplification
effect pointed out in \cite{kalkil}. At any rate, we can now use
the same tricks to solve it as \cite{DGP}. Ignoring the angular
variable by axial symmetry, and substituting $\varphi_k(\rho) =
D(k,\rho)F(k)$ where  $D$ obeys the Helmholtz equation
\be \big(\nabla_{\vec{y}}^2-k^2\big)D = \frac{2p}{2\pi (1-b)
M_6^4}\frac{\delta(\rho)}{\rho} \, , \label{eqn:dgreen} \ee
we then determine $F$ such that the ansatz $\varphi_k = DF$
correctly solves Eq. (\ref{fourier}). This yields $F =
\Bigl(1-\frac{M_4^2}{2p}k^2 D(k,0)\Bigr)^{-1}$, and therefore the
momentum space field $\varphi_k$ is
\be \varphi_k = \frac{D(k,\rho)}{1-\frac{M_4^2}{2p}k^2 D(k,0)} \,
. \label{momphi} \ee
The Helmholtz equation is defined on the full $2D$ cone, with the
source residing exactly on its vertex, and so we can write its
solution $D$ in terms of the modified Bessel function $K_0$, which
vanishes at infinity and hence describes properly gravity
localized to the brane as
\be D(k,\rho) = -\frac{p}{\pi (1-b) M_6^4} K_0(k \rho) \, .
\label{helmsoln} \ee
where we have explicitly used the fact that the other modified
Bessel function $I_0$ is equal to unity on the brane at $\rho =
0$, while it diverges at infinity. We point it out since it will
arise naturally later on, although here it plays no role yet.
Then, substituting this in $\varphi_k$ and Fourier transforming it
back to the transverse configuration space along the brane, and
integrating over the angular bulk variable in the usual
way\footnote{We use $f = \frac{1}{(2\pi)^2}\int
d^2\vec{k}\varphi_k \,e^{i \vec{k}\cdot \vec{x}_\bot} $ and
integrate over $\phi$ to get $f = \frac{1}{2\pi}\int_0^\infty
dk\,k\,\varphi_k \,J_0(k |\vec{x}_\bot|)$, as in any cylindrically
symmetric problem in potential theory.} we finally find the
solution for the shock wave profile on a thin 3-brane:
\be f = -\frac{p}{2\pi^2 (1-b) M_6^4} \int_0^\infty dk \frac{k
K_0(k \rho) J_0(k|\vec{x}_\bot|)}{1+\frac{M_4^2}{\pi (1-b)
M_6^4}k^2 K_0(0)} \, . \label{thinshock} \ee

This solution is very interesting - because it is pathological.
Its pathologies stem from the fact that $K_0$ diverges in the core
of the brane. Indeed, $K_0(\epsilon) \sim -\ln(\epsilon)
\rightarrow \infty$ as $\epsilon \rightarrow 0$. However, along
the brane, at $\rho = 0$, the divergence {\it precisely} cancels
between the numerator and the denominator in Eq.
(\ref{thinshock}). In this case, we find
\be f = -\frac{p}{2\pi M_4^2}\int_0^\infty dk \,
\frac{J_0(k|\vec{x}_\bot|)}{k} = \frac{p}{\pi M_4^2} \,
\ln\bigg(\frac{|\vec{x}_\bot|}{\ell}\bigg) \, , \label{aichelsexl}
\ee
which is exactly the $4D$ Aichelburg-Sexl solution - at {\it all}
distances along the 3-brane! This result would be incredibly
interesting, because it occurs despite the infinite bulk space
surrounding it, that could be explored by the graviton multiplet.
This looks like a very efficient mechanism to hide extra
dimensions from the gravitational exploration. However, if we move
off the brane by an even infinitesimal amount, to $\rho =\epsilon
\ne 0$, from (\ref{thinshock}) we find that the divergence of the
denominator forces the shock wave profile immediately to zero.
Thus the confinement of gravity to the brane is perfect - but it
happens because the Bessel function diverges in the core, and is
clearly a UV sensitive answer, as indeed noted in
\cite{DGP,highercod}. If we consider a physically more realistic
brane of finite thickness, the confinement may be far less than
perfect, leading to the reopening of extra dimensions to gravity
at some finite distance from the source \cite{highercod}.
Therefore, before we draw any conclusions about brane induced
gravity in codimension$\ge 2$, we first must regulate the short
distance singularities inside the brane.

Our shock wave example points us to a very simple way how to
regulate. The discontinuity of the solution is due to the
divergence in the function $K_0(\rho)$ at $\rho = 0$.
Mathematically, the divergence of $K_0$ on the source is the
familiar pathology from an ill-posed problem for an elliptic
Helmholtz equation for a pointlike source, which does not have a
regular exterior solution in $2D$. A well known prescription for
regulating this problem is to replace the pointlike source by a
ring-like one. We can then split the problem into a combination of
interior and exterior problems, which have well behaved solutions
for ring sources of a finite size. Since our pointlike source is
really a particle on a codimension-2 brane, a natural way to
replace it by a ring-like source is to consider a loop of string
moving on a cylinder. If we then want to derive the background
cylinder from a covariant action, the easiest way to do it is to
take a $4$-brane and wrap it around a circle of finite size. Such
tricks were used earlier in \cite{msled,pesota} and we employ it
next.

\section{Wrapped Branes and Truncated Cones}

\subsection{Making Vacua}

We wish to replace the thin brane vacuum (\ref{6dcone}) by one
where the brane has a core of finite thickness, but its metric
remains the same as (\ref{6dcone}) outside of the brane core. In
this case the thick brane vacuum will retain some of the
attractive features of the thin 3-brane, namely $4D$ flatness and
the presence of $4D$ General Relativity, for the sub-critical
case. Furthermore, we will find a simple model which will enable
us to see what goes on as the tension approaches and exceeds the
critical limit. Now, the reason why the metric Eq. (\ref{6dcone})
solved the field equations (\ref{fieldeqs}) was that the stress
energy tensor of a tensional 3-brane has the form $T^\mu{}_\nu = -
\lambda \delta^{(2)}(\vec y)$ along the brane and vanished away
from it. In component notation, this looks like $T^A{}_B = -
\lambda \delta^{(2)}(\vec y) \, {\rm diag}(1,1,1,1,0,0)$. A very
simple model of a 3-brane with a finite core is a cylinder: a
4-brane wrapped on a circle \cite{msled,pesota}. However, when the
4-brane has nonzero tension, with the vacuum action $S_{\rm
vacuum} = - \int d^5 x \sqrt{g_5} \lambda_5$, its stress energy
tensor will be $T^A{}_B = - \lambda_5 \delta^{}(\rho - r_0) \,
{\rm diag}(1,1,1,1,1,0)$, where we wound it around the circle of
radius $r_0$, with $\rho$ along the normal to the brane. Thus, to
wrap the 4-brane into a cylinder, we must cancel the pressure
$\propto \lambda_5$ in the compact direction.

A simple way to do it is to put an axion-like field $\Sigma$ on
the 4-brane, with a vacuum action
\be S_{vacuum} = - \int d^5 x \sqrt{g_5}\big(\lambda_5 +
\frac{1}{2} g^{ab}\partial_a \Sigma \partial_b \Sigma\big) \, ,
\label{4brvac} \ee
and look for solutions where $\Sigma$ breaks the translational
invariance in the compact direction, but so that its stress energy
tensor remains translationally symmetric. This is precisely the
method of Scherk and Schwarz \cite{schsch} for dimensionally
reducing supergravities to get massive lower dimensional theories.
The stress energy tensor obtained by varying Eq. (\ref{4brvac})
along the 4-brane is
\be T^a{}_b = - \lambda_5 \delta^a{}_b + \partial^a\Sigma
\partial_b \Sigma -\frac{1}{2} \delta^a{}_b g^{cd}\partial_c
\Sigma \partial_d \Sigma \, , \label{stress4br} \ee
where lower case latin indices $a,b, \ldots$ run over 4-brane
worldvolume values $0, 1, \ldots, 4$. Taking the fourth coordinate
to be the angle on the compact circle $\phi$ and substituting the
Scherk-Schwarz ansatz $\Sigma = q \phi$ we choose $q$ to precisely
cancel $T^\phi{}_\phi$. This requires
\be \lambda_5 = \frac{1}{2}q^2 g^{\phi \phi} \, , \label{tuning}
\ee
where $g^{\phi\phi} = \frac{1}{r_0^2}$ is the inverse radius
squared of the compact dimensions. Then the remaining components
of the 4-brane stress energy become precisely $T^\mu{}_\nu = -2
\lambda_5 \delta^\mu{}_\nu$. Thus the brane source now reads
$T^A{}_B = - 2\lambda_5 \delta(\rho - r_0) \, {\rm diag}(1, 1, 1,
1, 0, 0)$. The tensor structure is precisely the same as in the
stress energy tensor of a thin 3-brane. Now, since the thin brane
is axially symmetric from the bulk viewpoint, we can rewrite the
scalar $\delta$-function in polar coordinates as
$\delta^{(2)}(\vec y)  = \frac{1}{2\pi \rho} \delta(\rho)$. By
inspection, we see that if we shift the argument of the remaining
radial $\delta$-function to $\rho - r_0$, it becomes
$\delta^{(2)}_{\rm thick}(\vec y) = \frac{1}{2\pi r_0} \delta(\rho
- r_0)$, and so to model the thick brane by a wrapped 4-brane, we
need to identify $\lambda \delta^{(2)}_{\rm thick}(\vec y)$ with
$2 \lambda_5 \delta(\rho-r_0)$. This yields
\be
\lambda = 4\pi r_0 \lambda_5 \, ,
\label{lambdas}
\ee
relating the $5D$ and the effective $4D$ vacuum energies. The
effective $4D$ vacuum energy $\lambda$ contains the contributions
from the classical axion field $\Sigma$, doubling up its value,
because of the cancellation condition of Eq. (\ref{tuning}). This
is equivalent to the effective vacuum energy contributions in the
string landscape \cite{landscape}, which is also comprised of the
contributions from quantum fields (here, $\lambda_5$) and
classical fluxes (here $\propto q^2$). In the landscape, these
contributions need to be finely adjusted to yield a $4D$ Minkowski
vacuum.  Here, the situation is similar: because this stress
energy is completely localized on a codimension-1 cylindrical
surface of finite radius, one expects that the solution
representing this configuration should be obtained from patching
locally flat $6D$ Minkowski spaces inside and outside of the
cylinder. The exterior should have a deficit angle by the
construction of the thick brane stress energy, which shows that
our 4-brane source looks precisely like the thick 3-brane.

\begin{figure}[thb]
\vskip.6cm
\centerline{\includegraphics[width=0.45\hsize,width=0.45\vsize,angle=0]{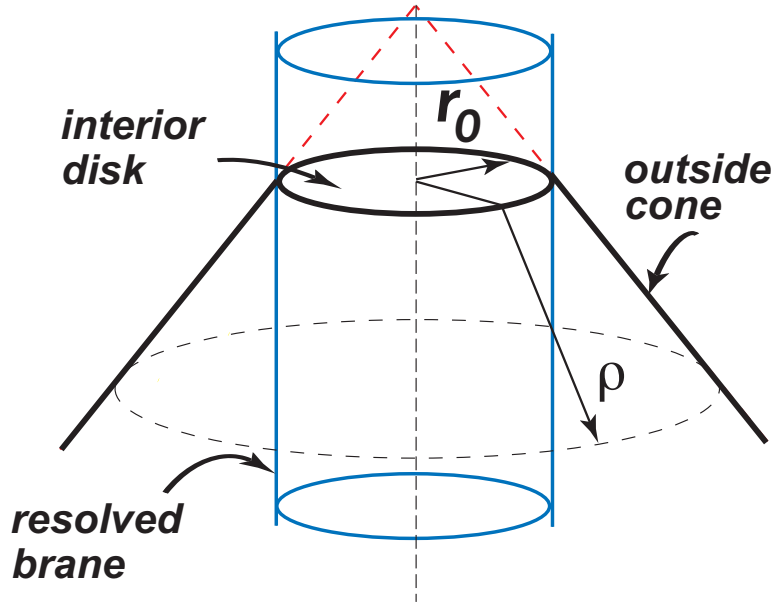}}
\caption{{$2D$ bulk geometry of the resolved brane.}
\label{figure}}
\vskip.5cm
\end{figure}

Indeed, the field equations for a 4-brane in a $6D$ bulk which
include brane localized gravity terms are \cite{DGP,highercod}, in
Gaussian-normal gauge, and with the 4-brane residing at $\rho = r_0$,
\be M_6^4 G_6{}^A{}_B + M_5^3 G_5{}^a{}_b \delta^A{}_a
\delta^b{}_B \delta(\rho-r_0) =  T^a{}_b \delta^A{}_a
\delta^b{}_B  \delta(\rho-r_0) \, ,
\label{thickfieldeqs}
\ee
which we apply to our cylindrical brane vacuum, imposing that the
stress energy tensor is covariantly conserved, that requires
$\partial^a \partial_a \Sigma = 0$. It is easy to check that by
axial symmetry the ansatz $\Sigma = q \phi$ trivially solves the
latter equation. Then, the tensor structure of the source,
$T^A{}_B = - 2\lambda_5 \delta(\rho-r_0) \, {\rm diag}(1, 1, 1, 1, 0,
0)$, guarantees that the flat $4D$ metric is a solution, if the
tension is again off-loaded into the bulk, like in the thin
3-brane case. Tracing the field equations (\ref{thickfieldeqs}),
and using (\ref{tuning}) we find that the condition for this is
that the metric in the remaining two dimensions, coordinatized by
the angular direction along the brane and the radial distance away
from it has curvature
\be
R_2 = \frac{4\lambda_5}{M_6^4}\delta(\rho-r_0) \, .
\label{twodthick}
\ee
Thus, by the arguments identical to those we invoked in the
construction of the thin 3-brane solutions, we can choose the
coordinates where the $2D$ metric is $ds_2^2 =
e^{-\vartheta}d\vec{z}^2$, where however the brane now resides at
$r_0 = |\vec z_0|$ around the origin. Substituting in
(\ref{twodthick}), we find $R_2 = e^{-\vartheta}\vec{\nabla}^2_2
\vartheta$. Comparing to (\ref{twodthick}) and using the fact that
$\vec{\nabla}^2_2\big[\Theta(r-r_0)\ln(r/r_0)\big] =
\delta(r-r_0)/r$, we infer that
\be
\vartheta = 2 b \Theta(r - r_0) \ln(\frac{r}{r_0}) \, \,
\label{varth}
\ee
where $\Theta(x)$ is the Heaviside step function, $\Theta(x) = 0$
for $x < 0$ and $\Theta(x) = 1$ for $x \ge 0$, and $b$ is given
precisely by combining Eqs. (\ref{tendef}) and (\ref{lambdas}),
\be
b = \frac{2 \lambda_5 r_0}{M_6^4} \, .
\label{newdef}
\ee
The solution is given by $ds_6{}^2 =  \eta_{\mu\nu} dx^\mu dx^\nu
+ e^{-2b\Theta(r-r_0)\ln(r/r_0)}\big(dr^2 + r^2 d\phi^2\big)$, in
polar coordinates. After defining the new radial coordinate $\rho$
by ${d\rho}/{dr} = ({r_0}/{r})^{b\Theta(r-r_0)}$, such that
\be \rho =
\frac{r}{1-b\Theta(r-r_0)}\bigg(\frac{r_0}{r}\bigg)^{b\Theta(r-r_0)}
-\frac{b r_0 \Theta(r-r_0)}{1-b \Theta (r-r_0)} \, ,
\label{eqn:rho} \ee
and $\Theta(r-r_0) = \Theta(\rho - r_0)$, we can rewrite the
metric as
\be ds_6{}^2 = \eta_{\mu\nu}dx^\mu dx^\nu + d\rho^2 +\bigg[
(1-b\Theta(\rho-r_0))\rho + b r_0 \Theta(\rho-r_0)\bigg]^2 d\phi^2 \, .
\label{thick6dvac} \ee
For $\rho < r_0$ the $2D$ part of the metric is $ds_2{}^2 =
d\rho^2 +\rho^2 d\phi^2$, i.e. a flat disk, while for $\rho > r_0$
it becomes $ds_2{}^2 = d\rho^2 + \big((1-b)\rho+b\,r_0\big)^2
d\phi^2 = d\rho^2 + (1-b)^2\big(\rho+\frac{b}{1-b}\,r_0\big)^2
d\phi^2$, precisely the metric on a cone, which we see by
comparing it to the thin 3-brane metric from the previous section.
Thus the combined solution represents a truncated cone, with a
flat mesa of radius $r_0$ on top, as depicted in Fig.
(\ref{figure}). The geometry of the brane is $4D$ Minkowski
$\times$ circle.

\begin{figure}[thb]
\vskip.6cm
\centerline{\includegraphics[width=0.4\hsize,width=0.4\vsize,angle=0]{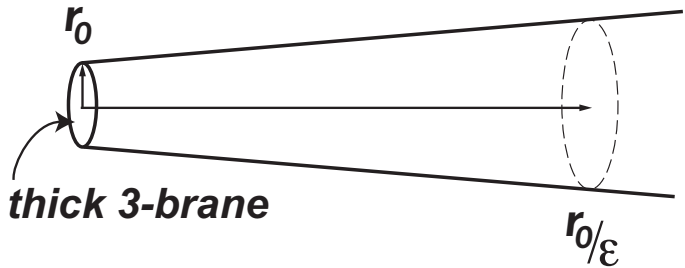}}
\caption{{$2D$ bulk geometry of the near-critical resolved 3-brane.}
\label{sliver}}
\vskip.5cm
\end{figure}

Looking at Eq. (\ref{thick6dvac}) for $\rho > r_0$ we see that the
critical string tension, where the deficit angle becomes $2\pi$,
is again given by $b=1$, which now corresponds to the tension
\be \lambda_{5 \, cr} = \frac{M_6^4}{2r_0} \, . \label{critten}
\ee
Comparing to the relation between the $5D$ tension $\lambda_5$ and
the effective $4D$ tension $\lambda$, given by Eq. (\ref{lambdas})
we find that the critical value of the effective $4D$ tension is,
not surprisingly, $\lambda_{cr} = 2\pi M_6^4$. We can now see what
happens as $\lambda \rightarrow \lambda_{cr}$ and beyond. Since we
are holding the brane core fixed, using its own internal dynamics,
all we need to do is analyze the region $\rho > r_0$ of Eq.
(\ref{thick6dvac}). When $b=1$, the metric outside of the brane
changes to
\be ds_6{}^2 = \eta_{\mu\nu}dx^\mu dx^\nu + d\rho^2 + r_0^2
d\phi^2 \, , \label{bulkcyl} \ee
which describes a cylinder $M_4 \times R \times S^1$, where the
radius of $S^1$ is the same as the brane radius. Now, when $b =
1-\epsilon$, it's easy to see that the bulk cone looks like a
sliver. In this limit, the exterior bulk metric is approximately
$ds_6{}^2 = \eta_{\mu\nu}dx^\mu dx^\nu + d\rho^2 + \big(\epsilon
(\rho - r_0)\,+\,r_0\big)^2\ d\phi^2$. This looks like the metric
on a cylinder for $r_0 \le \rho \la r_0/\epsilon$, because the
variation of the radius of the sliver, due to the change of radial
distance $\rho$, is very small compared to the radius of the brane
$r_0$. Near-critical branes reside inside deep bulk throats that
only asymptotically open into conical geometries \cite{kalwall},
see Fig (\ref{sliver}).

For supercritical branes, $b>1$, the exterior metric is just
\be ds_6{}^2 =  \eta_{\mu\nu}dx^\mu dx^\nu +  d\rho^2 +
(b-1)^2\big(\rho - \frac{b}{b-1}\,r_0 )^2 d\phi^2 \, ,
\label{tear} \ee
which describes a bulk geometry that near $\rho = r_0$ looks like
a flat disk. However now the metric has a singularity at $\rho =
\frac{b}{b-1} r_0 > r_0$, where the geometry tapers off to a
conical spike. Thus when $b >1$ the bulk spontaneously
compactifies on a $2D$ teardrop, precisely of the shape discussed
by \cite{gell}. Hence static flat branes in an infinite bulk will
only occur for sub-critical values of the tension. In light of the
discussions of \cite{naked}, one expects that the supercritical
brane solutions without bulk singularities exist, but that they
describe nonstationary backgrounds. Examples describing
topologically inflating domain walls on codimension-1 brane
induced gravity models have been discussed recently in
\cite{oriol}.

Note, that our construction is completely covariant from the bulk
point of view, since we can define the regulator by a variational
principle, from a well defined 4-brane action. Therefore we can
explicitly check the properties of brane localized gravity, and
calculate the crossover scale inferred in \cite{highercod}
exactly.

\subsection{Stress Energy Sources on Thick Branes}

Having constructed the thick brane vacua, we can turn to the
description of matter sources that reside on branes. For
simplicity we take the matter to be localized on the 4-brane, and
described by a matter sector action
\be S_{\rm matter} = - \int d^5 x \sqrt{g_5} \, {\cal L}_{\rm
matter} \, , \label{matter} \ee
where the matter couples to the induced metric on the 4-brane
$g_{5 \, ab}$. Since one of the dimensions along the brane is a
circle of radius $r_0$, and we are interested in the behavior of
the theory at length scales $r \gg r_0$, we dimensionally reduce
the brane-localized sector on this direction, representing all
brane fields by the standard Fourier expansion
\be \Phi_{\cal N} (x^\mu, r_0 \phi) = \sum^\infty_{n = - \infty}
\Phi_{{\cal N}, n}(x^\mu) \, e^{i n \phi} \, , \label{KK} \ee
where ${\cal N}$ represents the quantum numbers of any particular
representation on the brane. The Fourier coefficients then give
rise to the $4D$ KK towers of fields, with masses $M^2 = m^2 +
n^2/r_0^2$, where $m$ are the explicit mass terms from
(\ref{matter}), and the mass gap is $1/r_0$. This expansion also
applies to the brane induced metric and its curvature. Thus, at
distances $r \gg r_0$ we can always model the angular dependence
of any configuration  as a small, Yukawa-suppressed correction
from a KK momentum mode stretching along the compact circle.

When we consider the classical limit of the theory, to define the
gravitational sources that describe lumps of matter living on the
brane, we can model them as loops of string along the cylinder. At
distances $r \gg r_0$ these will look just like point particles on
a thin 3-brane. To determine the stress energy tensors of such
objects, that source their gravity, we replace (\ref{matter}) by a
Nambu-Goto action for a string,
\be
S_{\rm matter} = -\mu \int d^2\sigma \sqrt{\gamma} \, ,
\label{nambu}
\ee
where $\mu$ is the mass per unit length of the string,
$\sigma^\alpha$ are its two worldsheet coordinates, which we will
gauge fix to be $t$ and $r_0 \phi$, and $\gamma_{\alpha\beta}$ is
the induced worldsheet metric. We then vary this action with
respect to the metric of the target space on which the string
moves, in this case the metric of the background thick brane. We
then use the standard definition of the stress energy tensor
$\tau^{ab}$, given by the functional derivative equation $\delta
S_{\rm matter} = \frac{1}{2}\int d^5x\,\sqrt{g_5} \delta g_{ab}
\tau^{ab}$, to read off the stress energy tensor of the loop.
Following the standard tricks for pointlike sources, we first
rewrite (\ref{nambu}) as
\be S_m = -\mu \int d^2\sigma d^5 x \,\sqrt{\gamma}\,
\delta^{(5)}(x^M - x^M(\sigma)) \, , \ee
where the $\delta$-function puts the string on its shell (i.e.
enforces its equation of motion). Using $\gamma_{\alpha \beta} =
g_{ab}\frac{\partial x^a}{\partial \sigma^\alpha}\frac{\partial
x^b}{\partial \sigma^\beta}$ and $\delta \sqrt{\gamma} =
\frac{1}{2}\sqrt{\gamma}\gamma^{\alpha \beta}\delta \gamma_{\alpha
\beta}$, we find
\be \tau^{ab} = - \mu \int d^2
\sigma\,\frac{1}{\sqrt{g_5}}\delta^{(5)}(x^c - x^c(\sigma))
\gamma^{\alpha \beta}\frac{\partial x^a}{\partial
\sigma^\alpha}\frac{\partial x^b}{\partial \sigma^\beta} \, . \ee
Similar variation with respect to the target space coordinates of
the string leads to the string equation of motion, which is
$\nabla_\alpha (g_{ab} \nabla^\alpha x^b) = 0$, and is in fact
completely equivalent to the string stress energy conservation,
$\nabla_a \tau^{ab} = 0$.

With the gauge choice $\sigma^0 = x^0$ and $\sigma^1 = r_0 \phi$,
and using the vacuum solution (\ref{thick6dvac}) as the target
space, we get $\gamma_{\alpha\beta} = \eta_{\alpha\beta}$, and
hence the loop of string stress energy tensor becomes
\be \tau^{ab} = -\mu \int d^2\sigma \delta^{(5)}(x^c -
x^c(\sigma)) \eta^{\alpha \beta}\frac{\partial x^a}{\partial
\sigma^\alpha}\frac{\partial x^b}{\partial \sigma^\beta}.\\
\label{eqn:stringseint}
\ee
While this expression is only valid on the locally flat target
(\ref{thick6dvac}), as we will only be interested in the
linearized gravitational fields (which however are exact for
relativistic particles!) this is all we shall need. If we consider
static sources, the string representing only the lightest modes
must be translationally invariant in the compact direction,
because by the discussion above any inhomogeneities along the
string can be viewed as heavy KK states at distances $r \gg r_0$.
For such axially symmetric sources, only two components of
(\ref{eqn:stringseint}) survive:  $\tau^{00} = \mu
\delta^{(3)}(\vec{x})$, and $\tau^{\phi\phi} =  -\frac{\mu}{r_0}
\delta^{(3)}(\vec{x})$, which we can write more compactly as
\be \tau^a{}_b = -\mu \, \delta^{(3)}(\vec{x}) \,
\textrm{diag}(1,0,0,0,1) \, . \label{eqn:strinsediag} \ee
The parameter $\mu$ is the total  rest mass per unit length of the
configuration. To get the stress energy tensor of a relativistic
string, that moves along the cylindrical brane at the speed of
light, and has vanishing rest mass, we can simply boost
(\ref{eqn:strinsediag}) along the direction of motion
$x_\parallel$, go to the light cone coordinates  $x_\parallel =
v+u$, $t=v-u$ and take the limit of infinite boost parameter while
sending $\mu \rightarrow 0$ \cite{thooft}. Since the momentum of
the string $p$ is the length of the string $2\pi r_0$ multiplied
by the finite limit of $\mu \cosh \beta$ as $\beta$ diverges, for
sources composed of the lightest modes as in
(\ref{eqn:strinsediag}) this yields, thanks to the properties of
the vacuum (\ref{thick6dvac}),
\be \tau^\mu{}_\nu = \frac{2p}{2\pi r_0 \sqrt{g_4}}g_{4uv}
\delta(u)\delta^{(2)}(\vec{x}_\bot)\delta^\mu{}_v \delta^u{}_\nu
\, , ~~~~~~~~~~~~~ \tau^\phi{}_\phi = 0 \, , \label{shocktmunu}
\ee
which is identical in form to the relativistic stress energy on
thin branes, used in the previous section.

To find the gravitational fields of these sources, we need to
solve the brane induced gravity field equations
(\ref{thickfieldeqs}), in Gaussian-normal gauge with the brane at
$y=\rho-r_0 = 0$ and with $T^a{}_b$, which is given by the sum of
the vacuum contribution in Eq. (\ref{stress4br}) and $\tau^a{}_b$
as given in (\ref{eqn:strinsediag}) or (\ref{shocktmunu}), on the
right-hand side. It will turn out to be convenient to rewrite
these equations as the conditions on the Ricci tensor, which after
tracing over the indices and rearranging the scalar contributions,
become
\be M_6^4 R_6{}^A{}_B = \bigg[\bigg(T^a{}_b - M_5^3
\big(R_5{}^a{}_b - \frac{1}{2}\delta^a{}_b
R_5\big)\bigg)\delta^A{}_a \delta^b{}_B -\frac{1}{4}\delta^A{}_B
\big(T^a{}_a + \frac{3}{2}M_5^3 R_5\big)\bigg] \delta(\rho-r_0) \, .
  \label{riccieq}
\ee
We first consider the relativistic sources obeying
(\ref{shocktmunu}) and find the generalization of the shock wave
solution (\ref{shockedcone}), (\ref{thinshock}) on thick branes.

\section{Shocking Crossover Physics}

\subsection{Setting Up the Shock Wave}

As before, we determine the shock wave solution by introducing a
discontinuity in the metric of Eq. (\ref{thick6dvac}) according to
the prescription of \cite{thooft,shocks}, and demanding that the
wave profile solves Eq. (\ref{riccieq}) with (\ref{stress4br}) and
(\ref{shocktmunu}) as sources. We are working with sub-critical
branes, $b <1$, which admit static flat vacua. We can still,
however, consider the limit $b \rightarrow 1$. Using the null
coordinates $u, v$ we require that the source moves along $u=0$,
and that the wave profile explores all the transverse dimensions
available to gravity. By axial symmetry of the source, however, we
only need to look at the profiles $f(\vec{x}_\bot,\rho)\delta(u)$
that depend on the transverse coordinates along the brane
$\vec{x}_\bot$ and the transverse distance $\rho$ orthogonal to
the brane. Thus we seek the shock wave metric in the form
\be ds_6{}^2 = 4dudv - 4 \delta(u) f du^2 + d\vec{x}^2_\bot +
d\rho^2 + \bigg[(1-b\,\Theta(\rho-r_0))\rho+b\,r_0 \Theta(\rho
-r_0)\bigg]^2 d\phi^2 \, . \label{shockthick} \ee
Now, we substitute the sum of Eqs. (\ref{stress4br}) and
(\ref{shocktmunu}) along with the condition (\ref{tuning}) in Eq.
(\ref{riccieq}) and trace over the indices. We note that the
stress energy of a relativistic particle (\ref{shocktmunu}) is
traceless, reflecting that the source is scale-free. We can then
easily check that by the form of the shock wave metric
(\ref{shockthick}) the trace of the $6D$ Ricci tensor picks up
contributions only from the $\rho$-$\phi$ part of the metric, $R_6
= R_2$, leading to an equation identical to Eq. (\ref{twodthick}).
Thus as before this equation is trivially satisfied by Eq.
(\ref{shockthick}), with the deficit angle parameter  defined in
(\ref{newdef}), and so are all the tensor equations on the
truncated cone $\rho$-$\phi$.

Among the remaining tensor equations of (\ref{riccieq}), the only
one which contains nontrivial information, as before, is $\propto
R_6{}^v{}_u$.  Indeed, Eqns. (\ref{riccieq}) on the background
(\ref{shockthick}) reduce to
\be M_6^4 R_6{}^{\mu}{}_\nu + M_5^3 R_{5}{}^\mu{}_\nu \,
\delta(\rho-r_0)  = \tau^\mu{}_\nu \, \delta(\rho-r_0) \, ,
\label{riccishothi}
\ee
and the only components of the Ricci tensors which depends on the
shock wave profile are
\ba
&& R_6{}^v{}_u = \delta(u) \nabla_4{}^2 f \, , \nonumber \\
&& R_5{}^v{}_u = \delta(u) \nabla_2{}^2 f \, ,
\label{thickuvricci}
\ea
where we are explicitly using axial symmetry in the latter term
when we write the Laplacian in only the two transverse spatial
coordinates $\vec x_\bot$. Substituting this and
(\ref{shocktmunu}) in Eq. (\ref{riccishothi}), after a
straightforward algebra we finally get
\be \nabla_4{}^2 f + \frac{M_5^3}{M_6^4} \nabla_2{}^2 f \,
\delta(\rho- r_0 ) = \frac{2p}{2\pi r_0 M_6^4} \,
\delta^2(\vec{x}_\bot) \, \delta(\rho-r_0) \, . \label{lapeom} \ee
This equation is almost identical to Eq. (\ref{profile}) governing
the shock wave profile on a thin 3-brane. The main difference is
that the $\delta$-function is shifted from the origin and smeared
over a circle $\rho = r_0$, and that the Laplacian $\nabla_4{}^2$
is defined on the truncated cone (\ref{thick6dvac}), such that for
axially symmetric shocks, $\nabla_4{}^2 f = \nabla_{\vec
x_\bot}{}^2 f
+ f'' +\frac{1-b\Theta(\rho-r_0)}{(1-b\Theta(\rho-r_0))\rho
+ br_0\Theta(\rho-r_0)} f'$, where
the prime denotes the $\rho$-derivative. The latter term contains
a jump across the 4-brane.

\subsection{Solving for the Shock Wave Profile}

To solve (\ref{lapeom}), we again Fourier transform in the brane
transverse space, using Eq. (\ref{fourier}). This yields
\be \big(\nabla_{\vec{y}}^2 - k^2\big)\varphi_k =
\bigg(\frac{2p}{2\pi r_0 M_6^4} + \frac{M_5^3}{M_6^4} k^2
\varphi_k \bigg) \delta(\rho- r_0) \, . \label{varphithick} \ee
In contrast to Eq. (\ref{profile}) here we have a single
$\delta$-function. Of course, this is because this time the brane
is codimension-1. Hence we can use the standard methods for
finding bulk field wave functions in the codimension-1 brane
setups: we solve Eq. (\ref{varphithick}) inside $(-)$ and outside $(+)$ the
brane, and then use the boundary conditions, which are the
continuity of $\varphi_k$, the jump of $\varphi_k'$ as prescribed
by Gaussian pillbox integration of (\ref{varphithick}) around
the brane, and the regularity of the solution in the center of
(\ref{thick6dvac}) and at infinity:
\ba
\varphi^-_k(\, 0\,) &=& 0 \, , \nonumber \\
\varphi^+_k(\infty) &=& 0 \, , \nonumber \\
\varphi^+_k(r_0) &=& \varphi^-_k(r_0) \, , \nonumber \\
\varphi^+_k{}'(r_{0}) - \varphi^-_k{}'(r_{0})
&=& \bigg(\frac{2p}{2\pi r_0 M_6^4} + \frac{M_5^3}{M_6^4} k^2
\varphi_k(r_0) \bigg) \, . \label{bcs} \ea
Away from the brane the differential equation (\ref{varphithick})
gives
\be
\varphi_k{}''+\frac{1}{\rho
+ \frac{br_0\Theta(\rho-r_0)}{1-b\Theta(\rho-r_0)}}
\varphi_k{}' - k^2 \varphi_k = 0 \, ,
\ee
with Bessel functions $I_0\Bigl(k
(\rho+\frac{br_0\Theta(\rho-r_0)}{1-b\Theta(\rho-r_0)})\Bigr)$,
$K_0\Bigl(k
(\rho+\frac{br_0\Theta(\rho-r_0)}{1-b\Theta(\rho-r_0)})\Bigr)$ as
solutions. The first two of the boundary conditions in Eq.
(\ref{bcs}) then pick $\varphi_k^- \sim I_0$, $\varphi_k^+ \sim
K_0$, and the third condition sets $\varphi^-_k = A I_0\Bigl(k
\rho) K_0\Bigl(k\frac{r_0}{1-b} \Bigr)$ and $\varphi^+_k = A
I_0\Bigl(k r_0) K_0\Bigl(k(\rho+ \frac{br_0}{1-b}) \Bigr)$.
Finally the last of (\ref{bcs}) fixes the coefficient $A$,
yielding
\be A = -\frac{2p}{2\pi r_0 \Bigl[ M_6^4 k \Bigl(I_0(kr_0)
K_1(k\frac{r_0}{1-b}) + I_1(kr_0) K_0(k\frac{r_0}{1-b})\Bigr) +
M_5^3 k^2 I_0(kr_0)K_0(k\frac{r_0}{1-b}) \Bigr]} \, ,
\label{matchbcs} \ee
where we have used the recursion formulas for modified Bessel
functions $\partial_z I_0(z) = I_1(z)$ and $\partial_z K_0(z) = -
K_1(z)$. Substituting this and the form of the solutions
$\varphi_k$ back in the Fourier transform equations
(\ref{fourier}), and again using the axial symmetry in the $\vec
x_\bot$ plane to integrate over the transverse spatial angle,
yielding $f = \frac{1}{2\pi}\int_0^\infty dk \, k \varphi_k
J_0(k|\vec{x}_\bot|)$, we finally obtain the expression for the
shock wave profile on the truncated cone:
\be f(\vec x_\bot, \rho)  = -\frac{p}{2\pi^2 r_0} \int_0^\infty dk
\, \frac{I_0(k\rho_<) K_0(k\rho_>) J_0(k|\vec{x}_\bot|)}{M_6^4
\Bigl(I_0(kr_0) K_1(k\frac{r_0}{1-b}) + I_1(kr_0)
K_0(k\frac{r_0}{1-b})\Bigr) +M_5^3 k
I_0(kr_0)K_0(k\frac{r_0}{1-b}) } , \label{thickshosol1} \ee
where
\be I_0(k\rho_<) K_0(k\rho_>)  = \cases { I_0\Bigl(k\rho\Bigr)
K_0\Bigl(k\frac{r_0}{1-b}\Bigr) \, ,   & ~~~~ $\rho \le r_0$ \, ;
\cr I_0\Bigl(k r_0 \Bigr)  K_0\Bigl(k(\rho +
\frac{br_0}{1-b})\Bigr) \, ,  & ~~~~ $\rho \ge r_0$ \, . }
\label{thickshosol2} \ee
This solution is clearly not divergent away from the brane, unlike
(\ref{profile}), as long as $r_0 > 0$. Indeed, along the brane,
$\rho = r_0$, it reduces to
\be f(\vec x_\bot, r_0)  = -\frac{p}{2\pi^2 r_0} \int_0^\infty dk
\, \frac{I_0(kr_0) K_0(k\frac{r_0}{1-b})
J_0(k|\vec{x}_\bot|)}{M_6^4 \Bigl(I_0(kr_0) K_1(k\frac{r_0}{1-b})
+ I_1(kr_0) K_0(k\frac{r_0}{1-b})\Bigr) +M_5^3 k
I_0(kr_0)K_0(k\frac{r_0}{1-b}) } \, , \label{thickshosol1br} \ee
where all the individual contributions remain finite.

\subsection{Crossover Physics}

Now we turn to the physics of the solution
(\ref{thickshosol1})-(\ref{thickshosol1br}). It is subtle,
controlled by the deficit angle of the background
(\ref{thick6dvac}). Let us look at the field along the brane,
(\ref{thickshosol1br}). The shock wave profile in
(\ref{thickshosol1br}) is dominated by the modes with the momenta
$k \sim 1/|\vec x_\bot |$, due to the oscillatory nature of $J_0$.
The contributions of the modes with momenta  $k$ far from $1/|\vec
x_\bot |$ will interfere destructively. At transverse distances
much larger than the size of the compact dimension, $|\vec x_\bot|
\gg r_0$, we need to focus on the momenta for which $k r_0 \ll 1$.
Thus we can always replace the terms $\propto I_n(kr_0)$ by their
small argument expansion. For sub-critical branes, $1-b \sim {\cal
O}(1)$, and we can likewise replace any $K_n(k\frac{r_0}{1-b})$ by
their small argument expansion too. However, in the near-critical
limit the deficit angle approaches $2\pi$ and so $|1-b| \ll 1$. Hence
the argument of $K_0(k\frac{r_0}{1-b})$ may be very large even
when $kr_0 \ll 1$. Thus we must consider the near-critical branes
very carefully.

We start with sub-critical branes where $kr_0 \ll 1-b$, and we can
approximate all Bessel functions by their small argument form.
Using $I_\nu(z) \rightarrow
\frac{1}{\Gamma(\nu+1)}(\frac{z}{2})^\nu$, $K_\nu(z) \rightarrow
\frac{\Gamma(\nu)}{2}(\frac{2}{z})^\nu$ for $\nu > 0$ and $K_0(z)
\rightarrow \ln(2/z)$, the shock along the brane then becomes
\be f(\vec x_\bot, r_0) \simeq -\frac{p}{2\pi^2
M_6^4}\int_0^\infty dk\, k \,
\frac{\ln\Bigl[\frac{2(1-b)}{kr_0}\Bigr]
J_0(k|\vec{x}_\bot|)}{(1-b)+\frac{M_5^3 r_0}{M_6^4} \, k^2 \,
\ln\Bigl[\frac{2(1-b)}{kr_0}\Bigr]} \, . \label{approxshock} \ee
If $k \sim 1/|\vec x_\bot|$ is large enough for the second term in
the denominator to dominate, which - in this approximation -
happens when
\be k^2 > k^{2}_c(k) =\frac{ (1-b) M_6^4 }{M_5^3 r_0
\ln\Bigl[\frac{2(1-b)}{kr_0}\Bigr]} \, , \label{momcond} \ee
the logs in the integrand cancel, and the shock wave profile is
\be  f(\vec x_\bot, r_0) \simeq -\frac{p}{2\pi^2 M_5^3 r_0}
\int_{k_c}^\infty dk \frac{J_0(k|\vec{x}_\bot|)}{k} = \frac{p}{\pi
M_4^2} \ln\bigg(k_c |\vec{x}_\bot|\bigg) + \ldots \, .
\label{r0limit} \ee
where we have used Gauss law to normalize the lightest KK
modes in the expansion of the brane induced curvature on the
circle of radius $r_0$ to $M_4^2 = 2\pi M_5^3 r_0$. This is - to
the leading order - precisely the $4D$ Aichelburg-Sexl solution,
with the log profile normalized to the inverse of the critical
momentum $k_c$ given in (\ref{momcond}), as opposed to some
arbitrary short distance cutoff $\ell$.

For sub-critical branes when $1-b \sim {\cal O}(1)$,  this is the
end of the $4D$ story: all distances longer than $r_0$ fall in the
$k r_0 \ll 1$ regime, and the gravitational field looks $4D$, like
(\ref{r0limit}), only when $r < r_c = 1/k_c$. In this, generic,
case, the crossover scale is qualitatively {\it very different}
from the see-saw scale of \cite{highercod}. Indeed, using Gauss
law $M_5^3 = \frac{M_4^2}{2\pi r_0}$, we can rewrite $1/k_c^2$
from (\ref{momcond}) as
\be r_c^2(k) = \frac{M_4^2}{2\pi (1-b) M_6^4}
\ln\Bigl[\frac{2(1-b)}{kr_0}\Bigr] \, , \label{crossm4} \ee
or, ignoring the log, $r_c \sim  \frac{M_4}{M_{6 \, {\rm
eff}}^2}$, where $M_{6 \, {\rm eff}}^4 = (1-b) M_6^4$. Thus for
sub-critical branes the crossover scale is set by the `naive'
ratio of brane and effective bulk Planck scales, and depends on
the UV cutoff $r_0$ of the brane core only through the logarithmic
correction reminiscent of `running' in real space, which was
discussed for locally localized gravity in \cite{kalsorbo}. This
shows that the crossover weakly depends on the momentum, and is
very different from the see-saw mechanism of \cite{highercod}. We
also see that the bulk gravitational coupling $1/M_6^4$ is
amplified by the `lightning rod' factor $(1-b)^{-1}$ of
\cite{kalkil}. This `lightning rod' amplification is intuitively
clear, because gravitational field lines spread more slowly on a
cone than on a plane, at a fixed distance from a source, and hence
gravity must look stronger on a cone \cite{kalkil}. It plays an
even more important role in the near-critical limit, as we will
see below.

At distances larger than $r_c$, gravity eventually changes to
$6D$. For sub-critical branes, this transition actually looks like
a rapid `running' of the coupling of individual graviton modes in
the resonance that impersonates General Relativity. We can glimpse
this, for example, from Eq. (\ref{approxshock}), which shows that
the effective coupling controlling the transfer of the momentum
$k$ is
\be \frac{1}{M^4_{6 \, {\rm eff}}} = \frac{1}{(1-b) M^4_6}
\ln\Bigl[\frac{2(1-b)}{k r_0}\Bigr] \, . \label{effcoupl} \ee
In contrast to the $5D$ case \cite{shocks}, the running here is
logarithmic, because there are two extra dimensions as probed by
the localized masses on the wrapped brane. To recover the full
$6D$ form of the wave, we can take (\ref{thickshosol2}) or
(\ref{thickshosol1br}), and taking the limit of very large
distance, ignore the second term in the denominators of these
integrals. In this case,
\be f \simeq -\frac{p}{2\pi^2(1-b)M_6^4}\int_0^{\infty} dk \,k\,
I_0(kr_0)K_0(k(\rho+\frac{r_0}{1-b})) J_0(k|\vec{x}_\bot|) +
\ldots \, , \label{eqn:f1} \ee
where the integral can be done in closed form, leading to
\cite{kalkil}
\be f \simeq -\frac{p}{2\pi^2(1-b)M_6^4} \frac{1}{|\vec{x}_\bot|^2
+ \rho^2} + \ldots \, . \label{venez} \ee
This is the $6D$ gravitational shock wave, constructed in
\cite{higherdwaves}, with the `lightning rod' term accounting for
the brane tension as found in \cite{kalkil}.

Let us now consider the near-critical limit, $b \rightarrow 1$,
which requires particular care. In this limit, Eq. (\ref{crossm4})
suggests that the crossover scale diverges, indicating that
gravity remains $4D$ out to extremely large distances, regardless
of the brane thickness. However our discussion of the
near-critical limit in Sec. (3), in particular interpretation of
Eq. (\ref{bulkcyl}), has already taught us otherwise. As we have
seen there, in the near-critical limit the bulk around the brane
compactifies to a long cylinder, which opens up into a cone only
very far from the brane. Bulk gravity in the throat is in the $5D$
regime, which separates the $4D$ and $6D$ ones. Thus the crossover
scale beyond which gravity is not $4D$ is really the demarkation
between the $4D$ regime, and $5D$ gravity which lives inside the
throat. Looking at the shock wave solution (\ref{thickshosol1br}),
we see that while the small argument expansion for $I_n(z)$ always
suffices for large distances from the source, in the near-critical
limit we must treat the functions $K_n(z)$ carefully for the
momenta in the regime $1-b \ll kr_0 \ll 1$. Indeed,
this corresponds to length scales $\ell \la \frac{r_0}{1-b}$ which are
shorter than the length of the conical throat. So setting $I_0(kr_0)
\rightarrow 1$ and $I_1(kr_0) \rightarrow 0$ in
(\ref{thickshosol1br}), we notice that  where the terms $\propto
M_5^3$ dominate, gravity will remain $4D$ because the factors of
$K_0(k\frac{r_0}{1-b})$ exactly cancel in the integrand. This
happens for the momenta $k$ which satisfy  $k >
\frac{M_6^4}{M_5^3}\frac{K_1(k\frac{r_0}{1-b})}{K_0(k\frac{r_0}{1-b})}$.
In the near-critical limit this can occur when the argument of
Bessel functions is much larger than unity, so that the ratio
$K_1/K_0$ will be of the order of unity! Indeed, as we decrease
$kr_0$, going out to larger distances, $K_1$ will remain close to
$K_0$: since $K_1(z) = -
\partial_z K_0(z)$ and now, since $b \rightarrow 1$, we have large
values of the argument for which $K_0(z) \rightarrow {\sqrt
\frac{\pi}{2z}}e^{-z}$, so that $K_1 \rightarrow K_0$. The
transition can occur only once, because $K_n$ are monotonic, and
for our approximation to be valid it must happen when
$\frac{kr_0}{1-b} \ge {\cal O}(1)$ - otherwise the situation will
be essentially the same as in a generic sub-critical case. In that
case the crossover scale saturates to $k_c = \frac{M_6^4}{M_5^3}$,
or therefore
\be r_c = \frac{M_5^3}{M_6^4} \, . \label{crosseesaw} \ee
Clearly, this is valid as long as $r_c < \frac{r_0}{1-b}
$. The
scale $r_c$ is exactly the see-saw scale of \cite{highercod}:
again using Gauss's law for the $4D$ Planck mass, $M_4^2 = 2\pi
M^3_5 r_0$, we find
\be r_c = \frac{M_4^2}{2\pi M_6^4 r_0} \, . \label{seesaw} \ee
This demystifies the appearance of the see-saw effect, showing
that at least with our regularization, it does happen naturally,
but only for  near-critical branes whose deficit angle is very
close to $2\pi$, so that they still are flat and static.

As an explicit check that inside the conical throat gravity looks $5D$,
we can take the limit of our solution for a fixed distance $|\vec x_\bot|$
and formally take the limit $b \rightarrow 1$ while holding $r_0$ fixed.
Then if we take the solution in Eq. (\ref{thickshosol1br}), in the limit $b
\rightarrow 1$ we can replace $K_n$'s by their large argument
expansion for any finite momentum. After cancelling the like
terms, and taking the thin brane limit, we find, using footnote
(2),
\be f(\vec x_\bot, \rho)  = -\frac{p}{2\pi^2} \int_0^\infty dk \,
\frac{ J_0(k|\vec{x}_\bot|)}{M_6^4r_0  +  M_5^3 r_0 k } = - 2 p
\int_{-\infty}^\infty \frac{d^2\vec k}{(2\pi)^2} \, \frac{e^{i
\vec k \cdot \vec x_\bot}}{2\pi M_6^4r_0 \, k +  2\pi M_5^3 r_0 \,
k^2 } \, . \label{5dthickshosol1} \ee
This is {\it precisely} the shock wave profile on a flat thin
3-brane in $5D$, on the normal branch, with brane and bulk Planck
masses $M_{4 \, eff}^2 = 2\pi M^3_5 r_0$ and $M_{5 \, eff}^3 =
2\pi M^4_6 r_0$, respectively. It can be obtained as a limit of
the normal branch solutions of \cite{shocks} on a tensionless
brane.

So to sum up, we find that for distances $r_0 < r <r_c$ the shock wave
reduces to the Aichelburg-Sexl solution \cite{aichsexl}, and becomes
its higher-dimensional generalization when $r>r_c$. The crossover
scale for sub-critical and near-critical branes, beyond which gravity is
not $4D$ anymore, is given by
\be
r^2_c = \cases{ \frac{M_4^2}{2\pi (1-b) M_6^4}
\ln\Bigl[\frac{2(1-b)}{kr_0}\Bigr] \, ,
& \, \,  $b \la 1$  \, ; \cr
\frac{M_4^4}{4\pi^2 M_6^8 r_0^2} \, ,
& \, \,  $b \rightarrow 1 $ \, , }
\label{crosssum}
\ee
only in the latter case agreeing with the see-saw scale discussed
in \cite{highercod}. The origin of the see-saw mechanism is now
clear: in the near-critical limit, the brane resides in a very
deep throat generated by its own tension. Thus its localized $4D$
gravity first changes to $5D$ gravity residing inside the throat,
and only later to full $6D$ gravity \cite{kalwall},
at distances $\rho \gg r_0/(1-b)$ (see discussion of Eq. (\ref{bulkcyl})).
This shows {\it how} the
see-saw mechanism emerges.

\section{Linearized Gravity}

\subsection{Perturbations and Their Gauge Symmetries}

We can solve exactly the field equations (\ref{fieldeqs}) for
localized matter sources only in the relativistic limit. For
nonrelativistic sources, we can try to solve them perturbatively,
assuming that the solutions converge to the background vacuum far
from the sources. So starting with a background $ds^2_6 = g_{AB}
\, dx^A dx^B$ metric, we define the metric perturbations $h_{AB}$
as is usual, by the substitution $g_{AB} \rightarrow g_{AB} +
h_{AB}$. In addition, we must also perturb the brane location $r_0
\rightarrow r_0 + \xi$ and any nontrivial field configurations on
the brane, in our case the axion $\Sigma \rightarrow \Sigma +
\sigma$. Before substituting these expansions in the field
equations (\ref{fieldeqs}), we must account for the gauge
symmetries of the perturbations, to identify physically meaningful
variables which cannot be undone with diffeomorphisms.  So suppose
that we transform $x^A \rightarrow \bar x^A = x^A + \chi^A$. A
straightforward calculation then shows that the perturbations
transform according to
\be h_{AB} \rightarrow \bar h_{AB} = h_{AB} - \nabla_A \chi_B -
\nabla_B \chi_A \, , ~~~~~ \xi \rightarrow \bar \xi = \xi +
\chi^\rho|_{\rho = r_0} \, , ~~~~~ \sigma \rightarrow  \bar \sigma
= \sigma - \chi^A|_{\rho = r_0} \nabla_A \Sigma \, ,
\label{gaugetransfs} \ee
where the $\nabla_A$ and raising and lowering of indices are
defined with respect to the background metric $g_{AB}$. On the
static thick brane background (\ref{thick6dvac}), we can
immediately compute the transformations (\ref{gaugetransfs}),
splitting them up with respect to different representations of the
$4D$ Lorentz symmetry. They are
\ba \bar{h}_{\mu\nu} &= & h_{\mu\nu} - \partial_\mu \chi_\nu -
\partial_\nu \chi_\mu \, ,  \nonumber \\
\bar{h}_{\phi\phi} &=& h_{\phi\phi} - 2\partial_\phi \chi_\phi
- \alpha' \chi^\rho \, , \nonumber \\
\bar{h}_{\mu \phi} &=&
h_{\mu\phi}- \partial_\mu \chi_\phi - \partial_\phi \chi_\mu \, ,
\nonumber \\
\bar{h}_{\mu \rho} &=&
h_{\mu \rho }- \partial_\mu \chi^\rho
- \chi_\mu{}' \, , \nonumber \\
\bar{h}_{\rho\phi} &=& h_{\rho\phi} - \partial_\phi \chi^\rho -
\chi_\phi{}' + \chi_\phi \frac{\alpha'}{\alpha} \, ,
\nonumber \\
\bar{h}_{\rho\rho}  &=&
h_{\rho\rho}- 2 \chi^\rho{}'  \, , \nonumber \\
\bar{\xi} & = & \xi + \chi^\rho(r_0) \, , \nonumber \\
\bar \sigma &=& \sigma - q \chi^\phi(r_0) \, ,
 \label{diffeos}
\ea
where
\be \alpha=g_{\phi\phi} = ((1-b\Theta(\rho-r_0))\rho +
br_0\Theta(\rho-r_0))^2 \, , \label{alphadef} \ee
is the area of the disk at fixed $\rho$ in Eq. (\ref{thick6dvac}),
and $\chi^\rho(r_0) = \chi^\rho|_{\rho=r_0}$. We need to carefully
raise and lower the angular index $\phi$, since although the brane
is intrinsically flat, the full geometry isn't. Note, that while
$\alpha$ is continuous, %it's derivative is not:
$\frac{\alpha'}{\sqrt{\alpha}} = 2(1-b\Theta(\rho-r_0))$, so that
on the brane, the discontinuity, defined for any function as the
jump of its value across the brane, $\Delta f = f|_+ - f|_- =
f(r_0{}^+) - f(r_0{}^-)$, is $\Delta \frac{\alpha'}{\alpha} = -
4\frac{\lambda_5}{M_6^4}$, where we have used Eq. (\ref{newdef}).

Using these transformations we can {\it always} gauge fix the
metric perturbed by axially symmetric sources to brane-fixed
Gaussian-normal gauge, in the linear order. The key is to be
careful about what axial symmetry means: the vector field
$\partial_\phi$ must be a Killing vector of the metric,
unperturbed as well as perturbed. Hence any tensor, including the
metric, must be Lie-derived by $\partial_\phi$ to zero.  Now,
using (\ref{diffeos}), we can readily pick $\chi^\rho$ and
$\chi_\phi$ to set $\bar h_{\rho\rho}$ and $\bar h_{\rho\phi}$ to
zero, respectively, treating the corresponding transformation
rules in (\ref{diffeos}) as differential equations defining
$\chi^\rho$ and $\chi_\phi$. We can always choose the boundary
condition $\chi^\rho(r_0) = 0$, opting to keep the perturbation of
the brane location $\xi$, or brane bending, to remain unchanged,
for the moment. Next, we can pick a $\chi_\mu =
\chi_\mu(\rho,\phi)$ to set $\bar h_{\mu\rho}$ and $\bar
h_{\mu\phi}$ to zero too. This may seem to induce angular
dependence of $\bar h_{\mu\nu}$ by the first of Eqs.
(\ref{diffeos}). However, it's easy to check that in the
coordinates where $\bar h_{\rho\rho} = \bar h_{\rho\phi} = \bar
h_{\mu\phi} = \bar h_{\mu\rho} = 0$ the Killing conditions
enforcing axial symmetry imply that $\partial_\phi \bar h_{\mu\nu}
= \partial_\phi \bar h_{\phi\phi} = 0$. Thus the metric
perturbations $\bar h_{AB}$ are $\phi$-independent. In fact, this
proof extends beyond linear perturbation theory, as the reader can
check by consulting \cite{weinbook}. Notice that we have now
completely used up $\chi^\phi(r_0)$ - we can't change $\sigma$
without also changing $h_{\rho\phi}$, as is clear from
(\ref{diffeos}). This shows that the lightest in the tower of the
KK vectors $h_{\rho a}$ eats the brane-localized axion $\sigma$ to
become massive. Alternatively, when we pick the gauge $h_{\rho
a}=0$, as we did here, we are separating out the scalar $\sigma$
as the St\"uckelberg field of $h_{\rho a}$.

Dropping the overbars, note that in any Gaussian-normal gauge
$h_{\rho\rho} = h_{\rho\phi} = h_{\mu\phi} = h_{\mu\rho} = 0$, we
still have the residual gauge freedom generated by $\chi_\mu$,
which are independent of $\rho$ and $\phi$, and by
$\chi^\rho(r_0)$. These transformations do not change our gauge.
They do change $h_{\mu\nu}$, $h_{\phi\phi}$, $\xi$.
Transformations generated by $\chi^\mu$ are the standard $4D$
diffeomorphisms, and we will find their uses later on.
Transformations generated by $\chi^\rho(r_0)$ allow us to remove
the brane bending, nailing the brane straight to its background
value, $\rho = r_0$. In this gauge we can quickly find the field
equations for the perturbations, by directly substituting the
gauge-fixed perturbed metric into the field equations
(\ref{fieldeqs}) or (\ref{riccieq}).

Going to this gauge requires a little care, however. As we have
noted above, $\frac{\alpha'}{\alpha}$ is discontinuous on the
brane in the background (\ref{thick6dvac}). Hence when
$\chi^\rho(r_0) \ne 0$, the gauge transformation (\ref{diffeos})
changes the metric perturbation $h_{\phi\phi}$ {\it
discontinuously}. On the other hand, in brane-fixed
Gaussian-normal gauge the full field configuration has
discontinuities on the {\it straight} brane, proportional to its
the stress-energy. These must remain encoded in the
$\delta$-function sources in (\ref{riccieq}). Hence the metric
perturbations in brane-fixed Gaussian-normal gauge must be
continuous across the brane, and their derivatives may have at
most a finite jump there. In any other gauge found by using a
transformation (\ref{diffeos}) with $\chi^\rho(r_0) \ne 0$, the
metric perturbation $h_{\phi\phi}$ will appear discontinuous, so
that the brane matching conditions on the perturbations can only
be found by carefully implementing the Israel junction conditions
(for an illustration, see e.g. \cite{pesota}). So to get the field
equations, we choose to work in brane-fixed Gaussian-normal gauge,
where we get the brane matching conditions by using Gaussian
pillbox integration around the brane, and relating the normal
derivative discontinuities to $\delta$-sources in the field
equations. This discussion applies completely generally to {\it
any} model where a codimension-2 brane is resolved by a
cylindrical codimension-1 brane.

Nevertheless, it turns out the bulk field equations for
perturbations are easier to solve by stepping out of the
brane-fixed Gaussian-normal gauge, to a gauge where the
$\rho$-$\phi$ cone appears conformally flat, and explicitly
keeping the brane bending $\xi$. The latter is similar to what
happens in other codimension-1 cases \cite{gata}, except that here
the brane bending must be turned on even in the $4D$ vacuum
(\ref{thick6dvac}). The reason is that the $4D$ vacuum backgrounds
(\ref{thick6dvac}) are {\it not} codimension-1 vacua: the axion
flux breaks the symmetry between the longitudinal brane
directions. Brane bending probes this `vacuum' anisotropy. Using
the conformally flat $\rho$-$\phi$ metric, on the other hand,
simplifies the bulk field equations. For later use, we list here
how to relate such gauges. In a Gaussian-normal gauge the only
nonvanishing metric perturbation is $\hat h_{\phi\phi}$, now
denoted with the caret. Defining $\hat \Phi$ by $\hat h_{\phi\phi}
= \alpha \hat \Phi$, the perturbed metric is
\be ds_6{}^2 = (\eta_{\mu\nu} +  h_{\mu\nu}) \, dx^\mu dx^\nu +
d\rho^2 + (1+{\hat \Phi})\,\alpha(\rho) \, d\phi^2  \, ,
\label{gngperts} \ee
with $\alpha$ defined in (\ref{alphadef}). Let the brane bending
term $\xi$ in this gauge be non-zero. To go to a gauge where it is
zero, we apply (\ref{diffeos}), choosing $\chi^\rho(r_0) = - \xi$.
That takes us to the gauge $\tilde \xi = 0$, where the metric has
the same form as (\ref{gngperts}), but with
\be \tilde \Phi = \hat {\Phi} + \frac{\alpha'}{\alpha}\xi \, ,
\label{bfGn} \ee
instead of $\hat \Phi$. This shows how $\tilde \Phi$ `ate' $\xi$
in brane-fixed Gaussian-normal gauge. On the other hand, to
relate a general Gaussian-normal gauge to a conformal gauge, we
use the relationship
\be \hat{\Phi} = \Phi - \frac{\alpha'}{2\alpha}\int_{r_0}^\rho
d\rho \, \Phi \, , \label{conftrans} \ee
which is a gauge transformation of the form (\ref{diffeos}) with
$\chi^\rho = \frac12 \int_{r_0}^\rho d\rho \, \Phi$, and other
gauge parameters set to zero. For this gauge transformation,
$\chi^\rho(r_0) = 0$, and therefore it does not change the
discontinuities of the perturbations in different gauges. Now, to
see which gauge this transformation takes us from, we need to
invert this integral equation, and determine $\Phi$. That is
straightforward. By definition of $\chi^\rho$, we have $\Phi =
\frac12 {\chi^\rho}'$, so we can view (\ref{conftrans}) as a
differential equation for $\chi^\rho$, with the boundary condition
$\chi^\rho(r_0) = 0$. The equation is ${\chi^\rho}' -
\frac{\alpha'}{2\alpha} \chi^\rho = \hat \Phi$, and the solution
satisfying the boundary condition is $\chi^\rho = \sqrt{\alpha}
\int^\rho_{r_0} d\rho \, \frac{\hat \Phi}{\sqrt{\alpha}}$. Thus
\be \Phi = \frac12 \frac{d}{d\rho} \Bigl(\sqrt{\alpha}
\int^\rho_{r_0} d\rho \, \frac{\hat \Phi}{\sqrt{\alpha}}\Bigr)\, .
\label{confgdef} \ee
Then (\ref{conftrans})  and (\ref{diffeos}) imply that in the
original gauge $h_{\phi\phi} = \sqrt{\alpha} \Phi$. Furthermore,
from $\hat h_{\rho\rho} = 0$ and Eqs. (\ref{conftrans}) and $\Phi
= \frac12 {\chi^\rho}'$ we see that $h_{\rho\rho} = \sqrt{\alpha}
\Phi$ too. So indeed, the gauge transformation (\ref{diffeos})
defined by (\ref{conftrans}), (\ref{confgdef}) relates the general
Gaussian-normal perturbed metric \ (\ref{gngperts}) to the metric
\be ds_6{}^2 = (\eta_{\mu\nu} +  h_{\mu\nu}) \, dx^\mu dx^\nu
+(1+{\Phi}) \Bigl( d\rho^2 + \,\alpha(\rho) \, d\phi^2 \Bigr) \, ,
\label{confgperts} \ee
If in addition we want to relate brane-fixed Gaussian-normal
gauge perturbation to the $2D$ conformal gauge perturbation
(\ref{confgperts}), we can combine together the two cases. Then,
(\ref{bfGn}) and (\ref{conftrans}) yield
\be \tilde \Phi = \Phi - \frac{\alpha'}{2\alpha}\int_{r_0}^\rho
d\rho \, \Phi + \frac{\alpha'}{\alpha}\xi \, , \label{gntoconf}
\ee
relating the metric perturbation $\Phi$ in a $2D$ conformal gauge
with brane bending $\xi$ to the metric perturbation $\tilde \Phi$
in brane-fixed Gaussian-normal gauge. Since $\tilde \Phi$ must
be continuous on the brane, we see from (\ref{gntoconf}) that
$\Phi$ is not. Instead, it jumps across the brane by $\Delta \Phi
= - \Delta(\frac{\alpha'}{\alpha}) \xi = 4 \frac{\lambda_5}{M_6^4}
\xi$. These formulae will come in handy later on.

\subsection{Linearized Field Equations and Helicity Decomposition}

We find the field equations for perturbations with axially
symmetric sources by substituting the perturbed metric in the
brane-fixed Gaussian-normal gauge,
\be ds_6{}^2 = (\eta_{\mu\nu} +  h_{\mu\nu}) \, dx^\mu dx^\nu +
d\rho^2 + (1+{\tilde \Phi})\,\alpha(\rho) \, d\phi^2  \, ,
\label{gngbfix} \ee
with $\tilde \xi = 0$, into (\ref{riccieq}), and expanding to the
linear order in the perturbation $\tilde \Phi$. Since the field
equation $\nabla_a \nabla^a \Sigma = 0$ reduces to the source-free
$4D$ Klein-Gordon equation for the axion perturbation $\sigma$:
$\partial_4{}^2 \sigma = 0$, in the linear order the axion is not
sourced by axially symmetric matter distributions. We can
simply set $\sigma$ to zero from now on. This is self-consistent
at distances $|\vec x|, \rho \gg r_0$. In this limit we can
dimensionally reduce the brane theory to a set of KK towers with a
mass gap $\propto 1/r_0$. Since a background on the brane which is
not axially symmetric must involve higher-level KK states, to
encode the angular variation around the brane, at low energies $E
\ll 1/r_0$ we can only excite the lightest states, and the
backgrounds will automatically be axially symmetric.  In fact,
since the axion $\sigma$ really is the St\"uckelberg field of the
vector $h_{\rho a}$, it will only be turned on when the vector
field is turned on, which means when there is a non-trivial
current that can source $h_{\rho a}$.

Evaluating the curvature tensors and the stress-energy terms in
(\ref{riccieq}) we find the field equations for perturbations in
this limit. They are the momentum constraint, coming from the
$\rho$-$\mu$ components,
\be \partial_\lambda \big(h^\lambda{}_\mu' - \delta^\lambda{}_\mu
(h_4' + \tilde{\Phi}' + \frac{\alpha'}{2\alpha}\tilde{\Phi})\big)
= 0 \, , \label{momcons} \ee
the trace equation,
\be h_4'' + \tilde \Phi'' + \frac{\alpha'}{\alpha}\tilde \Phi' =
\frac{1}{2M_6^4}\bigg(\tau^a{}_a - \frac{3M_5^3}{2}\partial_4{}^2
\tilde \Phi + 3\lambda_5 \tilde \Phi +
\frac{3M_5^3}{2}\big(\partial^\lambda
\partial_\sigma h^\sigma{}_\lambda - \partial_4{}^2
h_4\big)\bigg)\delta(\rho-r_0) \, , ~~~
\label{trace}
\ee
the `radion field' equation, coming from the $\phi$-$\phi$
component,
\ba &&\partial_4{}^2 \tilde \Phi + \tilde \Phi'' +
\frac{\alpha'}{\alpha}\tilde \Phi' +
\frac{\alpha'}{2\alpha}h_4' = \nonumber \\
&& ~~~~~~~~~~ \frac{1}{2M_6^4}\bigg(\tau^a{}_a - 4\tau^\phi{}_\phi
- \frac{3M_5^3}{2}\partial_4{}^2 \tilde \Phi + 7\lambda_5 \tilde
\Phi - \frac{M_5^3}{2}\big(\partial^\lambda \partial_\sigma
h^\sigma{}_\lambda - \partial_4{}^2 h_4\big)\bigg)\delta(\rho-r_0) \, ,
~~~~~~ \label{phieqs} \ea
and finally the $4D$ tensor equations,
\ba &&\partial^\mu \partial_\lambda h^\lambda{}_\nu + \partial_\nu
\partial_\lambda h^{\mu\lambda} - \nabla_6{}^2 h^\mu{}_\nu -
\partial^\mu\partial_\nu h_4 -
\partial^\mu\partial_\nu \tilde \Phi = \nonumber \\
&& ~~~~
\frac{2}{M_6^4}\Bigg[\tau^\mu{}_\nu - \frac{1}{4}\tau^a{}_a
\delta^\mu{}_\nu + \frac{M_5^3}{2}\big(\partial^\mu\partial_\nu \tilde \Phi
- \frac{1}{4}\partial_4{}^2 \tilde \Phi \delta^\mu{}_\nu \big) +
\frac{\lambda_5}{4}\tilde \Phi \delta^\mu{}_\nu -
\label{prttensors}\\
&& ~~~~ \frac{M_5^3}{2}\Bigl(\partial^\mu \partial_\lambda
h^\lambda{}_\nu +
\partial_\nu \partial_\lambda h^{\mu\lambda} -
\partial_4{}^2 h^\mu{}_\nu - \partial^\mu\partial_\nu h_4
- \frac{1}{4}\delta^\mu{}_\nu\big(\partial^\lambda \partial_\sigma
h^\sigma{}_\lambda - \partial_4{}^2 h_4\big)\Bigr)\Bigg]
\delta(\rho-r_0) \, . ~~~~~ \nonumber
\ea
Here $h_4 = h^\mu{}_\mu$ is the trace of the $4D$ perturbation
$h_{\mu\nu}$, the sources $\tau^a{}_b$ describing the brane matter
perturbations naturally decompose as $\tau^\mu{}_\nu$ and
$\tau^\phi{}_\phi$, and $\tau^a{}_a = \tau^\mu{}_\mu +
\tau^\phi{}_\phi$ the full $5D$ trace of $\tau^a{}_b$. The
covariant d'Alembertian $\nabla_6{}^2$ is defined as before, with
respect to the full background metric (\ref{thick6dvac}).

This system contains redundancies, which we need to disentangle
away. First, we need to break down the perturbation
$h^{\mu}{}_{\nu}$ into irreducible $4D$ representations of the
Lorentz group, which is the symmetry of the background
(\ref{thick6dvac}). Using the flat limit of the decomposition
theorem in \cite{cgkp}, we can write
\be h^\mu{}_\nu = {\gamma }^\mu{}_\nu + \partial^\mu A_\nu +
\partial_\nu A^\mu + \partial^\mu{}\partial_\nu \Psi -
\frac{1}{4}\delta^\mu{}_\nu \,
\partial_4{}^2 \Psi + \frac{1}{4}\delta^\mu{}_\nu \, h_4 \, ,
\label{perthel} \ee
where $\gamma^\mu{}_\nu$ is a transverse-traceless ({\tt TT})
tensor satisfying $\partial_\mu \gamma^\mu{}_\nu =
\gamma^\mu{}_\mu = 0$, with 5 helicities, $A_\mu$ is a
Lorenz-gauge vector, $\partial_\mu A^\mu = 0$, with 3 helicities,
and $\Psi$ and $h_4 = h^\mu{}_\mu$ are two scalars. The total
number of degrees of freedom adds up to ten, which is the number
of independent components of a symmetric $4 \times 4 $ tensor.
Once we recall that these modes are all dependent on $\rho$ and,
in principle, could also depend on $\phi$, we really end up with
KK towers of degrees of freedom classified by their $4D$ masses in
addition to their helicities. Of course, not all of these modes
may end up being physical; some could be pure gauge modes, and one
has to carefully check that using the field equations and the
gauge symmetries of the problem. As we will only work with axially
symmetric perturbations, our task will be considerably simpler.

The field equations (\ref{momcons})-(\ref{prttensors}) for the
linearized theory also decompose by helicities due to $4D$ Lorentz
invariance of the background. A simple way to separate out the
equations is to define the  transverse-traceless tensor projection
operator $K_{\mu \nu \alpha \beta}$, such that
$K_{\mu\nu}{}^{\alpha \beta} h_{\alpha \beta} = \gamma_{\mu\nu}$.
This operator must satisfy $K_{\mu\nu\alpha\beta} =
K_{\nu\mu\alpha\beta} = K_{\mu\nu\beta\alpha}$ and
$K^\mu{}_{\mu\alpha\beta} = \partial^\mu K_{\mu\nu\alpha\beta}=
0$. The unique solution is
\ba K_{\mu\nu\alpha\beta} &=& - \,
\frac{1}{3}\Bigg[\eta_{\mu\nu}\eta_{\alpha\beta}
-\frac{3}{2}\bigg(\eta_{\mu\alpha}\eta_{\nu\beta} +
\eta_{\mu\beta}\eta_{\nu \alpha}\bigg) -
\eta_{\mu\nu}\frac{\partial_\alpha \partial_\beta}{\partial_4{}^2}
-  \eta_{\alpha\beta}\frac{\partial_\mu \partial_\nu}{\partial_4{}^2}
\nonumber \\
&& + \, \frac{3}{2}\bigg( \eta_{\nu \beta} \frac{\partial_\mu
\partial_\alpha}{\partial_4{}^2}+ \eta_{\mu\beta}
\frac{\partial_\nu \partial_\alpha}{\partial_4{}^2}
+ \eta_{\nu \alpha} \frac{\partial_\mu \partial_\beta}{\partial_4{}^2}
+ \eta_{\mu\alpha} \frac{\partial_\nu \partial_\beta}{\partial_4{}^2} \bigg)
- 2\frac{\partial_\mu \partial_\nu \partial_\alpha
\partial_\beta}{\partial_4{}^4}\Bigg] \, . \label{projop} \ea
Clearly, this operator annihilates the vector and scalar terms in
(\ref{perthel}), and when we apply it to Eq. (\ref{prttensors}),
it projects out the {\tt TT} contributions. Since the vacuum
(\ref{thick6dvac}) is $4D$-flat, stress energy conservation
implies $\partial_\mu \tau^\mu{}_\nu = 0$, and so
\be K_{\mu\nu}{}^{\alpha\beta}\tau_{\alpha\beta} = \tau^\mu{}_\nu
- \frac{1}{3}\bigg(\delta^\mu{}_\nu - \frac{\partial^\mu
\partial_\nu}{\partial_4{}^2}\bigg) \tau^\alpha{}_\alpha  \, .
\label{projsetensor} \ee
Hence our final form of the {\tt TT}-tensor field equation is
\be \nabla_6{}^2 \gamma^\mu{}_\nu +
\frac{M_5^3}{M_6^4}\partial_4{}^2 \gamma^\mu{}_\nu \, \delta(\rho-r_0) =
-\frac{2}{M_6^4}\bigg[\tau^\mu{}_\nu -
\frac{1}{3}\bigg(\delta^\mu{}_\nu - \frac{\partial^\mu
\partial_\nu}{\partial_4{}^2}\bigg) \tau^\alpha{}_\alpha
\bigg]\delta(\rho-r_0) \, . \label{tttensors} \ee

Subtracting the {\tt TT}-tensor equation from (\ref{prttensors})
leads to the system which contains only the vector and scalar
contributions. We can extract the vector in a similar way as the
{\tt TT}-tensor, by defining the vector projection operator
$V_{\alpha\beta}$, and acting on the momentum constraint 
(\ref{momcons}). Since the backgrounds (\ref{thick6dvac}) are
Lorentz-invariant, in the linear order of perturbation theory
vector sources vanish and the vector modes decouple from the
matter distribution. Hence we can set the vector field in
(\ref{perthel}) to zero.

This leaves us with scalars. In brane-fixed Gaussian-normal
gauge, they are $\tilde \Phi$, $\Psi$ and $h_4$, obeying the field
equations (\ref{momcons})-(\ref{phieqs}) and what remains of
(\ref{prttensors}) after we subtract out (\ref{tttensors}). To
pick out the minimal set of independent equations for scalars from
the system (\ref{momcons})-(\ref{prttensors}), we define $X =
\frac{1}{4} (h_4 - \partial_4{}^2 \Psi)$, write the scalar part of
the metric perturbation as
\be {}^s h^\mu{}_\nu = \partial^\mu \partial_\nu \Psi +
X\delta^\mu{}_\nu \, ,
\label{scalarpert} \ee
and substitute it into the difference of Eqs. (\ref{prttensors})
and (\ref{tttensors}). The result is a combination of two
independent $4D$ tensors, diagonal $\propto \delta^\mu{}_\nu$ and
off-diagonal $\propto \partial^\mu \partial_\nu$. Hence the
coefficients of these tensors must separately vanish. Likewise, we
could write the $4D$ trace of this equation. However it is a
trivial linear combination of the diagonal and off-diagonal terms,
and so we can ignore it immediately. Then, combining these
equations with (\ref{momcons})-(\ref{phieqs}), and substituting
(\ref{scalarpert}) for $h_{\mu\nu}$, we get
\ba \partial_\lambda \Bigl( 3X' + {\tilde \Phi}' +
\frac{\alpha'}{2\alpha} \tilde \Phi  \Bigr) = 0 \, ,
 ~~~~~~~~~~~~~~~~~~~~~~~~~~~~~~~~~~~~~~~~~~
 ~~~~~~~~~~~~~~~~~~~~~~~~~~~~~~~~~~~~&& \label{momconsa} \\
\tilde \Phi'' + \frac{\alpha'}{\alpha}\tilde \Phi' +
\partial_4{}^2 \Psi'' + 4X'' =
\frac{1}{2M_6^4}\bigg(\tau^\mu{}_\mu + \tau^\phi{}_\phi  +
3\lambda_5 \tilde \Phi - \frac{3M_5^3}{2}\partial_4{}^2 \big(
\tilde \Phi + 3X \big)  \bigg)\delta(\rho-r_0) \, , ~~
&& \label{tracea} \\
\partial_4{}^2 \tilde \Phi + \tilde \Phi'' +
\frac{\alpha'}{\alpha}\tilde \Phi' +
\frac{\alpha'}{2\alpha} \Bigl( \partial_4{}^2 \Psi' + 4X' \Bigr) =
~~~~~~~~~~~~~~~~~~~~~~~~~~~~
~~~~~~~~~~~~~~~~~~~~~~~~~~~~~~~~
&& \nonumber \\
 ~~~~~~~~~~~~~~= \frac{1}{2M_6^4}\bigg(\tau^\mu{}_\mu - 3\tau^\phi{}_\phi
+ 7\lambda_5 \tilde\Phi - \frac{3M_5^3}{2}\partial_4{}^2
\big(\tilde \Phi - X\big) \bigg)\delta(\rho-r_0) \, ,
~~~~~~~~~~~ && \label{phieqsa} \\
\Psi'' + \frac{\alpha'}{2\alpha} \Psi'  + \tilde \Phi + 2 X =
\frac{1}{M_6^4}\bigg(\frac{2}{3M_5^3} {\partial_4{}^{-2}}(
\tau^\mu{}_\mu)  - M_5^3 \big(\tilde \Phi + 2X\big) \bigg)\delta(\rho-r_0)
\, , ~~~~~~~~~~~~~~~~~~
&& \label{psia} \\
X'' + \frac{\alpha'}{2\alpha} X' + \partial_4{}^2 X = -
\frac{1}{4M_6^4}\bigg( \frac23 \big(\tau^\mu{}_\mu -
3\tau^\phi{}_\phi \big) +2 \lambda_5 \tilde \Phi - M_5^3
\partial_4{}^2 \big(\tilde \Phi - X \big)  \bigg)\delta(\rho-r_0)
\, , ~~~~~ &&\label{xa} \ea
where we have  decomposed the full d'Alembertian as $\nabla_6{}^2
= \partial_4{}^2 + \nabla_{\vec y}^2$ and used the formula for
axially symmetric fields $\nabla_{\vec y}^2 = \frac{d^2}{d\rho^2}
+ \frac{\alpha'}{2\alpha} \frac{d}{d\rho}$. We have replaced the
$5D$ stress-energy trace by $\tau^a{}_a= \tau^\mu{}_\mu +
\tau^\phi{}_\phi$. In Eq. (\ref{momconsa}), we can immediately
omit the $4D$ divergence $\partial_\lambda$ because the $4D$
background is flat Minkowski. In Eq. (\ref{psia}),
$\partial_4{}^{-2}$ acting on $\tau^\mu{}_\mu$ can be simply
understood as a multiplication of the Fourier components of
$\tau^\mu{}_\mu$ by $-1/k^2$, as we will do later on. Finally, we
note that the bulk terms in Eq. (\ref{xa}) depend only on $X$, and
hence can be readily integrated. Thus we should keep it in our
final set of independent field equations.

We find that if we add Eqs. (\ref{tracea}) and
(\ref{phieqsa}), and subtract from the result $\partial_4{}^2
\times$ Eq. (\ref{psia})  and $4 \times$ Eq. (\ref{xa}), the
result is a linear combination of the $\rho$-derivative of Eq.
(\ref{momconsa}) and Eq. (\ref{xa}). Next, we can similarly start
with Eq. (\ref{tracea}) but now subtract Eq. (\ref{phieqsa}),
$\partial_4{}^2 \times$ Eq. (\ref{psia})  and $4 \times$ Eq.
(\ref{xa}). The result is very simple:
\be \partial_4{}^2 \Bigl(\tilde \Phi + 3X \Bigr) +
\frac{\alpha'}{2\alpha} \Bigl(\partial_4{}^2 \Psi' + 4X' \Bigr) =
0 \, . \label{zeroeq} \ee
In this equation, the sources on the right hand-side completely
cancel, since they are localized on the brane. The reason is that
this is the perturbed gravitational equation in the radial
direction, which is sourceless in brane-fixed Gaussian-normal
gauge. This equation is particularly useful when we look for the
solutions in the bulk, as we will see below. Further, we can also
check, by using the $\rho$-derivative of (\ref{momconsa}), that
the bulk part of Eq. (\ref{psia}) reduces precisely to Eq.
(\ref{zeroeq}).

These observations teach us that in the bulk we can trade Eqs.
(\ref{tracea}), (\ref{phieqsa}) and (\ref{psia}) for one
independent combination, which, with wisdom after the fact, we
choose to be precisely Eq. (\ref{zeroeq}). This, however, is
slightly tricky on the brane itself. A  product of functions which
are discontinuous on the brane appears in Eq. (\ref{zeroeq}):
$\frac{\alpha'}{\alpha}$ and $\partial_4{}^2 \Psi' + 4X'$. Hence
although this equation contains no terms $\propto
\delta(\rho-r_0)$, it still gives some information about the
boundary behavior of the scalars, albeit weaker. It requires that
$\Delta\Bigl(\frac{\alpha'}{2\alpha} (\partial_4{}^2 \Psi' + 4X'
)\Bigr) = 0$. Although the factors are bounded, because they are
discontinuous this condition -- by itself -- is degenerate: it
relates the jumps of the functions to their values on different
sides of the brane. So using it alone we can't extract
unambiguously the junction conditions for $\partial_4{}^2 \Psi' +
4X'$. However, the degeneracies in the boundary information in
(\ref{zeroeq}) are completely artificial. They are easily lifted
by going back to any of the boundary conditions encoded in the
$\delta$-function terms in either of Eqs. (\ref{tracea}),
(\ref{phieqsa}) and (\ref{psia}). From examining these equations
it is clear that the most useful form of the boundary condition is
obtained from Eq. (\ref{psia}), because the boundary condition for
$\tilde \Phi$ is tied to the one for $X$ by Eq. (\ref{momconsa}).
Thus, our independent equations in the bulk are (\ref{momconsa}),
(\ref{xa}) and (\ref{zeroeq}), supplemented with the boundary
conditions obtained by Gaussian pillbox integration of
(\ref{psia}) and (\ref{xa}) around the brane. Further, we can
immediately integrate (\ref{momconsa}) once, with respect to
$x^\lambda$, and ignore the non-normalizable homogeneous mode on
the brane. With this, finally, our system of equations reduces to
the bulk system
\ba && 3X' + {\tilde \Phi}'
+ \frac{\alpha'}{2\alpha} \tilde \Phi  = 0 \, , \label{momconsaf} \\
&& \partial_4{}^2 \Bigl(\tilde \Phi + 3X \Bigr) +
\frac{\alpha'}{2\alpha} \Bigl(\partial_4{}^2 \Psi' + 4X' \Bigr) =
0 \, ,
\label{zeroeqf} \\
&& X'' + \frac{\alpha'}{2\alpha} X' + \partial_4{}^2 X = 0 \, ,
\label{xaf} \\
&& \Delta X' = -\frac{1}{4M_6^4}\bigg( \frac23 \big(\tau^\mu{}_\mu
- 3\tau^\phi{}_\phi \big) +2 \lambda_5 \tilde \Phi - M_5^3
\partial_4{}^2 \big(\tilde \Phi - X \big)  \bigg) \, ,
\label{Xbdy} \\
&& \Delta \Psi' = -\frac{M_5^3}{M_6^4}\big(\tilde \Phi + 2X\big) \, ,
\label{psibdy}
\ea
where the last two encode the boundary conditions, equating the
jumps of bulk derivatives to the brane stress-energy. This set of
equations is fairly straightforward to solve: Eq.
(\ref{momconsaf}) relates $\tilde \Phi$ to $X$, which is
determined by Eq. (\ref{xaf}) and boundary condition (\ref{Xbdy}).
Then Eqs. (\ref{zeroeqf}) and (\ref{psibdy}) determine the
remaining variable, $\Psi$.

\subsection{Solutions}

We now turn to finding the solutions of the system
(\ref{tttensors}), (\ref{momconsaf})-(\ref{psibdy}). The most
interesting solutions are those which describe the long range
fields of static sources on the brane. They are our key probe of
what kind of gravity the theory reduces to at long distances. One
could quite straightforwardly also determine the spectrum of the
theory around the vacuum, when the stress-energy sources
$\tau^a{}_b = 0$. We have done it, to find that the spectrum
contains a continuum of massive tensors and a scalar, fully
described by one independent variable $X$. In this section we will
give the solutions of linearized theory which apply to a general
case of arbitrary time dependent solutions. However what is really
important is to see how these modes couple to matter in order to
determine just how dangerous, or not, the extra modes may be at
long distances. For this reason, we will focus on the long range
static fields, completely determined by the static rings of matter
on the brane. We will also write down the formal solution for the
full metric perturbation $h^\mu{}_\nu$, which will enable us to
see the tensor structure of full long range gravity in the
linearized theory.

Let us start with the field equation governing the tensor modes,
Eq. (\ref{tttensors}). This equation can be solved very easily
using the standard Green's function techniques. Expanding out the
bulk d'Alembertian $\nabla_6{}^2$, and Fourier-transforming over
the $4D$ variables $x^\mu$, while keeping in mind that the fields
are axially symmetric, to $\gamma^\mu{}_\nu(x,\rho) = \int
\frac{d^4 k}{(2\pi)^4} \gamma^\mu{}_\nu(k,\rho) \exp(i k \cdot x)$
and  $\tau^a{}_b(x,\rho) = \int \frac{d^4 k}{(2\pi)^4}
\tau^a{}_b(k,\rho) \exp(i k \cdot x)$, we can rewrite the {\tt
TT}-tensor equation as
\be \gamma^\mu{}_\nu{}'' + \frac{\alpha'}{2\alpha}
\gamma^\mu{}_\nu{}' - k^2 \gamma^\mu{}_\nu - \frac{M_5^3}{M_6^4}
k^2  \gamma^\mu{}_\nu \, \delta(\rho-r_0) =
-\frac{2}{M_6^4}\bigg[\tau^\mu{}_\nu -
\frac{1}{3}\bigg(\delta^\mu{}_\nu - \frac{k^\mu k_\nu}{k^2}\bigg)
\tau^\alpha{}_\alpha \bigg]\delta(\rho-r_0) \, . ~~~
\label{tttensorsa} \ee
The solution of this equation can be written in terms of the
Green's function ${\cal G}_{\tt TT}(k,\rho)$, defined by the
equation
\be {\cal G}_{\tt TT}{}'' + \frac{\alpha'}{2\alpha} {\cal G}_{\tt
TT}{}' - k^2 {\cal G}_{\tt TT}  - \frac{M_5^3}{M_6^4} k^2 {\cal
G}_{\tt TT} \, \delta(\rho-r_0) = \frac{1}{M_6^4} \,
\delta(\rho-r_0) \, . \label{greens} \ee
Then the solution is simply
\be \gamma^\mu{}_\nu(k, \rho) = - 2 \, {\cal G}_{\tt TT}(k, \rho) \,
\bigg[ \tau^\mu{}_\nu - \frac{1}{3}\bigg(\delta^\mu{}_\nu -
\frac{k^\mu k_\nu}{k^2}\bigg) \tau^\alpha{}_\alpha \bigg] \, ,
\label{grttsoln} \ee
where we postpone the detailed determination of ${\cal G}_{\tt
TT}$ until later. For now, it is enough to see that it exists,
after we specify appropriate boundary conditions at bulk infinity
and inside the wrapped brane, which ensure the regularity of the
solution.

The scalar story is more intricate. The reason is that the
backgrounds (\ref{thick6dvac}), which we take as $4D$ vacua of the
theory, are not vacuum solutions from either the bulk or the
wrapped 4-brane theory points of view, because they contain a
non-trivial $\Sigma$ field contribution to stress-energy on the
brane. As we discussed, that is necessary to wrap the brane on a
circle and make it look like a codimension-2 object. However,
because of this, the brane bending term will always appear in
perturbation theory around the wrapped vacuum, even if the matter
perturbations $\tau^a{}_b$ vanish, in contrast to the simpler
codimension-1 cases studied in \cite{gata}. This is interesting
because it lends to a difference in the couplings of scalar metric
perturbations, and the helicity-0 modes from the massive graviton
multiplets at the same mass levels.

So let us construct the solution of
(\ref{momconsaf})-(\ref{psibdy}). We first Fourier-transform all
the fields over the $4D$ $x^\mu$ space, as in the case of {\tt
TT}-tensors above. Then we take a slightly roundabout approach,
due to the presence of the background flux of $\Sigma$, reflected
by the tension-dependent terms in (\ref{momconsaf})-(\ref{psibdy})
(explicit, as well as implicit in $\frac{\alpha'}{\alpha}$). We
need to carefully identify how they mix different scalar fields.
To this end it is useful to recall the gauge transformations
(\ref{gntoconf}), and explicitly restore the brane bending term
$\xi$ which keeps the brane-localized terms in check. Using
(\ref{gntoconf}) and the background equation
$(\frac{\alpha'}{\alpha})' = -\frac{\alpha'^2}{2\alpha^2} -
4\frac{\lambda_5}{M_6^4} \delta(\rho-r_0)$, we can check that
$\tilde \Phi' + \frac{\alpha'}{2\alpha} \tilde \Phi = \Phi'$.
Substituting this in Eq. (\ref{momconsaf}), we find $3X' + \Phi' =
4\frac{\lambda_5}{M_6^4} \xi \delta(\rho-r_0)$, which is very easy
to integrate over $\rho$: we get $3X + \Phi =
4\frac{\lambda_5}{M_6^4} \xi \Theta(y) + F$, where $F$ is a
Fourier-transform of some  function $F(x)$. However, we recall
that this equation was already integrated once over $x^\mu$, which
implies that $\partial_\mu F = 0$ for consistency, or $F = {\rm
const.}$. However, any such integration constant can be readily
removed by a residual $4D$ diffeomorphism of the form $\chi^\mu =
\frac{F}{2} x^\mu$ as one can immediately check. Thus we can set
$F=0$, to find
\be
\Phi = -3 X + \frac{4\lambda_5}{M_6^4}\xi\,\Theta(\rho - r_0) \, .
\label{phisoln}
\ee
Again using (\ref{gntoconf}) and noting the integral identity
$\int_{r_0}^\rho d\rho \, \Theta(\rho - r_0) = (\rho - r_0)
\Theta(\rho - r_0)$, we can write the solution for $\tilde \Phi$:
\be \tilde \Phi = -3X + \frac{3\alpha'}{2\alpha}\int_{r_0}^\rho
d\rho \,X + \frac{\alpha'}{\alpha}\xi +
\frac{4\lambda_5}{M_6^4}\xi
\,\Theta(\rho-r_0)\bigg[1-\frac{\alpha'}{2\alpha}\big(\rho-r_0\big)\bigg]
\, . \label{tPsoln} \ee
In particular, on the brane it is given by $\tilde \Phi(r_0) = -3X(r_0) + \frac{2}{r_0}\xi $.

Next we solve for $\Psi$, by substituting Eq. (\ref{tPsoln}) into
Eq. (\ref{zeroeqf}), and defining
\be \Upsilon'  = \Psi' + 3 \int_{r_0}^\rho d\rho \, X + 2\xi -
4\frac{\lambda_5}{M_6^4} \, \xi \, \Theta(\rho-r_0) \, .
\label{varphi} \ee
With this, Eq. (\ref{zeroeqf}) becomes
\be \frac{\alpha'}{2\alpha} \Bigl(\Upsilon' - \frac{4 X'}{k^2}
\Bigr) = - 4 \frac{\lambda_5}{M_6^4} \, \xi \, \Theta(\rho-r_0) \,
, \label{varphieq} \ee
These two equations are straightforward to integrate. After a
simple algebra we find
\be \Psi = \frac{4}{k^2}X - 3 \int_{r_0}^\rho d\rho
\int_{r_0}^{\rho} d\rho \, X(\rho) - 2 \rho \, \xi +
\frac{2\lambda_5}{M_6^4} \, \xi \, \int_{r_0}^\rho
\frac{d\rho}{(\frac{\alpha'}{\alpha})} \,  \Theta(\rho-r_0)  \,
\bigg(2(\rho-r_0)\frac{\alpha'}{\alpha} - 4\bigg) \, ,
\label{psisoln} \ee
where we have subtracted, using another residual $4D$
diffeomorphism, an arbitrary $4D$ wave which arises as the Fourier
transform of the constants of integrations that yield
(\ref{phisoln}). On the brane, this yields $\Psi = \frac{4}{k^2} X
- 2 r_0 \, \xi$. Before proceeding, note that both solutions for
$\tilde \Phi$ and $\Psi$, Eqs. (\ref{tPsoln}) and (\ref{psisoln}),
remain indifferent to the presence of brane matter sources, which
are completely encoded in the bulk field $X$ and the brane bending
$\xi$. Once we find solutions for $X$ and $\xi$ in the presence of
sources, we can just substitute them back into (\ref{tPsoln}) and
(\ref{psisoln}) and have $\tilde \Phi$ and $\Psi$ as well. This
greatly facilitates integration, and is the reason behind our
choice of Eqs. (\ref{momconsaf})-(\ref{psibdy}) for the
description of the scalar sector.

So let us determine $X$ and $\xi$. First, we solve for $\xi$ using
the boundary conditions (\ref{Xbdy}) and (\ref{psibdy}) and
solutions (\ref{tPsoln}) and (\ref{psisoln}). Using the
expression for $\tilde \Phi$ on the brane given above we can
eliminate $\tilde \Phi$ from the boundary conditions. Then we use
the solution (\ref{psibdy}) to relate $\Delta \Psi'$ to $\Delta
X'$, which gives
\be \Delta \Psi' = \frac{4}{k^2} \Delta X' - \frac{8
\lambda_5}{M_6^4 (\frac{\alpha'}{\alpha})|_+} \xi \, .
\label{psisolnjump} \ee
Finally, using (\ref{Xbdy}) to evaluate $\Delta X'$, we compare
(\ref{psibdy}) and (\ref{psisolnjump}), and  compute $\xi$ in
terms of the value of $X$ on the brane. A simple albeit tedious
algebra yields
\be \xi = \frac{3 (1-b) M_5^3 r^2_0}{2 b M_6^4} \, %\Bigl( 1 - b \Bigr) \,
\frac{k^2 + \frac{bM_6^4}{M_5^3 r_0}}{1-b + k^2 r_0^2}
\, X +  \frac{(1-b) r^2_0}{ b M_6^4} \, %\Bigl( 1 - b \Bigr) \,
\frac{\tau^\phi{}_\phi}{1-b + k^2 r_0^2} \, , \label{xisoln} \ee
where we have used $(\frac{\alpha'}{\alpha})|_+ = \frac{2}{r_0} -
\frac{4\lambda_5}{M_6^4}$, obtained from (\ref{alphadef}), and
resorted to Eq. (\ref{newdef}) to reintroduce $b =
\frac{2\lambda_5 r_0}{M_6^4}$. We notice, that in the thin brane
limit, $r_0 \rightarrow 0$, the brane bending term formally
vanishes, if we hold all other terms in Eq. (\ref{xisoln}) fixed.
Now we can finally go after $X$. Using the expressions for $\tilde
\Phi = -3X(r_0) + \frac{2}{r_0}\xi $ on the brane, and with $\xi$
given in Eq. (\ref{xisoln}), we can eliminate them from the
boundary condition (\ref{Xbdy}) in favor of $X$. This yields
\be
\Delta X' =  \frac{M_5^3}{M_6^4} \bigg(k^2
+ \frac{3b M_6^4}{4M_5^3r_0} - \beta(k^2) \bigg) X  -
 \frac{1}{6M_6^4} \,  \Bigg({\tau}^\alpha{}_{\alpha} +
\,  \frac{3M_5^3 r_0 k^2}{b M_6^4} \, \, \frac{1-b -
\frac{b M_6^4 r_0}{M_5^3}}{1-b +k^2 r_0^2}
\,{\tau}^\phi{}_{\phi} \Bigg) \, , \label{xbdysolved} \ee
where
\be
\beta(k^2) =  \frac{3M^3_5 r_0}{4 b M_6^4} \,
\frac{1-b}{1-b + k^2 r_0^2} \, \bigg(k^2 +
\frac{b M_6^4}{M_5^3 r_0}\bigg)^2 \, , \label{brscals}
\ee
is the correction to the brane kinetic term of the scalar $X$
originating from the brane bending contributions.  The field
equation in the bulk (\ref{xaf}) is very simple. If we
rewrite it with the boundary condition
(\ref{xbdysolved}) restored as a $\delta$-function term, in the
Schr\"odinger form, we can compare it to the {\tt
TT}-tensor field equations (\ref{tttensorsa}). The equation is
\ba && X'' + \frac{\alpha'}{2\alpha} X' - k^2 X -
\frac{M_5^3}{M_6^4} \bigg(k^2 + \frac{3b M_6^4}{4M_5^3r_0}  -
\beta(k^2) \bigg) X \, \delta(\rho-r_0) =
~~~~~~~~~~~~~~~~~~~~~~~~~~~~ \nonumber \\
&& ~~~~~~~~~~~~~~~~~~~~~~~~~~~~~~~ - \frac{1}{6M_6^4} \,
\Bigg({\tau}^\alpha{}_{\alpha} + \,  \frac{3M_5^3 r_0 k^2}{b M_6^4}
\, \, \frac{1-b - \frac{b M_6^4 r_0}{M_5^3}}{1-b
+k^2 r_0^2} \,{\tau}^\phi{}_{\phi} \Bigg) \, \delta(\rho-r_0) \, .
\label{xxaf} \ea
This equation is very similar to the {\tt TT}-tensor equation
(\ref{tttensorsa}), with the differences being in the
brane-localized terms. We will comment on them below. Here, we
note that we can now exploit the formal similarity with {\tt
TT}-tensor equation and use the Green's function methods again, to
find the solution. This time we define the scalar Green's function
${\cal G}_{\tt X}(k,\rho)$ by
\be {\cal G}_{\tt X}{}'' + \frac{\alpha'}{2\alpha} {\cal G}_{\tt
X}{}' - k^2 {\cal G}_{\tt X}  - \frac{M_5^3}{M_6^4} \bigg(k^2 +
\frac{3b M_6^4}{4M_5^3r_0}  - \beta(k^2) \bigg) \, {\cal G}_{\tt X}
\, \delta(\rho-r_0) = \frac{1}{M_6^4} \, \delta(\rho-r_0) \, .
\label{Xgreens} \ee
Formally the solution is
\be X(k, \rho) = - \frac16 \, {\cal G}_{\tt X}(k, \rho) \,
\Bigg[{\tau}^\alpha{}_{\alpha} + \,  \frac{3M_5^3 r_0 k^2}{ b M_6^4}
\, \, \frac{1-b - \frac{b M_6^4 r_0}{M_5^3}}{1-b
+k^2 r_0^2} \,{\tau}^\phi{}_{\phi} \Bigg] \, , \label{Xsoln} \ee
where again we defer the detailed determination of ${\cal G}_{\tt
X}$ until later.

Let us now briefly outline here the procedure for extracting the
long range physical fields from the solutions. We recall that by
axial symmetry of the configurations which we are exploring, at
distances $\ell > r_0$, we can dimensionally reduce the theory on
the compact circle. Hence the long distance dynamics will really
be described by a scalar-tensor theory, where the Brans-Dicke-like
scalar field, or the radion, is $g^{1/2}_{\phi\phi}$. In addition,
in the tensor sector there will be the helicity-0 mode resonance
as well as the helicity-2 ones. The radion $g^{1/2}_{\phi\phi}$
already renders the effective gravitational strength as evaluated
from the brane-induced gravity action field-dependent. Indeed, at
the linearized level about the vacuum, we see from Eq.
(\ref{gngbfix}) that since we are working in Gaussian-normal
brane-fixed gauge, the scalar field will be $\chi_{BD} =
(1+\frac{\tilde \Phi}{2}) \alpha^{1/2}(\rho)$, where
$\alpha^{1/2}(\rho)$ is its background value, and $
\alpha^{1/2}(\rho) \, \frac{\tilde \Phi}{2}$ the perturbation
sourced by the rings of matter on the wrapped brane. Clearly, the
background variation of $\chi_{BD}$ retains the memory of the
compactified dimension, which opens up as one moves away from the
cylindrical brane, as discussed recently in \cite{kalwall}.
Nevertheless, the description in terms of the reduced theory is
still perfectly justified at the linearized level with
axisymmetric sources. It is then useful to go to the analogue of
the effective `Einstein' frame for the dimensionally reduced
perturbation theory, in order to separate out graviton and scalar
forces, and compare the linearized fields to $4D$ General
Relativity. In doing so, we will ignore the prefactors
$\alpha^{1/2}$ as they are fully absorbed in the background
solution, and only remove the perturbations $\tilde \Phi$ from the
effective Planck scales, because we are most interested in the
effective theory along the brane. Thus the radial variation of the
effective bulk Planck scale is not important at the linearized
level of perturbation theory. So we proceed by noting that the
graviton kinetic terms in the full $6D$ theory come from
\be
S_{6D} \ni \int d^6 x \sqrt{g_6} \frac{M_6^4}{2} R_6 + \int_{\rho = r_0}
d^5 x \sqrt{g_5} \frac{M_5^3}{2} R_5 \, .
\label{actions}
\ee
Then dimensionally reducing on the circle changes these terms to
\be
S_{6D} \ni \int d^5 x \sqrt{g_5}  \,
\pi M_6^4 \alpha^{1/2}(\rho) \, \bigg(1+\frac{\tilde \Phi}{2}\bigg) \, R_5
+ \int_{\rho = r_0} d^4 x \sqrt{g_4} \, \pi M_5^3 r_0
 \bigg(1+\frac{\tilde \Phi}{2}\bigg) \,  R_4 \, .
\label{redactions}
\ee

To go to the effective `Einstein' frame, we absorb the radion
perturbation by conformally transforming the metric to $g^{\tt
E}_{AB} = \Omega^2 g_{AB}$. Although the graviton kinetic terms
live in different dimensions, we can render the conformal
rescalings of the two terms the same thanks to the fact that we
are working entirely in the Gaussian-normal brane-fixed gauge.
Since our starting metric satisfies $g_{\rho\rho} = 1, g_{\rho\mu}
=0$, we can change the radial coordinate to $\bar \rho= \int d\rho
\, \Omega$ to ensure that the bulk term changes under the
conformal transformation as $d\rho \, \sqrt{g_5} \, R_5 = d\rho'
\, \sqrt{g^{\tt E}_5} \, R^{\tt E}_5/ \Omega^2$, i.e. the same as
the brane term. We bear in mind that we will need to change
$x^\mu$ coordinates in order to keep the reduced metric in
Gaussian-normal brane fixed gauge. So to absorb the factors
$\propto \tilde \Phi$ from in front of the graviton kinetic terms
(\ref{redactions}) we need $\Omega^2 = 1+ \frac{\tilde \Phi}{2}$.
This defines our radial gauge transformation by $\bar \rho =
\int^\rho_{r_0} d\rho (1 + \frac{\tilde \Phi}{4}) + r_0 = \rho +
\frac14 \int^\rho_{r_0} d\rho \, \tilde \Phi$ where we have chosen
the integration constant to ensure that the brane remains fixed,
at $\bar \rho = r_0$. This transformation is of the form of
general diffeomorphisms (\ref{diffeos}) with $\chi^\rho =  \frac14
\int^\rho_{r_0} d\rho \, \tilde \Phi$. Then to keep the
perturbation $\bar h_{\mu\rho}$ from appearing and ensure that we
are still in the Gaussian-normal gauge, we need to pick $\chi_\mu'
= - \partial_\mu \chi^\rho$, or therefore $\chi_\mu = -   \frac14
\int^\rho_{r_0} d\rho \int^\rho_{r_0} d\rho \, \partial_\mu \tilde
\Phi$ (where we can ignore the overbars in these expressions to
the leading order).

But now, from these expressions we immediately see that the gauge
transformations {\it vanish} on the brane! Hence, the gauge fixing
on the brane is unaffected by the dimensional reduction, and when
calculating the fields along the brane we can just use the
expressions in the original coordinates, ignoring altogether their
behavior in the bulk\footnote{Forearmed with the knowledge that
the fields are localized near the brane, which will be ensured by
the asymptotic boundary conditions on Green's functions.}. Thus we
can use the conformal transformation $g^{\tt E}_{\mu\nu} =
(1+\frac{\tilde \Phi}{2} ) g_{\mu\nu}$ on the brane. Now after
Fourier transforming the decomposition theorem formula
(\ref{perthel}), and replacing $h_4$ by $X$ using Eq.
(\ref{scalarpert}), the full metric perturbation along the brane
in the original frame is
\be
h^\mu{}_\nu = \gamma^\mu{}_\nu - k^\mu k_\nu \Psi + X \delta^\mu{}_\nu \, ,
\label{hsolnf}
\ee
where on the brane $\Psi = \frac{4}{k^2} X - 2 r_0 \xi$. After our
conformal transformation to linear order in perturbation, we find
that $h_{\mu\nu}^{\tt E} = h_{\mu\nu} + \frac{\tilde \Phi}{2} \,
\eta_{\mu\nu}$, or therefore, using the expressions for $\tilde
\Phi$, $\Psi$ on the brane to express them all in terms of the
variables $X$ and $\xi$,
\be h^{\tt E}{\,}^\mu{}_\nu = \gamma^\mu{}_\nu -  4 \, \bigg( X -
\frac{r_0 k^2}{2} \xi \bigg) \, \frac{k^\mu k_\nu}{k^2} + \bigg(
\frac{\xi}{r_0} - \frac{X}{2} \bigg) \, \delta^\mu{}_\nu\, ,
\label{hsolnfef} \ee
where of course $\xi$ is given by (\ref{xisoln}). This is the
effective `Einstein' frame field of matter rings on the wrapped
brane, which may yet contain extra scalar-like contributions from
the helicity-0 modes. To see how it behaves we now need to find
out the explicit forms of Green's functions ${\cal G}_{\tt TT}$
and ${\cal G}_{\tt X}$.

\subsection{Fields of Static Sources Along the Brane}

Since we are looking for the static solutions, obeying $k^\mu =
(0, \vec k)$, which implies that all the momentum space Fourier
transforms are formally ${\cal G}(k) = 2\pi \delta(k^0) \, {\cal
G}(\vec k)$, we can factorize out the energy Green's function, set
$k^0=0$ in the formulas of the previous section, and drop the
integration $\int \frac{dk^0}{2\pi}$ from the Fourier integral. We
then solve the Green's function equations (\ref{greens}) and
(\ref{Xgreens}) using -- again, as in the construction of the
shock waves -- the standard technique of sewing together the
solutions of the homogeneous equation on either side of the brane,
regular in the center and far away.

Off the brane, both equations (\ref{greens}) and (\ref{Xgreens})
look the same. The solutions which are regular in the center and
at infinity are
\be {\cal G} =  \cases { A I_0\Bigl(k\rho\Bigr) \, ,   &
~~~~ $\rho \le r_0$ \, ; \cr B K_0\Bigl(k(\rho +
\frac{br_0}{1-b})\Bigr) \, ,  & ~~~~ $\rho \ge r_0$ \, , }
\label{ttgreens} \ee
where now $k = |\vec k|$. We sew them together at the brane by
using the boundary condition found with the Gaussian pillbox
integration of Eq. (\ref{greens}), that now gives different
results for the two Green's functions,
\be \Delta {\cal G}_{\tt TT} = \frac{M_5^3}{M_6^4} k^2 {\cal
G}_{\tt TT} + \frac{1}{M_6^4} \, ,  ~~~~~~~~ \Delta {\cal G}_{\tt
X} = \frac{M_5^3}{M_6^4} [k^2 + \frac{3b M_6^4}{4M_5^3 r_0} -
\beta(k^2)] {\cal G}_{\tt X} + \frac{1}{M_6^4} \, ,
\label{greensbcs} \ee
with $\beta(k^2)$ given in (\ref{brscals}), and then
setting the homogeneous
solution to zero. This yields
\be {\cal G}_{\tt TT} = \frac{1}{M_6^4\, k} \, \frac{ I_0(k\rho_<)
K_0(k \rho_>)}{I_0(kr_0)K_1(k\frac{r_0}{1-b}) + I_1(kr_0)
K_0(k\frac{r_0}{1-b})+\frac{M_5^3k}{M_6^4}I_0(kr_0)K_0(k\frac{r_0}{1-b})}
\, ,   \label{ttgrsoln} \ee
and
\be {\cal G}_{\tt X} = \frac{1}{M_6^4\, k} \, \frac{ I_0(k\rho_<)
K_0(k \rho_>)}{I_0(kr_0)K_1(k\frac{r_0}{1-b}) + I_1(kr_0)
K_0(k\frac{r_0}{1-b}) + \frac{M_5^3}{M_6^4 k} [k^2 +
\frac{3bM_6^4}{4M_5^3 r_0} - \beta(k^2)]
I_0(kr_0)K_0(k\frac{r_0}{1-b})} \, ,   \label{Xgrsoln} \ee
where, just as in Eq. (\ref{thickshosol2}),
\be
I_0(k\rho_<) K_0(k\rho_>)  = \cases { I_0\Bigl(k\rho\Bigr)
K_0\Bigl(k\frac{r_0}{1-b}\Bigr) \, ,   & ~~~~ $\rho \le r_0$ \, ;
\cr I_0\Bigl(k r_0 \Bigr)  K_0\Bigl(k(\rho +
\frac{br_0}{1-b})\Bigr) \, ,  & ~~~~ $\rho \ge r_0$ \, . }
\label{ttgrsolnum} \ee
In fact the Green's function ${\cal G}_{\tt TT}$ is {\it
identical} to the Green's function that appears in the shock wave
solution (\ref{thickshosol1}). This should hardly come as a
surprise, because it is obvious from Eq. (\ref{tttensorsa}) that
it has the same form as the shock wave equation when the stress
energy source is traceless. The only difference between these two
cases is in the domain of Fourier integration, which for
relativistic sources involves one integration fewer because of the
Lorenz boost of the longitudinal direction.

Since we wish to explore the fields along the brane, we set $\rho
= r_0$ in the Green's functions (\ref{ttgrsoln}) and
(\ref{Xgrsoln}). At distances $\ell \gg r_0$, where we can limit
our attention to axisymmetric configurations, for which the
solutions (\ref{ttgrsoln}) and (\ref{Xgrsoln}) are valid, $k r_0
\sim r_0/\ell \ll 1 $, we can always replace $I_0$ and $I_1$ by
their small argument expansion, and approximate them by $I_0
\rightarrow 1$, $I_1 \rightarrow 0$. We must be more cautious with
$K_0$ and $K_1$ since they depend on the deficit angle, and for
near critical branes $k r_0/(1-b)$ may be big even when $k r_0 \ll
1$. Thus we will need to consider the limits $b \la 1$ and $b
\rightarrow 1$ separately. We have already encountered this
previously, in the example with shock waves. With this in mind, we
see that the consistent approximation  of the Green's functions
(\ref{ttgrsoln}) and (\ref{Xgrsoln}) is
\ba {\cal G}_{\tt TT} &=&  \frac{1}{M_6^4 } \,
\frac{1}{\frac{K_1(k\frac{r_0}{1-b})}{K_0(k\frac{r_0}{1-b})} k +
\frac{M_5^3}{M_6^4} k^2 }
\, ,   \label{ttgrsolnbr} \\
{\cal G}_{\tt X} &=&  \frac{1}{M_6^4} \,
\frac{1}{\frac{K_1(k\frac{r_0}{1-b})}{K_0(k\frac{r_0}{1-b})} k
+\frac{M_5^3}{M_6^4} [k^2 + \frac{3bM_6^4}{4M_5^3 r_0} -
\beta(k^2)]} \, ,   \label{Xgrsolnbr} \ea

The {\tt TT}-tensor sector is particularly simple to understand
from (\ref{Xgrsoln}). As with shock waves, we see that the
behavior of the theory is governed by the ratio $k_c = \frac{M_6^4
K_1(k\frac{r_0}{1-b})}{M_5^3 K_0(k\frac{r_0}{1-b})}$, which
controls the denominator of ${\cal G}_{\tt TT}$ at low momenta.
For $k < k_c$, or distances $\ell > 1/k_c$, the dominant
contributions always come from $\propto
\frac{K_1(k\frac{r_0}{1-b})}{K_0(k\frac{r_0}{1-b})}$. In this
limit the long range fields manifestly reveal the extra
dimensions, since the scaling of the potentials will not be $4D$.
In the generic sub-critical cases, $b \la 1$ and for momenta $k
\ll 1/r_0$ we can replace the Bessel functions $K_\nu$ by their
small argument expansion along the brane, which yields the
crossover scale identical to Eq. (\ref{crossm4}), $r_c^2(k) =
\frac{M_4^2}{2\pi (1-b) M_6^4}
\ln\Bigl[\frac{2(1-b)}{kr_0}\Bigr]$. Beyond this distance, the
theory behaves as a $6D$ gravity on a cone.

In the near-critical limit, however, $b \rightarrow 1$, at
intermediate momenta we should instead approximate $K_\nu$ by
their large argument expansion, as in the shock wave analysis of
the near-critical models building up to Eq. (\ref{crosseesaw}).
In this case the crossover scale is $r_c = \frac{M^3_5}{M_6^4}$,
where gravity first changes into a $5D$ theory, because for
near-critical tensions one bulk dimension is efficiently
compactified on a circle, as discussed recently in \cite{kalwall},
and in our shock wave analysis. Eventually, the circle opens up
and the theory again turns into a $6D$ gravity on a cone. As a
result we see that the crossover scale, beyond which the theory
will not look $4D$, is exactly the same as in Eq.
(\ref{crosssum}), which was already revealed by the shock waves.
We again confirm that the see-saw mechanism of \cite{highercod}
is realized only in the near-critical limit, but the crucial
dynamics which manufactures the see-saw scale is the
compactification of one bulk dimension induced by a near-critical
brane \cite{kalwall}.

What remains is to check precisely what kind of $4D$ theory we
have below the crossover scale $r_c$, in the regime of distances
$r_0 \ll \ell \ll r_c$. For {\tt TT}-tensors, by previous
discussion we can neglect the terms in the denominator $\sim
\frac{K_1(k\frac{r_0}{1-b})}{K_0(k\frac{r_0}{1-b})} $. The Green's
function reduces to ${\cal G}_{\tt TT} \rightarrow \frac{1}{M_5^3
k^2 }$. The non-relativistic matter sources on the wrapped brane
are described by stress energy of Eq. (\ref{eqn:strinsediag}),
which Fourier-transforms to $\tau^a{}_b = - \mu \, {\rm
diag}(1,0,0,0,1)$. So the leading order solution for the {\tt
TT}-tensor field is
\be \gamma^\mu{}_\nu(k, \rho) = \frac{2 \mu}{3 M_5^3 k^2} \,
\cases{ 2 \, , ~~~~~ & $\mu = \nu = 0 \,$,\cr
 - \bigg(\delta^j{}_k -
\frac{k^j k_k}{k^2}\bigg) \, , ~~~~~ & $\mu = j, \, \nu = k \,$.}
\label{ttsolstat}
\ee
Next we need to evaluate the Fourier integrals
$\gamma^\mu{}_\nu(\vec x) = \int \frac{d^3\vec k}{(2\pi)^3}
\gamma^\mu{}_\nu(\vec k) e^{i \vec k \cdot \vec x}$. Since a
Fourier integral picks up the dominant contribution from the
momenta $k \sim 1/|\vec x|$, in the $4D$ regime, when $|\vec x| <
r_c$, we can approximate $\gamma^\mu{}_\nu(k)$ by
(\ref{ttsolstat}). The remaining integral is easy for
$\gamma^0{}_0$, and yields\footnote{Using $\int \frac{d^3\vec
k}{(2\pi)^3 } \frac{e^{i \vec k \cdot \vec x}}{\vec k^2}=
\frac{1}{4\pi |\vec x|} $.} $\gamma^0{}_0 = \frac{\mu}{3\pi M_5^3
|\vec x|}$ in this regime. Recalling that $\mu$ is the mass per
unit length of string (see Eq. (\ref{nambu})), so that $\mu =
\frac{{\cal M}}{2\pi r_0}$, and using as before Gauss law to trade
the $5D$ Planck scale $M_5$ for the $4D$ one, $M_5^3 =
\frac{M^2_4}{2\pi r_0}$, we finally obtain
\be
\gamma^0{}_0(\vec x) =  \frac{\cal M}{3\pi M_4^2 |\vec x|} \, .
\label{gamma00}
\ee
To find $\gamma^j{}_k$, we don't need to do the Fourier integrals
directly. Instead, using Lorentz invariance, we see from
(\ref{ttsolstat}) that $\gamma^j{}_k = \frac{{\cal C}_1}{|\vec x|}
\bigg(\delta^j{}_k  + {\cal C}_2 \frac{x^j x_k}{\vec x^2} \bigg)$.
Then since $\gamma^\mu{}_\nu$ is a {\tt TT}-tensor, and
$\gamma^j{}_0=0$, using transversality $\partial_j \gamma^j{}_k =
0$, we get ${\cal C}_2 =1$. Further vanishing trace
$\gamma^\mu{}_\mu=0$ implies $\gamma^k{}_k = - \gamma^0{}_0$,
which by comparing with Eq. (\ref{gamma00}) sets ${\cal C}_1 =
-\frac{\cal M}{12\pi M_4^2}$. Thus we find
\be
\gamma^j{}_k(\vec x) =  - \frac{\cal M}{12\pi M_4^2 |\vec x|}
\Bigg(\delta^j{}_k + \frac{x^j x_k}{\vec x^2}  \Bigg)
\, .
\label{gammajk}
\ee
It is convenient to rewrite the solutions by introducing the
effective $4D$ Newton's constant $G_{N \, eff} = \frac{1}{8\pi
M_4^2}$. Then, in the leading order for $|\vec x| < r_c$,
\ba \gamma^0{}_0(\vec x) &=&  \frac83 \, G_{N \, eff}
\, \frac{\cal M}{ |\vec x|} \, , \nonumber \\
\gamma^j{}_k(\vec x) &=&  - \frac23 \, G_{N \, eff} \, \frac{\cal M}{ |\vec x|} \,
\Bigg(\delta^j{}_k + \frac{x^j x_k}{\vec x^2}  \Bigg)
\, .
\label{gammaall}
\ea
If we were to ignore the scalar field $X$ for the moment, the
Newtonian potential $V_N = - h^0{}_0/2$ would be $V_N = -\frac43
\, G_{N \, eff} \, \frac{\cal M}{ |\vec x|}$, which is a factor of
$4/3$ larger than the usual formula of General Relativity. This is
a manifestation of the Iwasaki-van Dam-Veltman-Zakharov
discontinuity \cite{vdvz}, which signals the presence of the
helicity-0 modes in the graviton multiplet. The factor $4/3$
enhancement is precisely what one expects based on other examples
\cite{DGP}, and in this case shows that the extra helicity-0 modes
in the spin-2 multiplet mediate attractive force, just as the
helicity-2 modes. Thus, they are not ghosts.

The scalar Green's function ${\cal G}_{\tt X}$ is considerably
more involved because of the function $\beta(k^2)$ in the
denominator. It behaves differently for generic sub-critical and
for near-critical branes, and so we need to deal with it with some
care. First off, we note that the terms $k^2 +
\frac{3bM_6^4}{4M_5^3 r_0} - \beta(k^2)$ factorize as
\be k^2 + \frac{3bM_6^4}{4M_5^3 r_0} - \beta(k^2) = \frac{r_0^2
k^2}{1-b + r_0^2 k^2} \bigg[\bigg(1-\frac{3(1-b)M_5^3}{4bM_6^4
r_0}\bigg)k^2 + \bigg(\frac{3bM_6^4}{4M_5^3 r_0} -
\frac{1-b}{2r_0^2}\bigg) \bigg] \, . \label{factoriza} \ee
After substituting this in Eq. (\ref{Xgrsolnbr}) we find that
${\cal G}_{\tt X}$ reduces to
\be {\cal G}_{\tt X} = \frac{1}{M_6^4} \,
\frac{1}{\frac{K_1(k\frac{r_0}{1-b})}{K_0(k\frac{r_0}{1-b})} k
+\frac{M_5^3}{M_6^4} \frac{r_0^2 k^2}{1-b + r_0^2
k^2}\bigg[\bigg(1-\frac{3(1-b)M_5^3}{4bM_6^4 r_0}\bigg)k^2 +
\bigg(\frac{3bM_6^4}{4M_5^3 r_0} - \frac{1-b}{2r_0^2}\bigg)\bigg]}
\, . \label{Xgrapprgen} \ee
Now, we will generically require $r_0 \ll M_5^3/M_6^4$ in order to
have the crossover scale for gravity be much larger than the
radius of the compact dimension, $r_c \gg r_0$. Otherwise, we can
never get the theory to behave as a $4D$ gravity.

For generic sub-critical branes, $b \la 1$, at scales $k \ll
1/r_0$ the scalar Green's function (\ref{Xgrapprgen}) is
approximated by
\be
{\cal G}_{\tt X} = \frac{1}{M_6^4} \,
\frac{{r_0\ln[\frac{2(1-b)}{kr_0}]} }{{1-b} - \frac{3 M_5^6r^2_0}{4b M_6^8}
k^2 \bigg( k^2 + \frac{2bM_6^4}{3M_5^3 r_0}\bigg) {\ln[\frac{2(1-b)}{kr_0}]} } \, ,
\label{Xgrappr}
\ee
where we have approximated
$\frac{K_1(k\frac{r_0}{1-b})}{K_0(k\frac{r_0}{1-b})}$ with the
small argument expansion. Hence at very large distances, or low
momenta, the scalar Green's function depends on the momentum only
logarithmically, ${\cal G}_{\tt X} \simeq
\frac{{r_0\ln[\frac{2(1-b)}{kr_0}]} }{{(1-b)}M_6^4}$, implying that
the configuration space solution for the scalar field $X$ depends
on the distance as $X \sim {\cal M}/|\vec x|^3$ - i.e. as a field
in $6D$. However, because the momentum-dependent terms in the
denominator of ${\cal G}_{\tt X}$ are negative, as we move inwards
towards the source, the scalar force grows, and the scalar
eventually becomes strongly coupled as $k$ approaches the pole of
(\ref{Xgrappr}). This happens when $k \sim k_*$, where
\be k^2_* \simeq \frac{2(1-b)M_6^4}{M_5^3 r_0
\ln[\frac{2(1-b)}{kr_0}]} \, . \label{strcoupl} \ee
But this is essentially the same as the crossover momentum in the
{\tt TT}-tensor sector, as we can see by comparing $k_*$ with
$k_c$ given in Eq. (\ref{momcond}), after recalling that $M_4^2
\simeq M_5^3 r_0$. At first sight, it looks a little surprising
that for resolved codimension-2 branes the strongly coupled
scalars appear even around the vacuum, in contrast to the
codimension-1 brane induced gravity models, as discussed in
\cite{strongcouplings,niges,strong,Luty,nira}. However the $4D$
vacua aren't really vacuum solutions from the point of view of the
full brane worldvolume theory because they contain the flux of the
axion $\Sigma$ and hence source the brane bending term $\xi$ even
in the absence of localized matter sources, as seen in Eq.
(\ref{xisoln}). This in turn triggers the onset of strong
coupling.

Thus we see that for sub-critical branes with $b \la 1$, inside
the regime of length scales $\ell < r_*$ where
\be
r_*^2 \simeq \frac{M_4^2} {2(1-b)M_6^4}  \ln[\frac{2(1-b)}{kr_0}] \, ,
\label{strongcoupllength}
\ee
the {\tt TT}-tensor is $4D$, but with the wrong tensor structure,
as we have discussed following Eq. (\ref{gammaall}), however the
scalar sector is in fact strongly coupled. This makes the
perturbation theory around the vacuum on sub-critical branes
completely unreliable inside the regime where gravity might be
$4D$. The negative signs of the momentum-dependent terms in the
denominator of (\ref{Xgrappr}) might naively suggest presence of
ghosts in this regime, but at this level of the approximation we
cannot conclude that decisively. One exception to the pathological
behavior of the theory is the limit of relativistic sources, for
which $\tau^\mu{}_\mu = \tau^\phi{}_\phi = 0$, so that the scalars
are never sourced and the tensor sector reduces to the $4D$
Aichelburg-Sexl solution below the crossover scale. One needs some
means for exploring the scalar sector beyond linearized theory,
perhaps along the lines of \cite{takahiro}, or by seeking exact
solutions, before passing on the final verdict on the sub-critical
brane models. At the level of linearized perturbation theory
the gravitational effects below the
crossover scale are {\it not} calculable.

In the near-critical limit the scalar sector behaves very
differently. At distances $\ell \gg r_0$, where $k r_0 \simeq
\frac{r_0}{\ell} \ll 1$ the argument of the Bessel functions
$K_\nu$ will still be very large when $b \rightarrow 1$. We have
already noted this in the discussion of shock waves. Hence for
distances below the crossover scale we can use the large argument
expansion for $K_\nu$'s, which yields $K_1 \rightarrow K_0$, as
long as $r_0 k \gg 1-b$. Moreover, we can immediately neglect the
negative terms in the denominator of ${\cal G}_{\tt X}$ in Eq.
(\ref{Xgrapprgen}) because $\frac{(1-b)M_5^3}{M_6^4 r_0} = (1-b)
\frac{r_c}{r_0} < 1$. The scalar Green's function is therefore
\be
{\cal G}_{\tt X} = \frac{1}{M_6^4} \,
\frac{1}{k +\frac{M_5^3}{M_6^4}
 \frac{r_0^2 k^2}{1-b + r_0^2 k^2}(k^2
+ \frac{3}{4r_c r_0} )} \, .
\label{Xgrapprext1}
\ee
where $r_c = M^5_3/M_6^4$ is the {\tt TT}-tensor crossover scale,
as per Eq. (\ref{crosseesaw}). Now, it is convenient to define the
scale $r_{\tt vac}$ by
\be
1- b  = \frac{r_0^2}{r^2_{\tt vac}} \, ,
\label{rvac}
\ee
In the limit $b \rightarrow 1$ we have $r_{\tt vac} = \sqrt{1-b}
\frac{r_0}{1-b} \ll \frac{r_0}{1-b}$, and so the scale $r_{\tt
vac}$ is always smaller than the size of the conical throat
surrounding the brane. This is why the scale $r_{\tt vac}$ may
compete with $r_c$ for the control over the scalar sector. Both
$r_{\tt vac}$ and $r_c$ are smaller than the size of the throat,
and the details of sub-crossover dynamics depend on their ratio.
First, note that for momenta $k$ such that $r_{\tt vac} k > 1$, we
can expand $\frac{r_0^2k^2}{1-b + r_0^2 k^2} \simeq 1 + {\cal
O}(\frac{1-b}{r_0^2 k^2})$ and keep only the first term. The
scalar Green's function in this limit at momenta $r_0 k < 1-b$
reduces to
\be {\cal G}_{\tt X} = \frac{1}{M_5^3} \, \frac{1}{ \frac{k}{r_c}
+  k^2  + \frac{3}{4r_c  r_0} } \, . \label{Xgrapprext} \ee
Thus for momenta $k > {\rm max} \, (1/r_c,1/r_{\tt vac})$, the
theory looks $4D$, but the scalar has a mass term, $m^2_{\tt X} =
\frac{3}{4r_c r_0}$. Therefore its long range effects have a Yukawa
suppression $\propto \exp(- m_{\tt X} r)$. Hence in the range of
scales $\frac{1}{m_{\tt X}} < \ell < {\rm min} \, (r_c, r_{\tt
vac})$ the scalar field exchange will lead to negligible,
exponentially suppressed effects compared to the {\tt
TT}-gravitons. This mass term is completely analogous to a
brane-localized mass term of a bulk scalar. In such models the
scalar is repelled from the brane, since it prefers to reside in
the region of space where its inertia is minimized. This resembles
Meissner effect in superconductivity, where the mass
term pushes the magnetic field outside of the superconducting
medium. In our case however, the scalar is also pulled back to the
brane by the brane-localized kinetic terms, and the result is the
Yukawa suppression alone, which is manifest in (\ref{Xgrapprext})
in the same coupling $1/M_5^3$ as for the {\tt TT}-tensors below
the crossover scale, e.g. in Eq. (\ref{ttsolstat}). Similar phenomena
were also studied recently for gauge fields in gaugephobic models \cite{johncsaba}.

Although $X$ may be Yukawa-suppressed, so its direct long range
forces will be small compared to those arising from the {\tt
TT}-tensor sector, the brane bending term $\xi$ may yet yield a
long range tail, as is clear from Eq. (\ref{xisoln}). We see that
$\xi$ is proportional to $k^2 X$, and this will in general lead to
long range effects that may dominate the Yukawa-suppressed effect
from direct $X$-exchange. It turns out that the $\xi$-induced
effects will be parametrically smaller than the dominant {\tt
TT}-tensor fields, as we will show below.

At distances $\ell > r_c$, the linear term in $k$ in the
denominator of (\ref{Xgrapprext}) dominates the momentum transfer
in scalar exchange, modifying the theory to a $5D$ one, and
eventually, after resuming the corrections which we neglected in
setting $K_1/K_0 \simeq 1$, one would see how the theory changes
to $6D$. When $r_{\tt vac} > r_c$ the story of the scalar dynamics
ends here.

However, when $r_{\tt vac} < r_c$, in the regime of scales $r_{\tt
vac} < \ell < r_c$ below the crossover scale, we can approximate
the fraction in the denominator of (\ref{Xgrapprext1}) by
$\frac{r_0^2k^2}{1-b + r_0^2 k^2} \simeq \frac{r_0^2k^2}{1-b}$.
Thus, ${\cal G}_{\tt X}{}^{-1} \sim k +\frac{M_5^3 r_0^2}{(1-b)
M_6^4} k^2 (k^2 + \frac{3}{4r_c r_0} )$. Then, since the mass term
$m^2_{\tt X}$ is larger than the momenta in this regime, because
$k^2 < 1/r^2_{\tt vac}$ so that $\frac{k^2}{m^2_{\tt X}}<
\frac{4r_c r_0}{3 r_{\tt vac}^2} = (1-b) \frac{4r_c}{3 r_0} <1$, the
Green's function is approximately
\be {\cal G}_{\tt X} = \frac{1}{M_6^4} \, \frac{1}{k +
\frac{3r_{\tt vac}^2}{4r_0} k^2 } \, . \label{Xgrapprext2} \ee
What happened here is that due to the momentum
dependence of the effective coupling, the scalar field mass
term got overcompensated and the field became essentially massless
all the way out to a new scalar crossover scale, given by (using
Eq. (\ref{rvac}))
\be {\cal R}_c = \frac34 \frac{r^2_{\tt vac}}{r_0} = \frac34 \frac{r_0}{1-b} \,
, \label{scalcross} \ee
which is essentially equal to the length of the throat inside which
bulk gravity looks $5D$.
Below ${\cal R}_c$, the scalar propagator reduces to ${\cal
G}_{\tt X} \rightarrow \frac{4r_0}{3 M_6^4 r_{\tt vac}^2 k^2 } <
\frac{1}{M_5^3 k^2 }$, because $r_c < \frac{r_0}{1-b}$ in the
near-critical limit. Thus, when $r_{\tt vac} > r_c$, the scalar
remains massless all the way out to ${\cal R}_c$, but below the
crossover scale it couples more weakly than the {\tt TT}-tensor.

We can now put together the various contributions to the fields of
a static massive ring on the brane at distances below the
crossover scale. The scalar Green's function for near-critical
branes below the crossover scale is well approximated by
\be {\cal G}_{\tt X} = \cases{\frac{1}{M_5^3} \, \frac{1}{
k^2  + m^2_{\tt X} } \, ,  & $r_c < r_{\tt vac}, \, \frac{1}{r_c} < k$\, ,  \cr
\frac{4(1-b)}{3 M_6^4 r_0 } \, \frac{1}{ k^2 }\, ,
& $r_{\tt vac} < r_c, \,  \frac{1}{r_c} <k < \frac{1}{r_{\tt vac}} $ \, .}
\label{Xlightsc} \ee
The scalar field $X$ sourced by a ring of mass with
Fourier-transformed stress energy tensor $\tau^a{}_b = - \mu \,
{\rm diag}(1,0,0,0,1)$ is then, using the solution (\ref{Xsoln})
for this range of scales, and again replacing $\mu = \frac{\cal
M}{2\pi r_0}$ and $M_5^3 = \frac{M_4^2}{2\pi r_0}$ as in the
earlier discussion of {\tt TT}-tensors,
\be X = \cases{- \frac{\cal M}{3 M_4^2} \, \frac{1}{ k^2  +
m^2_{\tt X} } \, ,  & $r_c < r_{\tt vac}, \, \frac{1}{r_c} < k$\,
,  \cr
\frac{2{\cal M}}{9M_4^2 } \, \frac{(1-b)r_c}{r_0} \,
\frac{1}{ k^2 }\, , & $r_{\tt vac} < r_c, \,  \frac{1}{r_c} <k <
\frac{1}{r_{\tt vac}} $ \, ,} \label{Xlightscf} \ee
while the brane bending, normalized to $r_0$, is, from
(\ref{xisoln}),
\be \frac{\xi}{r_0} = \cases{- \frac{r_c r_0}{r^2_{\tt vac}} \, \frac{\cal M}{6 M_4^2} \,
 \bigg( \frac{10}{ k^2}  - \frac{1}{ k^2  +
m^2_{\tt X} } \bigg) \, ,  & $r_c < r_{\tt vac}, \, \frac{1}{r_c} < k$\,
,  \cr  \frac{(1-b)r_c}{r_0} \, \frac{\cal M}{3 M_4^2 } \,
\frac{1}{ k^2 } + {\rm constant} \, , & $r_{\tt vac} < r_c, \,
\frac{1}{r_c} <k < \frac{1}{r_{\tt vac}} $ \, .}
\label{Xlightscxi} \ee
The constant in the brane bending transforms to a gauge-dependent
contact term $\propto \delta^{(3)}(\vec x)$, which is ultralocal
and hence we can neglect it altogether. The configuration space
solutions\footnote{Which we find using $\int \frac{d^3\vec
k}{(2\pi)^3} \frac{e^{i \vec k \cdot \vec x}}{k^2 + m_{\tt X}^2} =
\frac{1}{4\pi |\vec x|} e^{-m_{\tt X} |\vec x|}$.} for the scalars
$X$ and $\xi$ are therefore
\be  X = \cases{- \frac{\cal M}{12 \pi M_4^2 |\vec x|} \,
e^{-m_{\tt X} |\vec x| } \, , & $|\vec x| < r_c < r_{\tt vac}$\, ,
\cr \frac{(1-b)r_c}{r_0} \,  \frac{\cal M}{18 \pi M_4^2 |\vec x|}
\, , & $r_{\tt vac} < |\vec x| <  r_c$ \, ,} \label{Xlightscfx}
\ee
and
\be  \frac{\xi}{r_0} = \cases{- \frac{r_c r_0}{r^2_{\tt vac}} \,
\frac{ \cal M}{24 \pi M_4^2 |\vec x|} \bigg(10 -  e^{-m_{\tt
X} |\vec x| } \bigg) \, , & $|\vec x| < r_c < r_{\tt vac}$\, , \cr
\frac{(1-b)r_c}{r_0} \,  \frac{\cal M}{12 \pi M_4^2 |\vec x|} \, ,
& $r_{\tt vac} < |\vec x| <  r_c$ \, .} \label{Xlightscxix} \ee

Now, in the case when $r_c < r_{\tt vac}$, the scalar field $X$
will have a very short range due to its mass. Indeed, since
$m_{\tt vec}^{-2} \sim r_0 r_c$ is the geometric mean between the
brane radius and the crossover scale, $m_{\tt vec}^{-1}$ will be
quite small. In fact, if the brane size is set by the $4D$ Planck
length and the crossover scale by the current horizon size, $r_0
\sim 1/M_4$, and $r_c \sim 1/H_0$ respectively, $m_{\tt X}^{-1}$
will be near the table top bounds of about $0.1 {\rm mm}$. Thus
$X$ is {\it de facto} decoupled at the scales where we normally
probe $4D$ gravity. On the other hand, the brane bending can still
simulate a long range effect which scales as $1/|\vec x|$ due to
the momentum dependence of the couplings. However, the strength of
its contributions is a factor $\frac{r_c r_0}{r_{\tt vac}^2} =
\frac{r_c}{r_{\tt vac}} \frac{r_0}{r_{\tt vac}} \ll 1$ down
compared to the {\tt TT}-tensor contributions in Eq.
(\ref{gammaall}). In the case when $r_{\tt vac} < r_c$, the
scalars remain essentially massless and source long range fields.
However, their strength is still suppressed, this time by
$\frac{(1-b)r_c}{r_0} \ll 1$, by definition of the near-critical
limit. While the scalar effects become more important as this
factor becomes smaller, so do the strong coupling effects which we
cannot neglect away from the near-critical limit. Indeed, as
$\frac{(1-b)r_c}{r_0} \rightarrow 1$, where the brane bending
effects could compete with the long range fields of {\tt TT}
tensors, the strong coupling effects become very important, and we
lose calculability in linearized perturbation theory below the
crossover scale.

Hence in either case the long range effects of the scalars in
linearized perturbation theory around the near-critical vacuum
cannot compete with the {\tt TT}-tensors, as long as the
perturbative treatment is valid. We can therefore ignore the
scalars in the formula (\ref{hsolnfef}). The long range fields are
determined by $\gamma^\mu{}_\nu$ alone, which are given in Eq.
(\ref{gammaall}). Clearly, these solutions do not look like $4D$
General Relativity. Instead, in the leading order they really
mimic a scalar tensor theory, due to the presence of the
helicity-0 contributions. To see this, we can compare the
solutions (\ref{gammaall}) with the spherically symmetric
linearized solution in the PPN approximation \cite{weinbook}. For
the purpose of this comparison, we define $\hat G_N = \frac43 G_{N
\, eff}$, so that
\ba \gamma^0{}_0 &=& 2 \hat G_N \frac{\cal M}{|\vec x|} \, , \nonumber \\
\gamma^j{}_k &=& - \frac12 \hat G_N \frac{\cal M}{|\vec x|} \delta^j{}_k
- \frac12 \hat G_N \frac{{\cal M} x^j x_k}{|\vec x|^3} \, .
\label{PPNsolns}
\ea
Comparing to \cite{weinbook} (taking into account overall sign
difference reflecting our conventions) we find that the solution
(\ref{PPNsolns}) mimics a Brans-Dicke theory with $\gamma =
\frac12$, or therefore, with the Brans-Dicke $\omega$ parameter
equal to zero, where the parameter $\omega$ is defined in the
usual way in the Brans-Dicke action as $S_{BD} = \int d^4x
\sqrt{g} (\Phi R - \omega (\nabla \Phi)^2/\Phi)$. This theory is
in conflict with observational data as it stands, and hence is not
a realistic description of our Universe.

However, it is possible that non-linear effects, which we have
neglected throughout this work, could play a role here. Indeed,
even in conventional General Relativity, nonlinearities start to
show up at distances comparable with the gravitational radii of
the sources. In brane induced gravity, and other frameworks that
strive to modify gravity in the IR, the nonlinear corrections will
become important even sooner than the nonlinearities in General
Relativity \cite{veinsh,strongcouplings,niges,strong,Luty,nira}.
Our results are found in linearized perturbation theory
around a near-critical vacuum, and so it is conceivable (although
not certain) that different strong coupling effects at higher
orders in perturbation theory might improve the behavior of long
range fields of masses far from the source but well below the
crossover scale. Alternatively, it is possible that additional
tweaks of the bulk theory, for example by curving the bulk
locally, could change how the helicity-0 mode couples to brane
matter. Such methods are widely used in the construction of string
landscapes \cite{landscape}, and might also be useful here.

We stress again that regardless of the ratio of $r_{\tt vac}/r_c$
the scalar couplings remain consistently {\it weak} for static
sources on near-critical branes with $b \rightarrow 1$. The static
solutions are governed by Euclidean
momenta, and so they are always finite in linearized theory
around near-critical vacua, in contrast to the fields of masses which
perturb generic sub-critical vacua. Thus linearized perturbation
theory remains under control on near-critical vacua. This as we
already mentioned does not guarantee that the linearized solutions
will remain dominant close to the masses that source the fields.
Further study of these issues, to ascertain how likely it is that
nonlinear phenomena may yield any additional screening of
helicity-0 modes, hence seems warranted. The near-critical vacua
at the very least provide us with a controllable new arena where
such phenomena could be studied.

The near-critical propagating solutions, on the other hand, also
remain under control, albeit their dynamics is more subtle. The
issue is that there may be new poles at Lorentzian momenta $k^2 =
- k_{\tt vac}^2 = - \frac{1-b}{r_0^2}$, in the solutions for $\xi$
and $X$, given respectively by Eqs. (\ref{xisoln}), (\ref{Xsoln})
and (\ref{Xgrsoln}), and their approximations below the crossover
scales. Clearly, these poles never play any role for static
configurations, which are controlled by Euclidean momenta for
which the fields remain finite. However one may worry if new
infinities could plague the linearized theory once a weak time
dependence is allowed. Now, on a generic sub-critical brane the
poles at $k^2 = - k^2_{\tt vac}$ reside practically on the $4D$
cutoff $\sim 1/r_0$,  deep in the strongly coupled regime of the
sub-critical linearized theory and hence can be completely ignored
at large distances as a UV mirage. On the other hand, in the
near-critical limit $b \rightarrow 1$ the scale $k_{\tt vac}$
becomes very low, and may be much lower than the crossover scale
$1/r_c$. However, as we will now explain, in this case these poles
are harmless because they merely point to a breakdown
of a gauge fixing of the linearized theory, which is fixed by an
interchange of scalar modes. The easiest way to see this is to
reconsider the formula for the $\xi$ field, Eq. (\ref{xisoln}),
and the boundary condition for $\Delta X'$, Eq.
(\ref{xbdysolved}), when Lorentzian momentum is $k^2 = - k^2_{\tt
vac} = - \frac{1-b}{r_0^2}$. Then, Eq. (\ref{xisoln}) indicates
that $\xi$ may diverge; however since the brane bending must be
bounded, this equation must be reinterpreted by multiplying it by
$1-b + r_0^2 k^2$ before setting $k^2 = - k^2_{\tt vac}$, and then
demanding that $\xi$ is finite, which means that the RHS must
vanish identically. This fixes $X$ on the brane to
\be X \bigg|_{k^2 = - k^2_{\tt vac}} = \frac{ 2r_0^2 } {3  M_5^3 (1-b  -
\frac{bM_6^4r_0}{M_5^3})} \,  \tau^\phi{}_\phi \bigg|_{k^2 = - k^2_{\tt vac}}
\, , \label{Xfix} \ee
while now $\xi$ is not determined by this boundary condition.
Since $X$ is now fixed by the matter source, it cannot be freely
chosen to satisfy the boundary condition for $\Delta X'$ in
(\ref{xbdysolved}). This is what fixes $\xi$: indeed, using
(\ref{xisoln}) and (\ref{brscals}) we can rewrite
(\ref{xbdysolved}) as
\be \Delta X'\bigg|_{k^2 = - k^2_{\tt vac}} =  \frac{M_5^3}{M_6^4} \Bigg[
\bigg(1-b - \frac{bM_6^4r_0}{M_5^3} \bigg) \, \frac{\xi}{2r^3_0}
- \frac{1-b}{1-b - \frac{bM_6^4r_0}{M_5^3} }\, \frac{{\tau}^\phi{}_{\phi}}{6M_5^3}
- \frac{{\tau}^\alpha{}_{\alpha}}{6M_5^3} \Bigg]\bigg|_{k^2 = - k^2_{\tt vac}}  \, ,
\label{xbdyforxi} \ee
so that for given sources and for $X$ fixed by (\ref{Xfix}) when
$k^2 = -\frac{1-b}{r_0^2}$ we can choose $\xi$ to satisfy
(\ref{xbdyforxi}). Therefore the Lorentzian poles at $k^2 = -
k^2_{\tt vac}$ are spurionic, and can be ignored in the
linearized perturbation theory on near-critical branes. The
linearized theory around the vacuum on such generic near-critical backgrounds remains
under control, without dramatic instabilities, and mimicking Brans-Dicke
theory with $\omega=0$, although we stress again that one has to
reevaluate it against the higher order corrections closer to the
source, because of the issues related to the Vainshtein scale and
strong coupling at higher orders.

\section{Conclusions}

Our main result is the observation that properly regulated brane
induced gravity theories, which yield calculable long distance
gravitational fields, really behave as a semiclassical landscape of vacua. Treating gravity 
classically, these models can impersonate $4D$ worlds between the UV cutoff that resolves
the core of the gravity-localizing defect and the crossover scale,
if their Scherk-Schwarz sector is carefully tuned to yield $4D$
Minkowski vacua. While we have explicitly worked with
codimension-2 defects, we feel that similar conclusions should
extend to any setups with codimension $\ge 2$. In the case of
codimension-2, we see that the brane vacua can remain completely
flat, while the bulk readjusts to absorb the tension as deficit
angle. Separating them there will be mismatched configurations,
where the tensional pressure in the compact direction may not be
precisely cancelled, which are either nonstationary, or never
approximate $4D$ behavior. Static supercritical branes
must be singular, just like supercritical local strings in $4D$
\cite{naked}. However in this case there should exist nonsingular
inflating solutions, that should really be the correct vacua for
supercritical branes.

We also find that the cosmological constant problem changes its
guise rather dramatically in codimension-2 setups. Although the
theory has many $4D$ flat vacua with Minkowski metric, the
effective $4D$ Planck scale, and for sub-critical tensions, the
crossover scale out to which gravity looks $4D$, are very
sensitive to the brane tension. These parameters depend on the
tension directly, as does the sub-critical crossover scale (see
Eq. (\ref{crosssum})) or through the compactification radius of
the wrapped 4-brane, as does the effective $4D$ Planck scale. So
if the theory is to mimic a weak $4D$ gravity over a large spatial
region, which looks flat and static, for a given brane and bulk
Planck scales $M_5$ and $M_6$ and a fixed brane radius $r_0$ one
must tune the brane tension precisely. Changes of brane tension
will generically lead to proportional changes of the crossover
scale and $4D$ gravitational coupling, requiring appropriate
retunings. However in the near-critical limit, where the bulk
compactifies to a cone so that $4D$ gravity first changes to a
$5D$ one, the crossover scale saturates at a value completely
independent of the tension, $r_c = M_5^3/M_6^4$. To get this to be
of the order of the current horizon size, $H_0^{-1} \sim 10^{28}
{\rm cm}$, and ensure that the rings of matter on the brane look
pointlike at energies below a ${\rm TeV}$, one needs $M_6 \ga {\rm
TeV}$, and therefore $M_5 \la 10^{19} {\rm GeV}$. Further to
reproduce the $4D$ Newton's constant below the crossover scale,
one needs to pick $r_0 \ga 10^{-19} {\rm GeV}^{-1}$. These numbers
are rather curious. Clearly, one must tune the theory to make sure
that the brane is very thin, with many hidden sector fields, to
get such hierarchies. We don't have much to add to the discussion
of how to do it, but merely note that this is in line with the
current philosophy of the brane induced gravity models \cite{DGP}.
While some work along the lines of embedding the theory into
string theory, that would provide the framework for its full
UV completion, has been pursued \cite{kiritsis,lowe,ignatios}, 
the existing constructions are really semiclassical models where
gravity is treated classically. 
Adopting this possibility, one may at least explore the low energy
consequences of such models.

Once we consider gravity of localized static masses on the brane,
we find that for sub-critical tensions there is a strong coupling
scale in the vacuum itself, which is essentially the same as the
crossover scale. Thus in the regime where gravity may appear $4D$,
linearized perturbation theory does not apply because the scalar
modes are strongly coupled. However in the near-critical limit the
linearized perturbation theory around the vacuum is under control.
We find that it does not contain instabilities, and generates long
range fields which are, to the leading order, the same as in the
Brans-Dicke theory with $\omega=0$. This would be in conflict with
tests of gravity in Solar System and beyond. However, throughout
this work we have taken the bulk to be locally flat. Adding bulk
fields and curvature will increase the diversity of possible
solutions and might yield additional effects screening the
helicity-0 mode in the spin-2 sector, perhaps similarly to what
happens in string landscape constructions \cite{landscape}.
Moreover, the nonlinear corrections might play some role too as
advocated in \cite{veinsh,strongcouplings}.

This discussion shows that many questions remain open. Clearly,
the most interesting phenomenological issues concern the
stabilization or decoupling of the helicity-0 modes. It is
interesting to explore the theory for dangerous instabilities
beyond the linear order. It also remains to see if the brane
induced gravity models can be consistently derived from
microscopic models that admit plausible UV completions, generating
$M_4 \gg M_6$. In other words, do such landscapes even exist
beyond a classical action, that one can write for them? Finally,
it would be interesting to see if the new guise of the
cosmological constant problem, and in particular the serendipitous
insensitivity of the crossover scale from brane tension in the
near-critical limit, might yield some new avenues for going around
the venerated Weinberg no-go theorem \cite{Wein}, that is still
the main obstruction to having a mechanism for protecting a small
$4D$ vacuum curvature from quantum radiative corrections. These
issues are outside of the scope of the present work, and we can
only hope that our sketch of such a diverse new landscape with a
different manifestation of the vacuum energy problem may motivate
the search for their resolutions.

\vskip1.5cm

{\bf \noindent Acknowledgements}

\vskip.3cm

We would like to thank Alberto Iglesias, Robert Myers, Minjoon Park, Oriol
Pujolas and Lorenzo Sorbo for interesting discussions. N.K. is
grateful to Galileo Galilei Institute, Florence, Italy, for kind
hospitality in the course of this work. This work was supported in
part by the DOE Grant DE-FG03-91ER40674, in part by the NSF Grant
PHY-0332258 and in part by a Research Innovation Award from the
Research Corporation.

\pagebreak


\begin{thebibliography}{99}

%\cite{Weinberg:1988cp}
\bibitem{Wein}
S.~Weinberg,
%``The cosmological constant problem,''
Rev.\ Mod.\ Phys.\  {\bf 61} , 1 (1989).
%%CITATION = RMPHA,61,1;%%

\bibitem{sne}
A.~G.~Riess {\it et al.},
%``Observational Evidence from Supernovae for an Accelerating Universe
%and a Cosmological Constant,''
Astron.\ J.\  {\bf 116}, 1009 (1998);
%{\tt [astro-ph/9805201]};
%%CITATION = ASTRO-PH 9805201;%%
S.~Perlmutter {\it et al.},
%``Measurements of Omega and Lambda from 42 High-Redshift Supernovae,''
Astrophys.\ J.\  {\bf 517}, 565 (1999);
%{\tt [astro-ph/9812133]}.
%%CITATION = ASTRO-PH 9812133;%%
J.~L.~Tonry {\it et al.},
%``Cosmological Results from High-z Supernovae,''
Astrophys.\ J.\  {\bf 594}, 1  (2003);
% [arXiv:astro-ph/0305008].
%%CITATION = ASTRO-PH 0305008;%%
R.~A.~Knop {\it et al.},
%``New Constraints on $\Omega_M$, $\Omega_\Lambda$, and w from an Independent
%Set of Eleven High-Redshift Supernovae Observed with HST,''
Astrophys.\ J.\  {\bf 598}, 102 (2003);
%% [arXiv:astro-ph/0309368]
%%CITATION = ASTRO-PH 0309368;%%
%%CITATION = ASTRO-PH 0309368;%%
A.~G.~Riess {\it et al.},  %[Supernova Search Team Collaboration],
%``Type Ia Supernova Discoveries at z>1 From the Hubble Space Telescope:
%Evidence for Past Deceleration and Constraints on Dark Energy Evolution,''
Astrophys.\ J.\  {\bf 607}, 665 (2004).
%  [arXiv:astro-ph/0402512].
%%CITATION = ASTRO-PH 0402512;%%

\bibitem{landscape}
%\cite{Bousso:2000xa}
%\bibitem{Bousso:2000xa}
R.~Bousso and J.~Polchinski,
%``Quantization of four-form fluxes and dynamical neutralization of the
%cosmological constant,''
JHEP {\bf 0006}, 006 (2000);
%[arXiv:hep-th/0004134].
%%CITATION = HEP-TH 0004134;%%
%\cite{Kachru:2003aw}
%\bibitem{Kachru:2003aw}
S.~Kachru, R.~Kallosh, A.~Linde and S.~P.~Trivedi,
%``De Sitter vacua in string theory,''
Phys.\ Rev.\ D {\bf 68}, 046005  (2003).
%  [arXiv:hep-th/0301240].
%%CITATION = HEP-TH 0301240;%%

%\cite{Linde:1984ir}
\bibitem{andrei}
A.~D.~Linde,
%``The Inflationary Universe,''
Rept.\ Prog.\ Phys.\  {\bf 47}, 925 (1984).
%%CITATION = RPPHA,47,925;%%

%\cite{Weinberg:1987dv}
\bibitem{weinanthr}
S.~Weinberg,
%``ANTHROPIC BOUND ON THE COSMOLOGICAL CONSTANT,''
Phys.\ Rev.\ Lett.\  {\bf 59}, 2607 (1987).
%%CITATION = PRLTA,59,2607;%%
%\cite{Vilenkin:1994ua}

\bibitem{vilenkin}
A.~Vilenkin,
%``Predictions from quantum cosmology,''
Phys.\ Rev.\ Lett.\  {\bf 74}, 846  (1995).
% [arXiv:gr-qc/9406010].
%%CITATION = GR-QC 9406010;%%

%\cite{Feng:2000if}
\bibitem{jmr}
J.~L.~Feng, J.~March-Russell, S.~Sethi and F.~Wilczek,
%``Saltatory relaxation of the cosmological constant,''
Nucl.\ Phys.\ B {\bf 602}, 307 (2001).
% [arXiv:hep-th/0005276].
%%CITATION = HEP-TH 0005276;%%

%\cite{Susskind:2003kw}
\bibitem{lenny}
L.~Susskind,
%``The anthropic landscape of string theory,''
{\tt  arXiv:hep-th/0302219}.
%%CITATION = HEP-TH 0302219;%%
%\cite{Banks:2003es}

\bibitem{tom}
T.~Banks, M.~Dine and E.~Gorbatov,
%``Is there a string theory landscape?,''
JHEP {\bf 0408}, 058 (2004).
%[arXiv:hep-th/0309170].
%%CITATION = HEP-TH 0309170;%%

%\cite{Rubakov:1983bz}
\bibitem{rusha}
V.~A.~Rubakov and M.~E.~Shaposhnikov,
%``Extra Space-Time Dimensions: Towards A Solution To The Cosmological
%Constant Problem,''
Phys.\ Lett.\ B {\bf 125}, 139 (1983).
%%CITATION = PHLTA,B125,139;%%

%\cite{Holdom:1983jh}
\bibitem{Holdom}
B.~Holdom, %``The Cosmological Constant And The Embedded Universe,''
ITP-744-STANFORD preprint, 1983.
%\href{http://www.slac.stanford.edu/spires/find/hep/www?r=itp-744-stanford}{SPIRES entry}

\bibitem{selftun}
%\cite{Arkani-Hamed:2000eg}
%\bibitem{Arkani-Hamed:2000eg}
N.~Arkani-Hamed, S.~Dimopoulos, N.~Kaloper and R.~Sundrum,
%``A small cosmological constant from a large extra dimension,''
Phys.\ Lett.\ B {\bf 480}, 193 (2000);
%[arXiv:hep-th/0001197].
%%CITATION = HEP-TH 0001197;%%
S.~Kachru, M.~B.~Schulz and E.~Silverstein,
%``Self-tuning flat domain walls in 5d gravity and string theory,''
Phys.\ Rev.\ D {\bf 62}, 045021 (2000).
%[arXiv:hep-th/0001206].
%%CITATION = HEP-TH 0001206;%%

\bibitem{nillest}
S.~Forste, Z.~Lalak, S.~Lavignac and H.~P.~Nilles,
%``A comment on self-tuning and vanishing cosmological constant in the  brane
%world,''
Phys.\ Lett.\ B {\bf 481}, 360 (2000).
%  [arXiv:hep-th/0002164].
%%CITATION = HEP-TH 0002164;%%

\bibitem{polstra}
%\cite{Polchinski:2000uf}
J.~Polchinski and M.~J.~Strassler,
%``The string dual of a confining four-dimensional gauge theory,''
{\tt  arXiv:hep-th/0003136}.
%%CITATION = HEP-TH 0003136;%%

\bibitem{DGP}
%\cite{Dvali:2000hr}
G.~R.~Dvali, G.~Gabadadze and M.~Porrati,
%``4D gravity on a brane in 5D Minkowski space,''
Phys.\ Lett.\ B {\bf 485}, 208 (2000);
% [arXiv:hep-th/0005016].
%%CITATION = HEP-TH 0005016;%%
%\cite{Dvali:2000xg}
%\bibitem{gigagia}
G.~R.~Dvali and G.~Gabadadze,
%``Gravity on a brane in infinite-volume extra space,''
Phys.\ Rev.\ D {\bf 63}, 065007 (2001).
% [arXiv:hep-th/0008054].
%%CITATION = HEP-TH 0008054;%%

\bibitem{shocks}
N.~Kaloper,
%``Brane-induced gravity's shocks,''
Phys.\ Rev.\ Lett.\  {\bf 94}, 181601 (2005)
[Erratum-ibid.\  {\bf 95}, 059901 (2005)];
%  [arXiv:hep-th/0501028].
 %%CITATION = HEP-TH 0501028;%%
%``Gravitational shock waves and their scattering in brane-induced gravity,''
Phys.\ Rev.\ D {\bf 71}, 086003 (2005) [Erratum-ibid: {\bf D71}
(2005) 086003].
%[arXiv:hep-th/0502035].
%%CITATION = HEP-TH 0502035;%%

%\cite{vanDam:1970vg}
\bibitem{vdvz}
%\cite{Iwasaki:1971uz}
Y.~Iwasaki,
%``Consistency Condition For Propagators,''
Phys.\ Rev.\ D {\bf 2}, 2255 (1970);
%%CITATION = PHRVA,D2,2255;%%
H.~van Dam and M.~J.~G.~Veltman,
%``Massive And Massless Yang-Mills And Gravitational Fields,''
Nucl.\ Phys.\ B {\bf 22}, 397 (1970);
%%CITATION = NUPHA,B22,397;%%
V.~I.~Zakharov, JETP Lett. {\bf 12} 312, (1970).

\bibitem{veinsh}
A.~I.~Vainshtein,
%``To The Problem Of non-vanishing Gravitation Mass,''
Phys.\ Lett.\ B {\bf 39} 393, (1972).
%%CITATION = PHLTA,B39,393;%%

\bibitem{strongcouplings}
C.~Deffayet, G.~R.~Dvali, G.~Gabadadze and A.~I.~Vainshtein,
%``Nonperturbative continuity in graviton mass versus perturbative
%discontinuity,''
Phys.\ Rev.\ D {\bf 65} 044026, (2002).
% [arXiv:hep-th/0106001].
%%CITATION = HEP-TH 0106001;%%

%\cite{Arkani-Hamed:2002sp}
\bibitem{niges}
N.~Arkani-Hamed, H.~Georgi and M.~D.~Schwartz,
%``Effective field theory for massive gravitons and gravity in theory space,''
Annals Phys.\  {\bf 305} 96, (2003).
% [arXiv:hep-th/0210184].
%%CITATION = HEP-TH 0210184;%%

\bibitem{strong}
%\cite{Rubakov:2003zb}
%\bibitem{Rubakov:2003zb}
V.~A.~Rubakov,
%``Strong coupling in brane-induced gravity in five dimensions,''
{\tt arXiv:hep-th/0303125}.
%%CITATION = HEP-TH 0303125;%%

\bibitem{Luty}
%\cite{Luty:2003vm}
M.~A.~Luty, M.~Porrati and R.~Rattazzi,
%``Strong interactions and stability in the DGP model,''
JHEP {\bf 0309}, 029 (2003).
%[arXiv:hep-th/0303116].
%%CITATION = HEP-TH 0303116;%%

%\cite{Nicolis:2004qq}
\bibitem{nira}
A.~Nicolis and R.~Rattazzi,
%``Classical and quantum consistency of the DGP model,''
JHEP {\bf 0406}, 059 (2004).
% [arXiv:hep-th/0404159].
%%CITATION = HEP-TH 0404159;%%

%\cite{Koyama:2005tx}
\bibitem{koyama}
K.~Koyama,
%``Are there ghosts in the self-accelerating brane universe?,''
Phys.\ Rev.\ D {\bf 72}, 123511 (2005);
% [arXiv:hep-th/0503191].
%%CITATION = HEP-TH 0503191;%%
%\cite{Gorbunov:2005zk}
D.~Gorbunov, K.~Koyama and S.~Sibiryakov,
%``More on ghosts in DGP model,''
Phys.\ Rev.\ D {\bf 73}, 044016 (2006).
 % [arXiv:hep-th/0512097].
%%CITATION = HEP-TH 0512097;%%

%\cite{Charmousis:2006pn}
\bibitem{cgkp}
C.~Charmousis, R.~Gregory, N.~Kaloper and A.~Padilla,
%``DGP specteroscopy,''
JHEP {\bf 0610}, 066 (2006).
%  [arXiv:hep-th/0604086].
%%CITATION = HEP-TH 0604086;%%
%\cite{Deffayet:2006wp}

\bibitem{giigl}
C.~Deffayet, G.~Gabadadze and A.~Iglesias,
%``Perturbations of self-accelerated universe,''
JCAP {\bf 0608}, 012 (2006).
%  [arXiv:hep-th/0607099].
%%CITATION = HEP-TH 0607099;%%

%\cite{Izumi:2006ca}
\bibitem{tanko}
K.~Izumi, K.~Koyama and T.~Tanaka,
%``Unexorcized ghost in DGP brane world,''
{\tt  arXiv:hep-th/0610282}.
%%CITATION = HEP-TH 0610282;%%
%\cite{Carena:2006yr}

\bibitem{minjoon}
M.~Carena, J.~Lykken, M.~Park and J.~Santiago,
%``Self-accelerating warped braneworlds,''
Phys.\ Rev.\  D {\bf 75}, 026009 (2007).
%[arXiv:hep-th/0611157].
%%CITATION = PHRVA,D75,026009;%%

%\cite{Dvali:2001ae}
\bibitem{highercod}
G.~Dvali, G.~Gabadadze, X.~r.~Hou and E.~Sefusatti,
%``See-saw modification of gravity,''
Phys.\ Rev.\ D {\bf 67}, 044019 (2003).
% [arXiv:hep-th/0111266].
%%CITATION = HEP-TH 0111266;%%

%\cite{Dubovsky:2002jm}
\bibitem{durub}
S.~L.~Dubovsky and V.~A.~Rubakov,
%``Brane-induced gravity in more than one extra dimensions: Violation of
%equivalence principle and ghost,''
Phys.\ Rev.\ D {\bf 67}, 104014 (2003).
%[arXiv:hep-th/0212222].
%%CITATION = HEP-TH 0212222;%%

%\cite{Kolanovic:2003am}
\bibitem{koporo}
M.~Kolanovic, M.~Porrati and J.~W.~Rombouts,
%``Regularization of brane induced gravity,''
Phys.\ Rev.\ D {\bf 68}, 064018 (2003).
%  [arXiv:hep-th/0304148].
%%CITATION = HEP-TH 0304148;%%

%\cite{Gabadadze:2003ck}
\bibitem{gashif}
G.~Gabadadze and M.~Shifman,
%``Softly massive gravity,''
Phys.\ Rev.\ D {\bf 69}, 124032 (2004).
%  [arXiv:hep-th/0312289].
%%CITATION = HEP-TH 0312289;%%

%\cite{Sundrum:1998ns}
\bibitem{raman}
R.~Sundrum,
%``Compactification for a three-brane universe,''
Phys.\ Rev.\ D {\bf 59}, 085010 (1999).
% [arXiv:hep-ph/9807348].
%%CITATION = HEP-PH 9807348;%%

%\cite{Nilles:2003km}
\bibitem{nilles}
H.~P.~Nilles, A.~Papazoglou and G.~Tasinato,
%``Selftuning and its footprints,''
Nucl.\ Phys.\ B {\bf 677}, 405 (2004);
%  [arXiv:hep-th/0309042].
%%CITATION = HEP-TH 0309042;%%
%\cite{Graesser:2004xv}
%\bibitem{graesser}
M.~L.~Graesser, J.~E.~Kile and P.~Wang,
%``Gravitational perturbations of a six dimensional self-tuning model,''
Phys.\ Rev.\ D {\bf 70}, 024008 (2004);
%[arXiv:hep-th/0403074].
%%CITATION = HEP-TH 0403074;%%
%\cite{Garriga:2004tq}
%\bibitem{massimo}
J.~Garriga and M.~Porrati,
%``Football shaped extra dimensions and the absence of self-tuning,''
JHEP {\bf 0408}, 028 (2004);
%[arXiv:hep-th/0406158].
%%CITATION = HEP-TH 0406158;%%
%\cite{Redi:2004tm}
M.~Redi,
%``Footballs, conical singularities and the Liouville equation,''
Phys.\ Rev.\ D {\bf 71}, 044006 (2005).
%[arXiv:hep-th/0412189].
%%CITATION = HEP-TH 0412189;%%
%%Cited 8 times in SPIRES-HEP

%\cite{Ponton:2000gi}
\bibitem{popo}
A.~Chodos and E.~Poppitz,
%``Warp factors and extended sources in two transverse dimensions,''
Phys.\ Lett.\ B {\bf 471}, 119 (1999);
%[arXiv:hep-th/9909199].
%%CITATION = HEP-TH 9909199;%%
E.~Ponton and E.~Poppitz,
%``Gravity localization on string-like defects in codimension two and the
%AdS/CFT correspondence,''
JHEP {\bf 0102}, 042 (2001).
%[arXiv:hep-th/0012033].
%%CITATION = HEP-TH 0012033;%%

%\cite{Cohen:1999ia}
\bibitem{cohkap}
A.~G.~Cohen and D.~B.~Kaplan,
%``Solving the hierarchy problem with noncompact extra dimensions,''
Phys.\ Lett.\ B {\bf 470}, 52 (1999).
%[arXiv:hep-th/9910132].
%%CITATION = HEP-TH 9910132;%%

%\cite{Gregory:1999gv}
\bibitem{ruth}
R.~Gregory,
%``Nonsingular global string compactifications,''
Phys.\ Rev.\ Lett.\  {\bf 84}, 2564 (2000);
% [arXiv:hep-th/9911015].
%%CITATION = HEP-TH 9911015;%%
%\cite{Gregory:2003xf}
%\bibitem{Gregory:2003xf}
% R.~Gregory,
%``Inflating p-branes,''
JHEP {\bf 0306}, 041 (2003).
%[arXiv:hep-th/0304262].
%%CITATION = HEP-TH 0304262;%%

%\cite{Chen:2000at}
\bibitem{luty}
J.~W.~Chen, M.~A.~Luty and E.~Ponton,
%``A critical cosmological constant from millimeter extra dimensions,''
JHEP {\bf 0009}, 012 (2000).
%[arXiv:hep-th/0003067].
%%CITATION = HEP-TH 0003067;%%

%\cite{Gherghetta:2000qi}
\bibitem{ghesha}
T.~Gherghetta and M.~E.~Shaposhnikov,
%``Localizing gravity on a string-like defect in six dimensions,''
Phys.\ Rev.\ Lett.\  {\bf 85}, 240 (2000);
%[arXiv:hep-th/0004014].
%%CITATION = HEP-TH 0004014;%%
%\cite{Gherghetta:2000jf}
%\bibitem{Gherghetta:2000jf}
T.~Gherghetta, E.~Roessl and M.~E.~Shaposhnikov,
%``Living inside a hedgehog: Higher-dimensional solutions that localize
%gravity,''
Phys.\ Lett.\ B {\bf 491}, 353 (2000).
%[arXiv:hep-th/0006251].
%%CITATION = HEP-TH 0006251;%%

\bibitem{ahddk}
N.~Arkani-Hamed, S.~Dimopoulos, G.~R.~Dvali and N.~Kaloper,
%``Infinitely large new dimensions,''
Phys.\ Rev.\ Lett.\  {\bf 84}, 586 (2000).
% [arXiv:hep-th/9907209].
%%CITATION = HEP-TH 9907209;%%

\bibitem{origami}
N.~Kaloper,
%``Origami world,''
JHEP {\bf 0405}, 061 (2004);
% [arXiv:hep-th/0403208].
%%CITATION = HEP-TH 0403208;%%
%``Origami world,''
AIP Conf.\ Proc.\  {\bf 743}, 318 (2005).
%%CITATION = APCPC,743,318;%%

%\cite{Kanti:2001vb}
\bibitem{kmo}
P.~Kanti, R.~Madden and K.~A.~Olive,
%``A 6-D brane world model,''
Phys.\ Rev.\ D {\bf 64}, 044021 (2001).
% [arXiv:hep-th/0104177].
%%CITATION = HEP-TH 0104177;%%

%\cite{Aghababaie:2003ar}
\bibitem{msled}
Y.~Aghababaie {\it et al.},
%``Warped brane worlds in six dimensional supergravity,''
JHEP {\bf 0309}, 037 (2003);
%[arXiv:hep-th/0308064].
%%CITATION = HEP-TH 0308064;%%
%\cite{Burgess:2004yq}
%\bibitem{Burgess:2004yq}
C.~P.~Burgess, J.~Matias and F.~Quevedo,
%``MSLED: A minimal supersymmetric large extra dimensions scenario,''
Nucl.\ Phys.\ B {\bf 706}, 71 (2005).
%[arXiv:hep-ph/0404135].
%%CITATION = HEP-PH 0404135;%%

%\cite{Cline:2003ak}
\bibitem{cline2}
J.~M.~Cline, J.~Descheneau, M.~Giovannini and J.~Vinet,
%``Cosmology of codimension-two braneworlds,''
JHEP {\bf 0306}, 048 (2003).
%[arXiv:hep-th/0304147].
%%CITATION = HEP-TH 0304147;%%

%\cite{Bostock:2003cv}
\bibitem{bgsn}
O.~Corradini, A.~Iglesias, Z.~Kakushadze and P.~Langfelder,
%``Gravity on a 3-brane in 6D bulk,''
Phys.\ Lett.\ B {\bf 521}, 96 (2001);
%[arXiv:hep-th/0108055].
%%CITATION = HEP-TH 0108055;%%
P.~Bostock, R.~Gregory, I.~Navarro and J.~Santiago,
%``Einstein gravity on the codimension 2 brane?,''
Phys.\ Rev.\ Lett.\  {\bf 92}, 221601 (2004).
%[arXiv:hep-th/0311074].
%%CITATION = HEP-TH 0311074;%%

%\cite{Vinet:2004bk}
\bibitem{clinvin}
J.~Vinet and J.~M.~Cline,
%``Can codimension-two branes solve the cosmological constant problem?,''
Phys.\ Rev.\ D {\bf 70}, 083514 (2004);
%[arXiv:hep-th/0406141].
%%CITATION = HEP-TH 0406141;%%
%``Codimension-two branes in six-dimensional supergravity and the cosmological
%constant problem,''
Phys.\ Rev.\ D {\bf 71}, 064011 (2005).
%[arXiv:hep-th/0501098].
%%CITATION = HEP-TH 0501098;%%

%\cite{deRham:2005ci}
\bibitem{derhato}
C.~de Rham and A.~J.~Tolley,
%``Gravitational waves in a codimension two braneworld,''
JCAP {\bf 0602}, 003 (2006).
%[arXiv:hep-th/0511138].
%%CITATION = JCAPA,0602,003;%%

%\cite{Tolley:2005nu}
\bibitem{tbha}
 A.~J.~Tolley, C.~P.~Burgess, D.~Hoover and Y.~Aghababaie,
%``Bulk singularities and the effective cosmological constant for higher
%co-dimension branes,''
JHEP {\bf 0603}, 091 (2006)
%  [arXiv:hep-th/0512218].
%%CITATION = JHEPA,0603,091;%%

%\cite{Navarro:2004di}
\bibitem{nasa}
H.~M.~Lee and G.~Tasinato,
%``Cosmology of intersecting brane world models in Gauss-Bonnet gravity,''
JCAP {\bf 0404}, 009 (2004);
%[arXiv:hep-th/0401221].
%%CITATION = HEP-TH 0401221;%%
I.~Navarro and J.~Santiago,
%``Gravity on codimension 2 brane worlds,''
JHEP {\bf 0502}, 007 (2005);
%[arXiv:hep-th/0411250].
%%CITATION = HEP-TH 0411250;%%
%\cite{Papantonopoulos:2005ma}
%\bibitem{papani}
E.~Papantonopoulos and A.~Papazoglou,
%``Brane-bulk matter relation for a purely conical codimension-2 brane
%world,''
JCAP {\bf 0507}, 004 (2005);
%[arXiv:hep-th/0501112].
%%CITATION = HEP-TH 0501112;%%
%\cite{Charmousis:2005ey}
%\bibitem{chaze}
C.~Charmousis and R.~Zegers,
%``Matching conditions for a brane of arbitrary codimension,''
JHEP {\bf 0508}, 075 (2005);
%[arXiv:hep-th/0502170].
%%CITATION = HEP-TH 0502170;%%
%\cite{Charmousis:2005ez}
%\bibitem{Charmousis:2005ez}
% C.~Charmousis and R.~Zegers,
%``Einstein gravity on an even codimension brane,''
Phys.\ Rev.\ D {\bf 72}, 064005 (2005);
% [arXiv:hep-th/0502171].
%%CITATION = HEP-TH 0502171;%%
%\cite{Kofinas:2005py}
%\bibitem{Kofinas:2005py}
G.~Kofinas,
%``On braneworld cosmologies from six dimensions, and absence thereof,''
Phys.\ Lett.\  B {\bf 633}, 141 (2006).
%[arXiv:hep-th/0506035].
%%CITATION = PHLTA,B633,141;%%

%\cite{Kaloper:2006ek}
\bibitem{kalkil}
N.~Kaloper and D.~Kiley,
%``Exact black holes and gravitational shock waves on codimension-2 branes,''
JHEP {\bf 0603}, 077 (2006).
% [arXiv:hep-th/0601110].
%%CITATION = HEP-TH 0601110;%%

%\cite{Peloso:2006cq}
\bibitem{pesota}
M.~Peloso, L.~Sorbo and G.~Tasinato,
%``Standard 4d gravity on a brane in six dimensional flux compactifications,''
Phys.\ Rev.\ D {\bf 73}, 104025 (2006).
%  [arXiv:hep-th/0603026].
%%CITATION = HEP-TH 0603026;%%

\bibitem{papakoba}
%\cite{Papantonopoulos:2006dv}
%\bibitem{Papantonopoulos:2006dv}
E.~Papantonopoulos, A.~Papazoglou and V.~Zamarias,
%``Regularization of conical singularities in warped six-dimensional
%compactifications,''
{\tt arXiv:hep-th/0611311};
 %%CITATION = HEP-TH/0611311;%%
T.~Kobayashi and M.~Minamitsuji,
%``Gravity on an extended brane in six-dimensional warped flux
%compactifications,''
{\tt  arXiv:hep-th/0703029}.
  %%CITATION = HEP-TH/0703029;%%

%\cite{Scherk:1978ta}
\bibitem{schsch}
J.~Scherk and J.~H.~Schwarz,
%``Spontaneous Breaking Of Supersymmetry Through Dimensional Reduction,''
Phys.\ Lett.\ B {\bf 82}, 60 (1979);
%%CITATION = PHLTA,B82,60;%%
%\cite{Scherk:1979zr}
%\bibitem{Scherk:1979zr}
%J.~Scherk and J.~H.~Schwarz,
%``How To Get Masses From Extra Dimensions,''
Nucl.\ Phys.\ B {\bf 153}, 61 (1979).
%%CITATION = NUPHA,B153,61;%%

\bibitem{naked}
%\cite{Gott:1984ef}
%\bibitem{Gott:1984ef}
J.~R.~I.~Gott,
%``Gravitational lensing effects of vacuum strings: Exact solutions,''
Astrophys.\ J.\  {\bf 288} 422, (1985);
%%CITATION = ASJOA,288,422;%%
%\cite{Laguna:1989rx}
%\bibitem{Laguna:1989rx}
P.~Laguna and D.~Garfinkle,
%``SPACE-TIME OF SUPERMASSIVE U(1) GAUGE COSMIC STRINGS,''
Phys.\ Rev.\ D {\bf 40}, 1011 (1989);
%%CITATION = PHRVA,D40,1011;%%
%\cite{Linet:1990fk}
%\bibitem{Linet:1990fk}
B.~Linet,
%``On the supermassive U(1) gauge cosmic strings,''
Class.\ Quant.\ Grav.\  {\bf 7}, L75 (1990);
%%CITATION = CQGRD,7,L75;%%
%\cite{Ortiz:1990tn}
%\bibitem{Ortiz:1990tn}
M.~E.~Ortiz,
%``A New look at supermassive cosmic strings,''
Phys.\ Rev.\ D {\bf 43}, 2521 (1991);
%%CITATION = PHRVA,D43,2521;%%
%\cite{Cho:1998xy}
%\bibitem{Cho:1998xy}
I.~Cho,
%``Inflation and nonsingular spacetimes of cosmic strings,''
Phys.\ Rev.\ D {\bf 58}, 103509 (1998).
%[arXiv:gr-qc/9804086].
%%CITATION = GR-QC 9804086;%%

%\cite{Gell-Mann:1984mu}
\bibitem{gell}
M.~Gell-Mann and B.~Zwiebach,
%``Curling Up Two Spatial Dimensions With SU(1,1) / U(1),''
Phys.\ Lett.\ B {\bf 147}, 111 (1984);
%%CITATION = PHLTA,B147,111;%%%\cite{Gell-Mann:1985if}
%\bibitem{Gell-Mann:1985if}
%M.~Gell-Mann and B.~Zwiebach,
%``Dimensional Reduction Of Space-Time Induced By Nonlinear Scalar Dynamics
%And Noncompact Extra Dimensions,''
Nucl.\ Phys.\ B {\bf 260}, 569 (1985).
%%CITATION = NUPHA,B260,569;%%

%\cite{Dvali:2006if}
\bibitem{oriol}
G.~Dvali, G.~Gabadadze, O.~Pujolas and R.~Rahman,
%``Domain walls as probes of gravity,''
{\tt arXiv:hep-th/0612016}.
%%CITATION = HEP-TH 0612016;%%

\bibitem{aichsexl}
P.~C.~Aichelburg and R.~U.~Sexl,
%``On The Gravitational Field Of A Massless Particle,''
Gen.\ Rel.\ Grav.\ {\bf 2}, 303 (1971).
%%CITATION = GRGVA,2,303;%%


\bibitem{higherdwaves}
V.~Ferrari, P.~Pendenza and G.~Veneziano,
%``Beamlike Gravitational Waves And Their Geodesics,''
Gen.\ Rel.\ Grav.\  {\bf 20}, 1185 (1988);
%%CITATION = GRGVA,20,1185;%%
%\bibitem{devega}
H.~de Vega and N.~Sanchez,
%``Quantum String Scattering In The Aichelburg-Sexl Geometry,''
Nucl.\ Ph.\ B {\bf 317}, 706 (1989).
%%CITATION = NUPHA,B317,706;%%

\bibitem{kalwall}
N.~Kaloper, {\tt arXiv:hep-th/0702206}.

\bibitem{kiritsis}
%\cite{Kiritsis:2001bc}
E.~Kiritsis, N.~Tetradis and T.~N.~Tomaras,
%``Thick branes and 4D gravity,''
JHEP {\bf 0108}, 012 (2001).
%[arXiv:hep-th/0106050].
%%CITATION = JHEPA,0108,012;%%
  
%\cite{Corley:2001hg}
\bibitem{lowe}
S.~Corley, D.~A.~Lowe and S.~Ramgoolam,
%``Einstein-Hilbert action on the brane for the bulk graviton,''
JHEP {\bf 0107}, 030 (2001).
%[arXiv:hep-th/0106067].
%%CITATION = JHEPA,0107,030;%%
    
%\cite{Antoniadis:2002tr}
\bibitem{ignatios}
I.~Antoniadis, R.~Minasian and P.~Vanhove,
%``Non-compact Calabi-Yau manifolds and localized gravity,''
Nucl.\ Phys.\  B {\bf 648}, 69 (2003).
% [arXiv:hep-th/0209030].
%%CITATION = NUPHA,B648,69;%%
    
%\cite{Kaloper:2005wq}
\bibitem{kalsorbo}
N.~Kaloper and L.~Sorbo,
%``Locally localized gravity: The inside story,''
JHEP {\bf 0508}, 070 (2005).
 % [arXiv:hep-th/0507191].
%%CITATION = HEP-TH 0507191;%%

\bibitem{weinbook}
S.~Weinberg, {\it Gravitation and cosmology}, J. Wiley and Sons, 1972 (see sec. 13.5 for
the discussion of maximally symmetric subspaces, and sec. 9.9 for PPN limit in Brans-Dicke).

\bibitem{thooft}
T.~Dray and G.~'t Hooft,
%``The Gravitational Shock Wave Of A Massless Particle,''
Nucl.\ Phys.\ B {\bf 253}, 173 (1985);
%%CITATION = NUPHA,B253,173;%%
%T.~Dray and G.~'t Hooft,
%``The Gravitational Effect Of Colliding Planar Shells Of Matter,''
Class.\ Quant.\ Grav.\  {\bf 3}, 825 (1986).
%%CITATION = CQGRD,3,825;%%

\bibitem{kostas}
K.~Sfetsos,
%``On gravitational shock waves in curved space-times,''
Nucl.\ Phys.\ B {\bf 436}, 721 (1995).
% [arXiv:hep-th/9408169].
%%CITATION = HEP-TH 9408169;%%

%\cite{Garriga:1999yh}
\bibitem{gata}
J.~Garriga and T.~Tanaka,
%``Gravity in the brane-world,''
Phys.\ Rev.\ Lett.\  {\bf 84}, 2778 (2000).
%  [arXiv:hep-th/9911055].
%%CITATION = PRLTA,84,2778;%%

\bibitem{takahiro}
%\cite{Tanaka:2003zb}
%\bibitem{Tanaka:2003zb}
T.~Tanaka,
%``Weak gravity in DGP braneworld model,''
Phys.\ Rev.\  D {\bf 69}, 024001 (2004).
%  [arXiv:gr-qc/0305031].
%%CITATION = PHRVA,D69,024001;%%

\bibitem{johncsaba}
%\cite{Cacciapaglia:2006mz}
%\bibitem{Cacciapaglia:2006mz}
G.~Cacciapaglia, C.~Csaki, G.~Marandella and J.~Terning,
%``The gaugephobic Higgs,''
JHEP {\bf 0702}, 036 (2007).
%  [arXiv:hep-ph/0611358].
%%CITATION = JHEPA,0702,036;%%

\end{thebibliography}
\end{document}